\documentclass[iop]{emulateapj}

\usepackage{natbib,aas_macros,amsmath}
\usepackage{multirow,color}
\usepackage{natbib}

\newcommand{\carcsec}{$\!\!\arcsec$}
\newcommand{\m}[1]{\mathrm{#1}}

\newcommand{\redc}[1]{\textcolor{black}{#1}}
\newcommand{\redcr}[1]{\textcolor{black}{#1}}
\newcommand{\redcrr}[1]{\textcolor{black}{#1}}

\begin{document}
\shortauthors{Harikane et al.}
\slugcomment{ApJ in Press}

\shorttitle{ISM Properties of Ly$\m{\alpha}$ Emitters}

\title{
SILVERRUSH. V. Census of Ly$\alpha$, {\sc [Oiii]$\lambda$5007}, 
H$\alpha$, and {\sc [Cii]}158$\mu$\lowercase{m} Line Emission\\
with $\sim1000$ LAEs at $\lowercase{z}=4.9-7.0$ Revealed with Subaru/HSC
%
%
%
}

\email{hari@icrr.u-tokyo.ac.jp}
\author{
Yuichi Harikane\altaffilmark{1,2},
Masami Ouchi\altaffilmark{1,3},
Takatoshi Shibuya\altaffilmark{1,4},
Takashi Kojima\altaffilmark{1,2},
Haibin Zhang\altaffilmark{1,2},
Ryohei Itoh\altaffilmark{1,2},
Yoshiaki Ono\altaffilmark{1},
Ryo Higuchi\altaffilmark{1,2},
Akio K. Inoue\altaffilmark{5},
Jacopo Chevallard\altaffilmark{6,18},
Peter L. Capak\altaffilmark{7,8},
Tohru Nagao\altaffilmark{9},
Masato Onodera\altaffilmark{10},
Andreas L. Faisst\altaffilmark{8},
Crystal L. Martin\altaffilmark{11},
Michael Rauch\altaffilmark{12},
Gustavo A. Bruzual\altaffilmark{13},
Stephane Charlot\altaffilmark{14},
Iary Davidzon\altaffilmark{8},
Seiji Fujimoto\altaffilmark{1,15},
Miftahul Hilmi\altaffilmark{1,15},
Olivier Ilbert\altaffilmark{16},
Chien-Hsiu Lee\altaffilmark{10},
Yoshiki Matsuoka\altaffilmark{9},
John D. Silverman\altaffilmark{3},
and 
Sune Toft\altaffilmark{17}
}

\altaffiltext{1}{
Institute for Cosmic Ray Research, The University of Tokyo, 5-1-5 Kashiwanoha, Kashiwa, Chiba 277-8582, Japan
}
\altaffiltext{2}{
Department of Physics, Graduate School of Science, The University of Tokyo, 7-3-1 Hongo, Bunkyo, Tokyo, 113-0033, Japan
}
\altaffiltext{3}{
Kavli Institute for the Physics and Mathematics of the Universe (Kavli IPMU, WPI), University of Tokyo, Kashiwa, Chiba 277-8583, Japan
}
\altaffiltext{4}{
Department of Computer Sciences, Kitami Institute of Technology, 165 Koen-cho, Kitami, Hokkaido 090-8507, Japan
}
\altaffiltext{5}{
Department of Environmental Science and Technology, Faculty of Design Technology, Osaka Sangyo University, 3-1-1, Nagaito, Daito, Osaka 574-8530, Japan
}
\altaffiltext{6}{
Scientific Support Office, Directorate of Science and Robotic Exploration, ESA/ESTEC, Keplerlaan 1, 2201 AZ Noordwijk, The Netherlands
}
\altaffiltext{7}{
California Institute of Technology, MC 105-24, 1200 East California Blvd., Pasadena, CA 91125, USA
}
\altaffiltext{8}{
Infrared Processing and Analysis Center, California Institute of Technology, MC 100-22, 770 South Wilson Ave., Pasadena, CA 91125, USA
}
\altaffiltext{9}{
Research Center for Space and Cosmic Evolution, Ehime University, Bunkyo-cho, Matsuyama, Ehime 790-8577, Japan
}
\altaffiltext{10}{
Subaru Telescope, National Astronomical Observatory of Japan, 650 North A’ohoku Place, Hilo, HI 96720, USA
}
\altaffiltext{11}{
Department of Physics, University of California, Santa Barbara, CA, 93106, USA
}
\altaffiltext{12}{
Carnegie Observatories, 813 Santa Barbara Street, Pasadena, CA 91101, USA
}
\altaffiltext{13}{
Centro de Radioastronomia y Astrofisica (CRyA), UNAM, Campus Morelia Apartado Postal 3-72, 58089 Morelia, Michoacan, Mexico
}
\altaffiltext{14}{
Sorbonne Universites, UPMC-CNRS, UMR7095, Institut d’Astrophysique de Paris, F-75014, Paris, France
}
\altaffiltext{15}{
Department of Astronomy, Graduate School of Science, The University of Tokyo, 7-3-1 Hongo, Bunkyo, Tokyo 113-0033, Japan
}
\altaffiltext{16}{
Aix Marseille Univ, CNRS, LAM, Laboratoire d’Astrophysique de Marseille, Marseille, France
}
\altaffiltext{17}{
Cosmic Dawn Center (DAWN), Niels Bohr Institute, Juliane Mariesvej 30, DK-2100 Copenhagen, Denmark
}
\altaffiltext{18}{
ESA Research Fellow
}

\begin{abstract}
We investigate Ly$\alpha$, {\sc [Oiii]$\lambda$5007}, H$\alpha$, and {\sc [Cii]}158$\mu$\lowercase{m} emission from $1,124$ galaxies at $z=4.9-7.0$.
Our sample is composed of $1,092$ Ly$\m{\alpha}$ emitters (LAEs) at $z=4.9$, $5.7$, $6.6$, and $7.0$ identified by Subaru/Hyper Suprime-Cam (HSC) narrowband surveys covered by {\it Spitzer} large area survey with Subaru/HSC (SPLASH) and 34 galaxies at $z=5.148-7.508$ with deep ALMA {\sc [Cii]}158$\mu$m data in the literature.
Fluxes of strong rest-frame optical lines of [{\sc Oiii}] and H$\m{\alpha}$ (H$\beta$) are constrained by significant excesses found in the SPLASH $3.6$ and $4.5\ \m{\mu m}$ photometry.
At $z=4.9$, we find that the rest-frame H$\alpha$ equivalent width and  the Ly$\alpha$ escape fraction $f_{\rm Ly\alpha}$ positively correlate with the rest-frame Ly$\alpha$ equivalent width $EW^0_\m{Ly\alpha}$.
The $f_{\rm Ly\alpha}-EW^0_\m{Ly\alpha}$ correlation is similarly found at $z\sim 0-2$, suggesting no evolution of the correlation over $z\simeq 0-5$.
The typical ionizing photon production efficiency of LAEs is  $\m{log}\xi_\m{ion}/\m{[Hz\ erg^{-1}]}\simeq25.5$ significantly (60-100\%) higher than those of LBGs at a given UV magnitude.
At $z=5.7-7.0$, there exists an interesting turn-over trend that the [{\sc Oiii}]$/\m{H\alpha}$ flux ratio increases in $EW^0_\m{Ly\alpha}\simeq 0-30\ \m{\AA}$, and then decreases out to $EW^0_\m{Ly\alpha}\simeq130\ \m{\AA}$.
We also identify an anti-correlation between a [{\sc Cii}] luminosity to star-formation rate ratio ($L_\m{[CII]}/SFR$) and $EW^0_\m{Ly\alpha}$ at the $>99\%$ confidence level.
We carefully investigate physical origins of the correlations with stellar-synthesis and photoionization models, and find that a simple anti-correlation between $EW_\m{Ly\alpha}^0$ and metallicity explains self-consistently all of the correlations of Ly$\alpha$, H$\alpha$, [{\sc Oiii}]$/\m{H\alpha}$, and [{\sc Cii}] identified in our study, indicating detections of metal-poor ($\sim0.03 Z_\odot$) galaxies with $EW^0_\m{Ly\alpha}\simeq200\m{\AA}$. 
\end{abstract}

\keywords{%
galaxies: formation ---
galaxies: evolution ---
galaxies: high-redshift 
}

\section{Introduction}\label{ss_intro}
Probing physical conditions of the inter-stellar medium (ISM) is fundamental in understanding star formation and gas reprocessing in galaxies across cosmic time. 
Recent ALMA observations are uncovering interesting features of the ISM in high-redshift galaxies.
Early observations found surprisingly weak [{\sc Cii}]158$\mu\mathrm{m}$ emission in Ly$\alpha$ emitters (LAEs) at $z\sim6-7$ \citep[{\sc [Cii]} deficit; e.g.,][]{2013ApJ...778..102O,2014ApJ...792...34O,2015A&A...574A..19S,2015MNRAS.452...54M}.
On the other hand, recent studies detected strong [{\sc Cii}] emission in galaxies at $z=5-7$, whose [{\sc Cii}] luminosities are comparable to local star-forming galaxies \citep[e.g.,][]{2015Natur.522..455C,2016ApJ...829L..11P,2017ApJ...836L...2B}.
A theoretical study discusses that the {\sc [Cii]} deficit can be explained by very low metallicity ($0.05\ Z_\odot$) in the ISM \citep[][]{2015ApJ...813...36V,2017ApJ...846..105O}.
Thus estimating metallicities of the high-redshift galaxies is crucial to our understanding of the origin of the {\sc[Cii]} deficit.

The ISM property is also important for cosmic reionization.
Observations by the Planck satellite and high redshift UV luminosity functions (LFs) suggest that faint and abundant star-forming galaxies dominate the reionization process \citep[e.g.,][]{2015ApJ...802L..19R}.
Furthermore, \citet{2018ApJ...854...73I} claim that the ionizing photon budget of star-forming galaxies is sufficient for reionizing the universe with the escape fraction of ionizing photons of $f_\m{esc}=0.15^{+0.06}_{-0.02}$ and the faint limit of the UV LF of $M_\m{trunc}>-12.5$ for an assumed constant ionizing photon production efficiency of $\m{log}\xi_\m{ion}/[\m{Hz\ erg^{-1}}]=25.34$, which is the number of Lyman continuum photons per UV ($1500\ \m{\AA}$) luminosity \citep[see also][]{2016ApJ...829...99F}.
On the other hand, \citet{2015A&A...578A..83G} argue that faint AGNs are important contributors to the reionization from their estimates of number densities and ionizing emissivities \citep[c.f.,][]{2015ApJ...813L...8M,2018MNRAS.474.2904P}.
One caveat in these two contradictory results is that properties of ionizing sources (i.e., $f_\m{esc}$ and $\xi_\m{ion}$) are not guaranteed to be the same as the typically assumed values.
\redc{Various studies constrain ionizing photon production efficiencies of star forming galaxies to be $\m{log}\xi_\m{ion}/[\m{Hz\ erg^{-1}}]=24.8-25.3$ at $z\sim0-2$ (e.g., \citealt{2017MNRAS.465.3637M,2017MNRAS.467.4118I,2017arXiv171100013S}; see also \citealt{2018MNRAS.tmp..369S})}.
Recently, \citet{2016ApJ...831..176B} report $\m{log}\xi_\m{ion}/[\m{Hz\ erg^{-1}}]=25.3-25.8$ for Lyman break galaxies (LBGs) at $z\sim4-5$, relatively higher than the canonical value \citep[i.e., 25.2;][]{2015ApJ...802L..19R}.
\citet{2016ApJ...831L...9N} also estimate $\xi_\m{ion}$ of 15 LAEs at $z=3.1-3.7$, which is $0.2-0.5\ \m{dex}$ higher than those of typical LBGs at similar redshifts.
Since the faint star-forming galaxies are expected to be strong line emitters, it is important to estimate $\xi_\m{ion}$ of LAEs at higher redshift, as their ISM properties are likely more similar to the ionizing sources.

\begin{figure*}
\begin{center}
  \includegraphics[clip,bb=40 30 600 325,width=0.9\hsize]{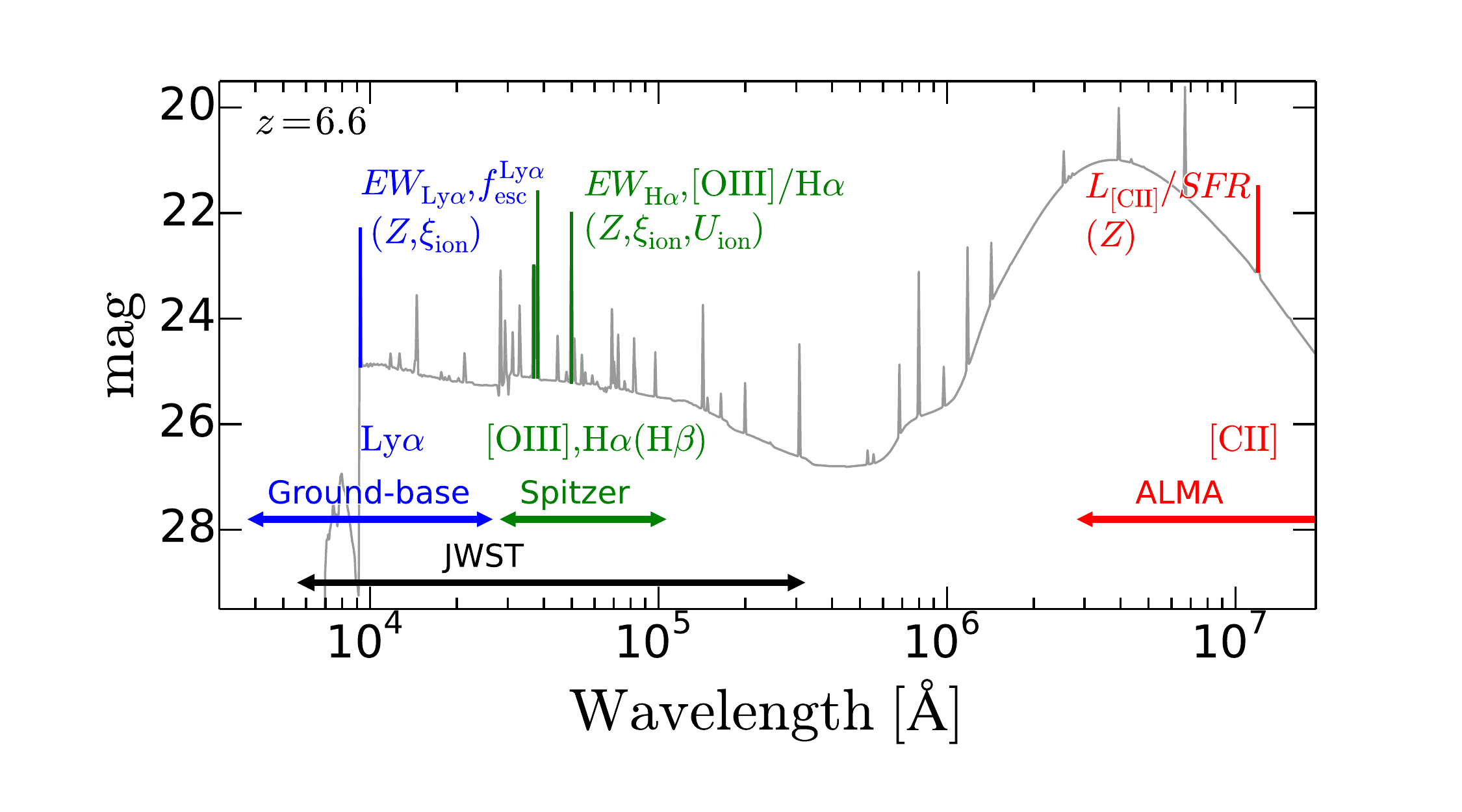}
 \end{center}
   \caption{Schematic view of the strategy of this study.
   We measure the Ly$\m{\alpha}$, {\sc[Oiii]}$\lambda$5007 and H$\alpha$ (H$\beta$), and {\sc [Cii]} emission line strengths to investigate the Ly$\alpha$ equivalent widths ($EW_\m{Ly\alpha}$) and Ly$\alpha$ escape fractions ($f_\m{esc}^\m{Ly\alpha}$), the H$\alpha$ equivalent widths ($EW_\m{H\alpha}$) and [{\sc Oiii}]$/\m{H\alpha}$ ratios, the ratios of the {[\sc Cii]} luminosity to SFR ($L_\m{[CII]}/SFR$), respectively.
   These quantities are related to the metallicity ($Z$), the ionizing photon production efficiency ($\xi_\m{ion}$), and the ionization parameter ($U_\m{ion}$).
   The redshifted wavelengths of the Ly$\alpha$, [{\sc Oiii}] and H$\alpha$ (H$\beta$), and [{\sc Cii}] emission lines are covered by ground-based telescopes (e.g., Subaru, VISTA, UKIRT), Spitzer, and ALMA, respectively (and in near future by JWST).
   \redc{The gray curve shows a model spectal energy distribution (SED) of a star forming galaxy with $\m{log}(Z_\m{neb}/Z_\odot)=-1.0$, $\m{log}U_\m{ion}=-2.4$, and $\m{log(Age/yr)}=8$ generated by BEAGLE (see Section \ref{ss_sed}).}
   \label{fig_sedEM}}
\end{figure*}

Metallicities and ionizing photon production efficiencies of galaxies can be estimated from rest-frame optical emission lines such as H$\m{\alpha}$, H$\m{\beta}$, [{\sc Oiii}]$\lambda\lambda$4959,5007, and [{\sc Oii}]$\lambda\lambda$3726,3729.
However at $z\gtrsim4$, some of these emission lines are redshifted into the mid-infrared, where they cannot be observed with ground-based telescopes.
Thus we need new future space telescopes (e.g., JWST) to investigate rest-frame optical emission lines of high redshift galaxies.
On the other hand, recent studies reveal that the redshifted emission lines significantly affect infrared broad-band photometry \citep[e.g.,][]{2013ApJ...763..129S,2014ApJ...784...58S,2015ApJ...801..122S,2016MNRAS.461.3886R,2016ApJ...821..122F,2016ApJ...823..143R,2017ApJ...839...73C}.
Thus, infrared broad-band photometry can be useful to estimate the rest-frame optical emission line fluxes which are not accessible with the ground-based telescopes before the JWST era.

The Subaru/Hyper Suprime-Cam Subaru strategic program (HSC-SSP) survey started in early 2014, and its first data release took place in 2017 February (\citealt{2012SPIE.8446E..0ZM} and  \citealt{2018PASJ...70S...8A,2018PASJ...70S...4A}; see also \citealt{Miyazaki2017hsc,Komiyama2017hsc,Furusawa2017hsc,Kawanomoto2017hsc}).
The HSC-SSP survey provides a large high-redshift galaxy sample, especially LAEs selected with the narrow-band (NB) filters.
The $NB816$ and $NB921$ imaging data are already taken in the HSC-SSP survey.
In addition, the $NB718$ and $NB973$ data are taken in the Cosmic HydrOgen Reionization Unveiled with Subaru (CHORUS) project (PI: A. K. Inoue; A. K. Inoue et al. in prep), which is an independent program of the HSC-SSP survey. 
{\it Spitzer} large area survey with Subaru/HSC (SPLASH; PI: P. Capak; P. Capak et al. in preparation.)\footnote{http://splash.caltech.edu} has obtained the {\it Spitzer}/Infrared Array Camera (IRAC) images overlapped with these NB data, which allow us to conduct statistical studies of the rest-frame optical emission in the high redshift LAEs.
Furthermore, the number of galaxies observed with ALMA is increasing, which will improve our understanding of the {\sc [Cii]} deficit.
Thus in this study, we investigate the ISM properties of high redshift galaxies by measuring the Ly$\alpha$, [{\sc Oiii}]$\lambda$5007, H$\alpha$, H$\beta$, and {[\sc Cii]} emission line strength (Figure \ref{fig_sedEM}).

This paper is one in a series of papers from twin programs devoted to scientific results on high redshift galaxies based on the HSC-SSP survey data.
One program is our LAE study with the large-area narrow-band images complemented by spectroscopic observations, named Systematic Identification of LAEs for Visible Exploration and Reionization Research Using Subaru HSC \citep[SILVERRUSH;][]{2018PASJ...70S..13O,2018PASJ...70S..14S,2018PASJ...70S..15S,2018PASJ...70S..16K,2018arXiv180100067I,2018arXiv180100531H}.
The other one is a luminous Lyman break galaxy (LBG) studies, named Great Optically Luminous Dropout Research Using Subaru HSC \citep[GOLDRUSH;][]{2018PASJ...70S..10O,2018PASJ...70S..11H,2018PASJ...70S..12T}.

This paper is organized as follows.
We present our sample and imaging datasets in Section \ref{ss_sample}, and describe methods to estimate line fluxes in Section \ref{ss_method}.
We show results in Section \ref{ss_result}, discuss our results in Section \ref{ss_discussion}, and summarize our findings in Section \ref{ss_summary}.
Throughout this paper we use the recent Planck cosmological parameter sets constrained with the temperature power spectrum, temperature-polarization cross spectrum, polarization power spectrum, low-$l$ polarization, CMB lensing, and external data \citep[TT, TE, EE+lowP+lensing+ext result; ][]{2016A&A...594A..13P}:
$\Omega_\m{m}=0.3089$, $\Omega_\Lambda=0.6911$, $\Omega_\m{b}=0.049$, $h=0.6774$, and $\sigma_8=0.8159$.
We assume a \citet{2003PASP..115..763C} initial mass function (IMF).
All magnitudes are in the AB system \citep{1983ApJ...266..713O}.

\section{Sample}\label{ss_sample}
\subsection{LAE Sample}\label{ss_laesample}
\begin{figure*}
\begin{center}
  \includegraphics[clip,bb=0 20 550 260,width=0.9\hsize]{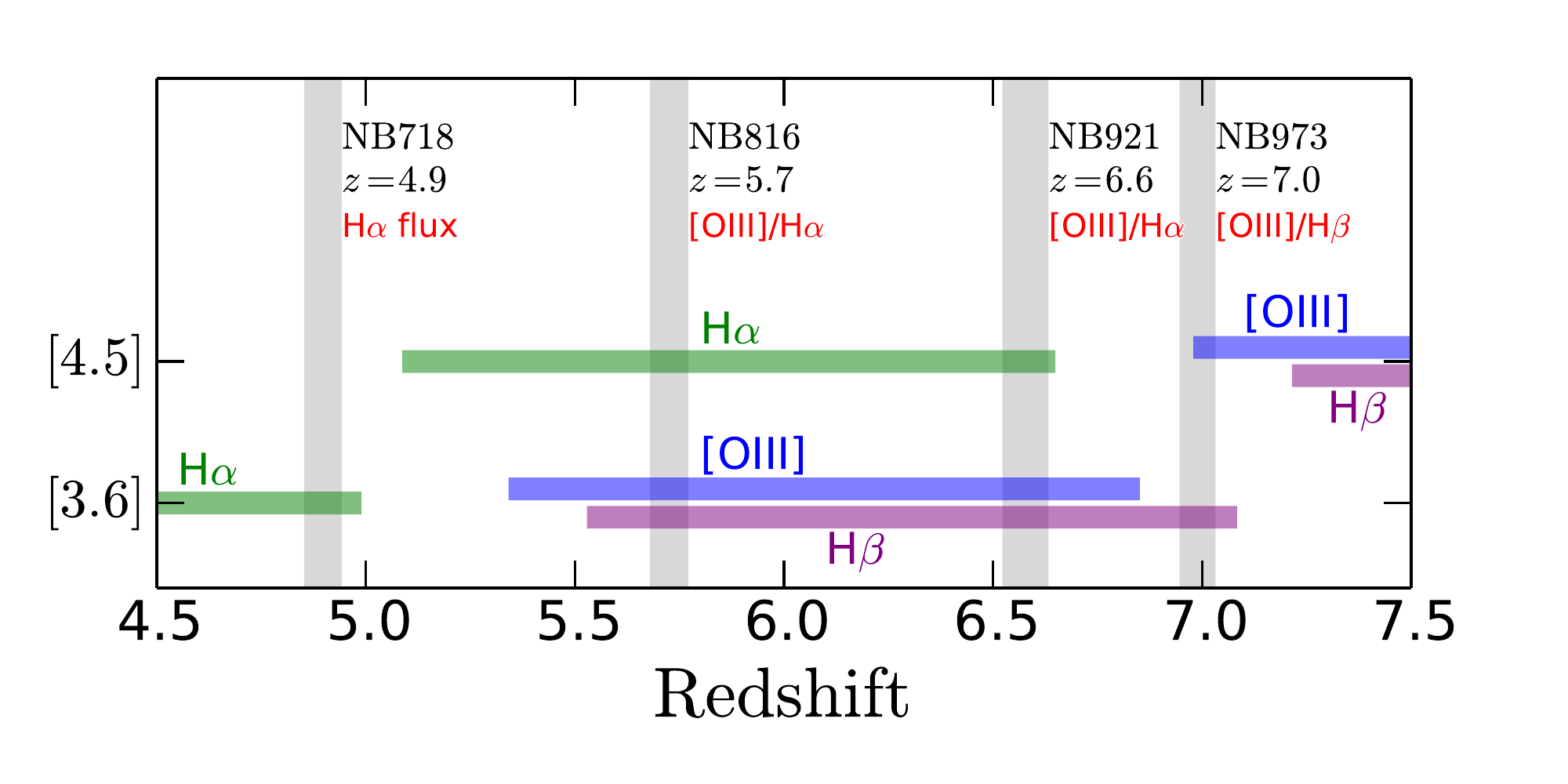}
 \end{center}
   \caption{Contributions of the strong emission lines to the {\it Spitzer}/IRAC filters.
   The green, blue, and purple bands show redshift windows where the H$\m{\alpha}$, {\sc[Oiii]}$\lambda$5007, and H$\m{\beta}$ lines enter in the $[3.6]$ and $[4.5]$ bands, respectively.
   We can estimate the H$\m{\alpha}$ flux, {\sc[Oiii]}/H$\m{\alpha}$ ratio, and {\sc[Oiii]}/H$\m{\beta}$ ratio in LAEs at z=$4.9$, $5.7$ and $6.6$, and $7.0$, respectively, assuming the Case B recombination ($\m{H\alpha /H\beta}=2.86$) after correction for dust extinction (see Section \ref{ss_lineflux}).
   \label{fig_ch1ch2}}
\end{figure*}

We use LAE samples at $z=4.9$, $5.7$, $6.6$, and $7.0$ selected with the NB filters of $NB718$, $NB816$, $NB921$, and $NB973$, respectively.
Figure \ref{fig_ch1ch2} shows redshift windows where strong rest-frame optical emission lines enter in the {\it Spitzer}/IRAC $3.6\ \m{\mu m}$ ($[3.6]$) and $4.5\ \m{\mu m}$ ($[4.5]$) bands.
At $z=4.9$, the $\m{H\alpha}$ line is redshifted into the $[3.6]$ band, while no strong emission line into the $[4.5]$ band.
Thus we can estimate the $\m{H\alpha}$ flux at $z=4.9$ from the IRAC band photometry.
At $z=5.7$ and $6.6$, since the {\sc [Oiii]}$\lambda5007+\m{H\beta}$ and $\m{H\alpha}$ lines affect the $[3.6]$ and $[4.5]$ band, respectively, the IRAC photometry can infer the {\sc[Oiii]}$/\m{H\alpha}$ ratio.
At $z=7.0$, we can estimate the ratio of {\sc[Oiii]}$\lambda5007$ to $\m{H\beta}$, which enter the $[3.6]$ and $[4.5]$ bands, respectively.

\begin{deluxetable*}{ccccccccccccccc}
\setlength{\tabcolsep}{0.35cm}
\tablecaption{Summary of Imaging Data Used in This Study}
\tablehead{\colhead{} & \multicolumn{9}{c}{Subaru} & \multicolumn{3}{c}{VISTA/UKIRT} & \multicolumn{2}{c}{\it Spitzer}\\
\colhead{Field} & \colhead{$g$} & \colhead{$r$} & \colhead{$i$} & \colhead{$z$} & \colhead{$y$} & \colhead{$NB718$}  & \colhead{$NB816$} & \colhead{$NB921$} & \colhead{$NB973$} & \colhead{$J$}& \colhead{$H$} & \colhead{$K_s/K$} & \colhead{$[3.6]$} & \colhead{$[4.5]$}
}
\startdata
\multicolumn{13}{c}{$5\sigma$ limiting magnitude\tablenotemark{a}}\\
\hline
UD-COSMOS & 27.13 & 26.84 & 26.46 & 26.10 & 25.28 & 26.11 & 25.98 & 26.17 & 25.05 & 25.32 & 25.05 & 25.16 & 25.11 & 24.89 \\
UD-SXDS & 27.15 & 26.68 & 26.53 & 25.96 & 25.15 & \nodata & 25.40 & 25.36 & \nodata & 25.28 & 24.75 & 25.01 & 25.30 & 24.88\\
\hline
\multicolumn{13}{c}{Aperture correction\tablenotemark{b}}\\
\hline
UD-COSMOS & 0.31 & 0.22 & 0.23 & 0.20 & 0.37 & 0.28 & 0.25 & 0.23 & 0.23 & 0.27 & 0.20 & 0.18 & 0.52 & 0.55 \\
UD-SXDS       & 0.24 & 0.25 & 0.23 & 0.28 & 0.19 & \nodata & 0.14 & 0.34 & \nodata & 0.15 & 0.15 & 0.15 & 0.52 & 0.55
\enddata
\tablenotetext{a}{$5\sigma$ limiting magnitudes measured in $1.\carcsec5$, $2.\carcsec0$, and $3.\carcsec0$ diameter apertures in $grizyNB718NB816NB921NB973$, $JHK_s(K)$, and $[3.6][4.5]$ images, respectively.}
\tablenotetext{b}{Aperture corrections of $2\arcsec$ and $3\arcsec$ diameter apertures in the $grizyNB718NB816NB921NB973JHK_s(K)$ and $[3.6][4.5]$ images, respectively.
Values in the $[3.6]$ and $[4.5]$ bands are taken from \citet{2010ApJ...724.1524O}.}
\label{table_data}
\end{deluxetable*}

In this study, we use LAEs in the UD-COSMOS and UD-SXDS fields, where the deep optical to mid-infrared imaging data are available.
These two fields are observed with $grizyNB816NB921$ in the UltraDeep (UD) layer of the HSC-SSP survey.
The HSC data are reduced by the HSC-SSP collaboration with {\tt hscPipe} \citep{2017arXiv170506766B} that is the HSC data reduction pipeline based on the Large Synoptic Survey Telescope (LSST) pipeline \citep{2008arXiv0805.2366I,2010SPIE.7740E..15A,2015arXiv151207914J}.
The astrometric and photometric calibration are based on the data of the Panoramic Survey Telescope and Rapid Response System (Pan-STARRS) 1 imaging survey \citep{2013ApJS..205...20M,2012ApJ...756..158S,2012ApJ...750...99T}.
In addition, $NB718$ and $NB973$ imaging data taken in the CHORUS project are available in the UD-COSMOS field.
The UD-COSMOS and UD-SXDS fields are covered in the $JHK_s$ and $JHK$ bands with VISTA/VIRCAM and UKIRT/WFCAM in the UltraVISTA survey \citep{2012A&A...544A.156M} and UKIDSS/UDS project \citep{2007MNRAS.379.1599L}, respectively.
Here, we utilize the second data release (DR2) of UltraVISTA and the tenth data release (DR10) of UKIDSS/UDS.
The SPLASH covers both UD-COSMOS and UD-SXDS fields in the IRAC $[3.6]$ and $[4.5]$ bands (P. Capak et al. in preparation; V. Mehta et al. in preparation).
The total area coverage of the UD-COSMOS and UD-SXDS fields is $4\ \m{deg}^2$.
Table \ref{table_data} summarizes the imaging data used in this study.

The LAE samples at $z=5.7$ and $6.6$ are selected in \citet{2018PASJ...70S..14S} based on the HSC-SSP survey data in both the UD-COSMOS and UD-SXDS fields.
A total of 426 and 495 LAEs are selected at $z=5.7$ and $6.6$, respectively, with the following color criteria:\\
$z=5.7:$
\begin{eqnarray}
NB816<NB816_{5\sigma}\ \m{and}\ i-NB816>1.2\ \m{and}\notag\\
g>g_{3\sigma}\ \m{and}\ \left[(r\leq r_{3\sigma}\ \m{and}\  r-i\geq1.0)\ \m{or}\ r>r_{3\sigma}\right],
\end{eqnarray}
$z=6.6:$
\begin{eqnarray}
NB921<NB921_{5\sigma}\ \m{and}\ z-NB921>1.0\ \m{and}\notag\\
g>g_{3\sigma}\ \m{and}\ r>r_{3\sigma}\ \m{and}\notag\\
\left[(z\leq z_{3\sigma}\ \m{and}\  i-z\geq1.3)\ \m{or}\ z>z_{3\sigma}\right].
\end{eqnarray}
The subscripts ``$5\sigma$" and ``$3\sigma$" indicate the 5 and 3 magnitude limits for a given filter, respectively,
\redc{Since our LAEs are selected based on the HSC data, our sample is larger and brighter than previous Subaru/Suprime-Cam studies such as \citet{2010ApJ...724.1524O}.}

We use LAE samples at $z=4.9$ and $7.0$ selected based on the $NB718$ and $NB973$ images in the CHORUS project and the HSC-SSP survey data in the UD-COSMOS field.
A total of $141$ and $30$ LAEs are selected at $z=4.9$ and $7.0$, respectively, with the following color criteria:\\
$z=4.9:$
\begin{eqnarray}
NB718<NB718_{5\sigma}\ \m{and}\notag\\ 
ri-NB718>\m{max}(0.7,3\sigma(ri-NB718))\ \m{and}\notag\\
r-i>0.8\ \m{and}\ g>g_{2\sigma},
\end{eqnarray}
$z=7.0:$
\begin{eqnarray}
NB973<NB973_{5\sigma}\ \m{and}\notag\\
(y<y_{3\sigma}\ \m{and}\ y-NB973>1)\ \m{or}\ y>y_{3\sigma}\ \m{and}\notag\\
\left[(z<z_{3\sigma}\ \m{and}\ z-y>2)\ \m{or}\ z>z_{3\sigma}\right]\ \m{and}\notag\\
g>g_{2\sigma}\ \m{and}\ r>r_{2\sigma}\ \m{and}\ i>i_{2\sigma},
\end{eqnarray}
where $ri$ is the magnitude in the $ri$ band whose flux is defined with $r$ and $i$ band fluxes as $f_{ri}=0.3f_{r}+0.7f_{i}$, and $3\sigma(ri-NB718)$ is the $3\sigma$ error of the $ri-NB718$ color.
Details of the sample selection will be presented in H. Zhang et al. in preparation for NB718 and R. Itoh et al. in preparation for NB973.

\begin{deluxetable*}{cccccccccc}
\setlength{\tabcolsep}{0.35cm}
\tablecaption{Examples of Spectroscopically Confirmed LAEs Used in This Study}
\tablehead{\colhead{ID} & \colhead{$\mathrm{R.A.\ (J2000)}$} & \colhead{$\mathrm{decl.\ (J2000)}$} & \colhead{$z_\m{spec}$} & \colhead{$EW_\m{Ly\alpha}^0$} & \colhead{$M_\m{UV}$} & \colhead{$[3.6]$} & \colhead{$[4.5]$} & \colhead{$[3.6]-[4.5]$} & \colhead{Ref.} \\
\colhead{(1)}& \colhead{(2)}& \colhead{(3)}& \colhead{(4)} &  \colhead{(5)}& \colhead{(6)}& \colhead{(7)}& \colhead{(8)}& \colhead{(9)}& \colhead{(10)}}
\startdata
HSC J021828-051423 & 02:18:28.87 & -05:14:23.01 & 5.737 & $198.4_{-63.5}^{+160.6}$ & $-20.4\pm0.5$ & $>26.0$ & $>25.9$ & \nodata & H18\\
HSC J021724-053309 & 02:17:24.02 & -05:33:09.61 & 5.707 & $75.0_{-13.8}^{+19.9}$ & $-21.3\pm0.2$ & $25.3\pm0.3$ & $25.4\pm0.3$ & $-0.1\pm0.4$ & H18\\
HSC J021859-052916 & 02:18:59.92 & -05:29:16.81 & 5.674 & $14.7_{-1.7}^{+1.8}$ & $-22.6\pm0.1$ & $24.6\pm0.2$ & $25.1\pm0.3$ & $-0.5\pm0.3$ & H18\\
HSC J021827-044736 & 02:18:27.44 & -04:47:36.98 & 5.703 & $178.8_{-61.2}^{+172.9}$ & $-20.2\pm0.6$ & $25.6\pm0.4$ & $>25.9$ & $<-0.3$ & H18\\
HSC J021830-051457 & 02:18:30.53 & -05:14:57.81 & 5.688 & $154.3_{-49.9}^{+124.2}$ & $-20.4\pm0.5$ & $>26.0$ & $>25.9$ & \nodata & H18\\
HSC J021624-045516 & 02:16:24.70 & -04:55:16.55 & 5.706 & $75.5_{-17.4}^{+28.4}$ & $-20.9\pm0.3$ & $>26.0$ & $>25.9$ & \nodata & H18\\
HSC J100058+014815 & 10:00:58.00 & +01:48:15.14 & 6.604 & \redc{$211.0_{-20.0}^{+20.0}$} & $-22.4\pm0.2$ & $24.0\pm0.1$ & $25.3\pm0.3$ & $-1.3\pm0.3$ & S15\\
HSC J021757-050844 & 02:17:57.58 & -05:08:44.63 & 6.595 & \redc{$78.0_{-6.0}^{+8.0}$} & $-21.4\pm0.5$ & $24.7\pm0.2$ & $25.4\pm0.3$ & $-0.7\pm0.4$ & O10\\
HSC J100109+021513 & 10:01:09.72 & +02:15:13.45 & 5.712 & $214.3_{-46.1}^{+78.6}$ & $-20.7\pm0.3$ & $23.3\pm0.0$ & $22.7\pm0.0$ & $0.6\pm0.1$ & M12\\
HSC J100129+014929 & 10:01:29.07 & +01:49:29.81 & 5.707 & $82.0_{-11.8}^{+15.7}$ & $-21.4\pm0.2$ & $24.8\pm0.2$ & $25.7\pm0.5$ & $-0.9\pm0.5$ & M12\\
HSC J100123+015600 & 10:01:23.84 & +01:56:00.46 & 5.726 & $91.5_{-20.3}^{+33.3}$ & $-20.8\pm0.3$ & $>26.1$ & $>25.9$ & \nodata & M12\\
HSC J021843-050915 & 02:18:43.62 & -05:09:15.63 & 6.510 & $20.2_{-6.2}^{+9.8}$ & $-22.0\pm0.3$ & $25.0\pm0.2$ & $25.5\pm0.4$ & $-0.5\pm0.4$ & This\\
HSC J021703-045619 & 02:17:03.46 & -04:56:19.07 & 6.589 & $34.0_{-12.9}^{+30.6}$ & $-21.4\pm0.5$ & $>26.0$ & $>25.9$ & \nodata & O10\\
HSC J021702-050604 & 02:17:02.56 & -05:06:04.61 & 6.545 & $68.9_{-35.6}^{+83.0}$ & $-20.5\pm0.7$ & $>26.0$ & $>25.9$ & \nodata & O10\\
HSC J021819-050900 & 02:18:19.39 & -05:09:00.65 & 6.563 & $49.5_{-25.0}^{+60.0}$ & $-20.8\pm0.7$ & $>26.0$ & $>25.9$ & \nodata & O10\\
HSC J021654-045556 & 02:16:54.54 & -04:55:56.94 & 6.617 & $29.8_{-13.7}^{+36.9}$ & $-21.2\pm0.6$ & $>26.0$ & $>25.9$ & \nodata & O10\\
HSC J095952+013723 & 09:59:52.13 & +01:37:23.24 & 5.724 & $72.7_{-15.5}^{+24.2}$ & $-20.9\pm0.2$ & $>26.1$ & $>25.9$ & \nodata & M12\\
HSC J095952+015005 & 09:59:52.03 & +01:50:05.95 & 5.744 & $33.6_{-5.2}^{+6.5}$ & $-21.5\pm0.1$ & $25.6\pm0.3$ & $25.5\pm0.4$ & $0.1\pm0.5$ & M12\\
HSC J021737-043943 & 02:17:37.96 & -04:39:43.02 & 5.755 & $60.4_{-14.1}^{+22.4}$ & $-21.0\pm0.3$ & $25.1\pm0.2$ & $25.9\pm0.5$ & $-0.8\pm0.6$ & H18\\
HSC J100015+020056 & 10:00:15.66 & +02:00:56.04 & 5.718 & $92.5_{-24.1}^{+44.7}$ & $-20.5\pm0.3$ & $>26.1$ & $>25.9$ & \nodata & M12\\
HSC J021734-044558 & 02:17:34.57 & -04:45:58.95 & 5.702 & $45.1_{-9.8}^{+14.4}$ & $-21.2\pm0.2$ & $25.6\pm0.4$ & $>25.9$ & $<-0.2$ & O08\\
HSC J100131+023105 & 10:01:31.08 & +02:31:05.77 & 5.690 & $91.5_{-25.2}^{+47.9}$ & $-20.5\pm0.4$ & $>26.1$ & $>25.9$ & \nodata & M12\\
HSC J100301+020236 & 10:03:01.15 & +02:02:36.04 & 5.682 & $14.7_{-2.3}^{+2.5}$ & $-22.0\pm0.1$ & $24.3\pm0.1$ & $24.1\pm0.1$ & $0.2\pm0.1$ & M12\\
HSC J021654-052155 & 02:16:54.60 & -05:21:55.52 & 5.712 & $127.2_{-48.0}^{+129.1}$ & $-20.1\pm0.6$ & $>26.0$ & $>25.9$ & \nodata & O08\\
HSC J021748-053127 & 02:17:48.46 & -05:31:27.02 & 5.690 & $54.4_{-13.4}^{+21.9}$ & $-20.9\pm0.3$ & $>26.0$ & $>25.9$ & \nodata & O08\\
HSC J100127+023005 & 10:01:27.77 & +02:30:05.83 & 5.696 & $49.9_{-10.3}^{+15.0}$ & $-21.0\pm0.2$ & $25.4\pm0.3$ & $>25.9$ & $<-0.5$ & M12\\
HSC J021745-052936 & 02:17:45.24 & -05:29:36.01 & 5.688 & $>112.1$ & $>-19.4$ & $25.6\pm0.4$ & $>25.9$ & $<-0.3$ & O08\\
HSC J095954+021039 & 09:59:54.77 & +02:10:39.26 & 5.662 & $45.9_{-10.0}^{+14.7}$ & $-21.0\pm0.2$ & $>26.1$ & $>25.9$ & \nodata & M12\\
HSC J095919+020322 & 09:59:19.74 & +02:03:22.02 & 5.704 & $152.6_{-61.9}^{+157.0}$ & $-19.8\pm0.6$ & $>26.1$ & $>25.9$ & \nodata & M12\\
HSC J095954+021516 & 09:59:54.52 & +02:15:16.50 & 5.688 & $60.9_{-15.5}^{+26.5}$ & $-20.7\pm0.3$ & $>26.1$ & $>25.9$ & \nodata & M12\\
HSC J100005+020717 & 10:00:05.06 & +02:07:17.01 & 5.704 & $118.6_{-43.4}^{+120.2}$ & $-20.0\pm0.6$ & $>26.1$ & $>25.9$ & \nodata & M12\\
HSC J021804-052147 & 02:18:04.17 & -05:21:47.25 & 5.734 & $22.7_{-5.4}^{+7.0}$ & $-21.4\pm0.2$ & $>26.0$ & $>25.9$ & \nodata & H18\\
HSC J100022+024103 & 10:00:22.51 & +02:41:03.25 & 5.661 & $26.7_{-5.6}^{+7.3}$ & $-21.3\pm0.2$ & $25.7\pm0.4$ & $>25.9$ & $<-0.1$ & M12\\
HSC J021848-051715 & 02:18:48.23 & -05:17:15.45 & 5.741 & $29.8_{-7.5}^{+10.9}$ & $-21.2\pm0.2$ & $>26.0$ & $>25.9$ & \nodata & H18\\
HSC J100030+021714 & 10:00:30.41 & +02:17:14.73 & 5.695 & $104.5_{-40.7}^{+109.1}$ & $-19.9\pm0.6$ & $>26.1$ & $>25.9$ & \nodata & M12\\
HSC J021558-045301 & 02:15:58.49 & -04:53:01.75 & 5.718 & $87.6_{-36.1}^{+94.5}$ & $-20.1\pm0.6$ & $>26.0$ & $>25.9$ & \nodata & H18\\
HSC J100131+014320 & 10:01:31.11 & +01:43:20.50 & 5.728 & $77.2_{-26.6}^{+65.3}$ & $-20.2\pm0.5$ & $26.1\pm0.5$ & $>25.9$ & $<0.2$ & M12\\
HSC J095944+020050 & 09:59:44.07 & +02:00:50.74 & 5.688 & $57.9_{-17.7}^{+34.6}$ & $-20.4\pm0.4$ & $25.6\pm0.4$ & $>25.9$ & $<-0.3$ & M12\\
HSC J021709-050329 & 02:17:09.77 & -05:03:29.18 & 5.709 & $80.5_{-33.1}^{+87.3}$ & $-20.1\pm0.6$ & $>26.0$ & $>25.9$ & \nodata & H18\\
HSC J021803-052643 & 02:18:03.87 & -05:26:43.45 & 5.747 & $>69.8$ & $>-19.4$ & $>26.0$ & $>25.9$ & \nodata & H18\\
HSC J021805-052704 & 02:18:05.17 & -05:27:04.06 & 5.746 & $>68.4$ & $>-19.4$ & $>26.0$ & $>25.9$ & \nodata & O08\\
HSC J021739-043837 & 02:17:39.25 & -04:38:37.21 & 5.720 & $119.3_{-58.4}^{+134.6}$ & $-19.6\pm0.7$ & $>26.0$ & $>25.9$ & \nodata & H18\\
HSC J021857-045648 & 02:18:57.32 & -04:56:48.88 & 5.681 & $120.2_{-59.3}^{+136.5}$ & $-19.5\pm0.7$ & $>26.0$ & $>25.9$ & \nodata & H18\\
HSC J021639-051346 & 02:16:39.89 & -05:13:46.75 & 5.702 & $108.9_{-53.7}^{+122.9}$ & $-19.6\pm0.7$ & $>26.0$ & $>25.9$ & \nodata & O08\\
HSC J021805-052026 & 02:18:05.28 & -05:20:26.90 & 5.742 & $44.5_{-16.0}^{+33.1}$ & $-20.5\pm0.4$ & $>26.0$ & $>25.9$ & \nodata & H18\\
HSC J100058+013642 & 10:00:58.41 & +01:36:42.89 & 5.688 & $>72.9$ & $>-19.2$ & $>26.1$ & $>25.9$ & \nodata & M12\\
HSC J100029+015000 & 10:00:29.58 & +01:50:00.78 & 5.707 & $84.2_{-35.9}^{+91.9}$ & $-19.8\pm0.6$ & $>26.1$ & $>25.9$ & \nodata & M12\\
HSC J021911-045707 & 02:19:11.03 & -04:57:07.48 & 5.704 & $>53.3$ & $>-19.5$ & $>26.0$ & $>25.9$ & \nodata & H18\\
HSC J021628-050103 & 02:16:28.05 & -05:01:03.85 & 5.691 & $>43.5$ & $>-19.4$ & $>26.0$ & $>25.9$ & \nodata & H18\\
HSC J100107+015222 & 10:01:07.35 & +01:52:22.88 & 5.668 & $38.1_{-15.6}^{+35.1}$ & $-20.2\pm0.5$ & $>26.1$ & $>25.9$ & \nodata & M12
\enddata
\tablecomments{(1) Object ID same as \citet{2018PASJ...70S..15S}.
(2) Right ascension.
(3) Declination.
(4) Spectroscopic redshift of the Ly$\m{\alpha}$ emission line, Lyman break feature, or rest-frame UV absorption line.
(5) Rest-frame Ly$\m{\alpha}$ EW or its $2\sigma$ lower limit.
(6) Absolute UV magnitude or its $2\sigma$ lower limit.
(7)-(8) Total magnitudes in the $[3.6]$ and $[4.5]$ bands. The lower limit is $2\sigma$.
(9) $[3.6]-[4.5]$ color.
(10) Reference (O08: \citealt{2008ApJS..176..301O}, O10: \citealt{2010ApJ...723..869O}, M12: \citealt{2012ApJ...760..128M}, S15: \citet{2015ApJ...808..139S}, S17: \citealt{2018PASJ...70S..15S}, H18: \citealt{2018arXiv180100531H}, This: this work, see Section \ref{ss_laesample}).}
\label{tab_spec}
\end{deluxetable*}

\begin{deluxetable*}{ccccccc}
\setlength{\tabcolsep}{0.35cm}
\tablecaption{List of Galaxies Used in Our {\sc [Cii]} Study}
\tablehead{\colhead{Name} & \colhead{$z_\m{spec}$} & \colhead{$EW_\m{Ly\alpha}^0$} & \colhead{$EW_\m{Ly\alpha}^\m{0,int}$} & \colhead{$\m{log}L_\m{[CII]}$} & \colhead{$\m{log}SFR_\m{tot}$} & \colhead{Ref.} \\
\colhead{(1)}& \colhead{(2)}& \colhead{(3)}& \colhead{(4)} &  \colhead{(5)}& \colhead{(6)}& \colhead{(7)}}
\startdata
HCM6A & 6.56 & $25.1$ & $35.2$ & $<7.81$ & $1.00$ & K13, H02\\
IOK-1 & 6.965 & $43.0$ & $63.9$ & $<7.53$ & $1.38$ & O14, O12\\
z8-GND-5296 & 7.508 & $8.0$ & $27.6$ & $<8.55$ & $1.37$ & S15, F12\\
BDF-521 & 7.109 & $64.0$ & $120.3$ & $<7.78$ & $0.78$ & M15, V11\\
BDF-3299 & 7.008 & $50.0$ & $75.8$ & $<7.30$ & $0.76$ & M15, V11\\
SDF46975 & 6.844 & $43.0$ & $63.1$ & $<7.76$ & $1.19$ & M15, O12\\
A1689-zD1 & 7.5 & $<27.0$ & $<93.1$ & $<7.95$ & $1.07^{+0.15}_{-0.08}$ & W15\\
HZ1 & 5.690 & $5.3^{+2.6}_{-4.1}$ & $5.3^{+2.6}_{-4.1}$ & $8.40\pm0.32$ & $1.38^{+0.11}_{-0.05}$ & C15, M12\\
HZ2 & 5.670 & $6.9\pm2.0$ & $6.9\pm2.0$ & $8.56\pm0.41$ & $1.40^{+0.09}_{-0.03}$ & C15\\
HZ3 & 5.546 & $<3.6$ & $<3.6$ & $8.67\pm0.28$ & $1.26^{+0.19}_{-0.07}$ & C15\\
HZ4 & 5.540 & $10.2^{+0.9}_{-4.4}$ & $10.2^{+0.9}_{-4.4}$ & $8.98\pm0.22$ & $1.71^{+0.46}_{-0.15}$ & C15, M12\\
HZ6 & 5.290 & $8.0^{+12.1}_{-7.3}$ & $8.0^{+12.1}_{-7.3}$ & $9.15\pm0.17$ & $1.69^{+0.39}_{-0.11}$ & C15, M12\\
HZ7 & 5.250 & $9.8\pm5.5$ & $9.8\pm5.5$ & $8.74\pm0.24$ & $1.32^{+0.10}_{-0.04}$ & C15\\
HZ8 & 5.148 & $27.1^{+12.9}_{-14.7}$ & $27.1^{+12.9}_{-14.7}$ & $8.41\pm0.18$ & $1.26^{+0.12}_{-0.05}$ & C15, M12\\
HZ9 & 5.548 & $14.4^{+6.8}_{-5.4}$ & $14.4^{+6.8}_{-5.4}$ & $9.21\pm0.09$ & $1.83^{+0.19}_{-0.13}$ & C15, M12\\
HZ10 & 5.659 & $24.5^{+9.2}_{-11.0}$ & $24.5^{+9.2}_{-11.0}$ & $9.13\pm0.13$ & $2.23^{+0.08}_{-0.07}$ & C15, M12\\
CLM1 & 6.176 & $50.0$ & $59.2$ & $8.38\pm0.06$ & $1.57\pm0.05$ & W15, C03\\
WMH5 & 6.076 & $13.0\pm4.0$ & $14.8\pm4.6$ & $8.82\pm0.05$ & $1.63\pm0.05$ & W15, W13\\
A383-5.1 & 6.029 & $138.0$ & $154.9$ & $6.95\pm0.15$ & $0.51$ & K16, St15\\
SXDF-NB1006-2 & 7.215 & $>15.4$ & $>38.4$ & $<7.92$ & $2.54^{+0.21}_{-0.35}$ & I16, S12\\
COSMOS13679 & 7.154 & $15.0$ & $30.9$ & $7.87\pm0.10$ & $1.38$ & P16\\
NTTDF6345 & 6.701 & $15.0$ & $21.7$ & $8.27\pm0.07$ & $1.18$ & P16\\
UDS16291 & 6.638 & $6.0$ & $8.6$ & $7.86\pm0.10$ & $1.20$ & P16\\
COSMOS24108 & 6.629 & $27.0$ & $38.7$ & $8.00\pm0.10$ & $1.46$ & P16\\
RXJ1347-1145 & 6.765 & $26.0\pm4.0$ & $37.8\pm5.8$ & $7.18^{+0.06}_{-0.12}$ & $0.93^{+0.30}_{-0.05}$ & B16\\
COS-3018555981 & 6.854 & $<2.9$ & $<4.3$ & $8.67\pm0.05$ & $1.37^{+0.44}_{-0.02}$ & S17, L17\\
COS-2987030247 & 6.816 & $16.2^{+5.2}_{-5.5}$ & $23.7^{+7.6}_{-8.0}$ & $8.56\pm0.06$ & $1.52^{+0.63}_{-0.06}$ & S17, L17\\
CR7 & 6.604 & $211.0\pm20.0$ & $301.6\pm28.6$ & $8.30\pm0.09$ & $1.65\pm0.02$ & M17, So15\\
NTTDF2313 & 6.07 & $0$ & $0$ & $<7.65$ & $1.08$ & C17\\
BDF2203 & 6.12 & $3.0$ & $3.5$ & $8.10\pm0.09$ & $1.20$ & C17\\
GOODS3203 & 6.27 & $5.0$ & $6.2$ & $<8.08$ & $1.26$ & C17\\
COSMOS20521 & 6.36 & $10.0$ & $12.8$ & $<7.68$ & $1.15$ & C17\\
UDS4821 & 6.561 & $48.0$ & $67.3$ & $<7.83$ & $1.11$ & C17\\
Himiko & 6.595 & $78.0^{+8.0}_{-6.0}$ & $111.2^{+11.4}_{-8.6}$ & $8.08\pm0.07$ & $1.31\pm0.03$ & C18, O13
\enddata
\tablecomments{(1) Object Name.
(2) Redshift determined with Ly$\alpha$, Lyman break, rest-frame UV absorption lines, or {[\sc Cii]}158$\m{\mu m}$.
(3) Rest-frame Ly$\alpha$ EW not corrected for the inter-galactic medium (IGM) absorption.
(4) Rest-frame Ly$\alpha$ EW corrected for the IGM absorption with Equations (\ref{eq_IGM_57})-(\ref{eq_IGM_73}).
(5) {[\sc Cii]}158$\m{\mu m}$ luminosity or its $3\sigma$ upper limit in units of $L_\odot$.
(6) Total SFR in units of $M_\odot\ \m{yr^{-1}}$.
(7) Reference (H02: \citealt{2002ApJ...568L..75H}, C03: \citealt{2003A&A...405L..19C}, V11: \citealt{2011ApJ...730L..35V}, O12: \citealt{2012ApJ...744...83O}, S12: \citealt{2012ApJ...752..114S}, M12: \citealt{2012ApJ...760..128M}, W13: \citealt{2013AJ....145....4W}, K13: \citealt{2013ApJ...771L..20K}, F13: \citealt{2013Natur.502..524F}, O13: \citealt{2013ApJ...778..102O}, O14: \citealt{2014ApJ...792...34O}, S15: \citealt{2015A&A...574A..19S}, St15: \citealt{2015MNRAS.450.1846S}, M15: \citealt{2015MNRAS.452...54M}, W15: \citealt{2015Natur.519..327W}, C15: \citealt{2015Natur.522..455C}, W15: \citealt{2015ApJ...807..180W}, So15: \citealt{2015ApJ...808..139S}, K16: \citealt{2016MNRAS.462L...6K}, I16: \citealt{2016Sci...352.1559I}, P16: \citealt{2016ApJ...829L..11P}, B17: \citealt{2017ApJ...836L...2B}, S17: \citealt{2017arXiv170604614S}, L17: \citealt{2017ApJ...851...40L}, M17: \citealt{2017ApJ...851..145M}, \redc{C17: \citealt{2017arXiv171203985C}, C18: \citealt{2018ApJ...854L...7C}}).}
\label{tab_cii}
\end{deluxetable*}

\begin{figure*}
 \begin{center}
  \includegraphics[clip,bb=0 0 720 280,width=1\hsize]{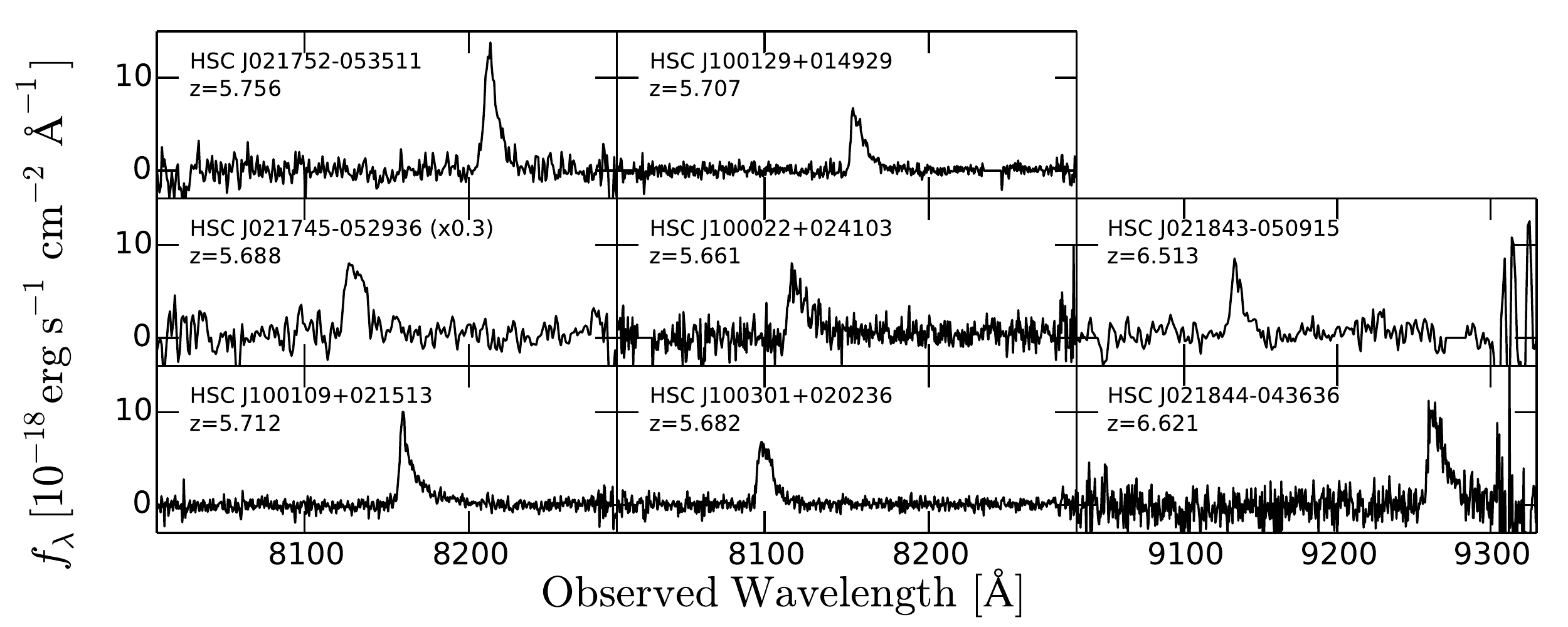}
 \end{center}
   \caption{Examples of the spectra of our LAEs.
   We show spectra of HSC J021752-053511 \citep{2018PASJ...70S..15S}, HSC J021745-052936 \citep{2008ApJS..176..301O}, HSC J100109+021513, HSC J100129+014929, HSC J100022+024103, and HSC J100301+020236 \citep{2012ApJ...760..128M}, \redc{HSC J021843-050915 (this work; see Section \ref{ss_laesample})} and HSC J021844-043636 \citep{2010ApJ...723..869O}.
   For a panel in which a factor is shown after the object ID, multiply the flux scale by this factor to obtain a correct scale.
   The units of the vertical axes in HSC J021843-050915 and HSC J021844-043636 are arbitrary.
         \label{fig_spec}}
\end{figure*}

Out of $1,092$ LAEs in the sample, 805 LAEs are covered with the $JHK_s(K)[3.6][4.5]$ images, and 96 LAEs are spectroscopically confirmed with Ly$\m{\alpha}$ emission, Lyman break features, or rest-frame UV absorption lines \citep{2018PASJ...70S..15S}.
\redc{In addition to the confirmed LAEs listed in \citet{2018PASJ...70S..15S}, we spectroscopically identified HSC J021843-050915 at $z=6.513$ in our Magellan/LDSS3 observation in October 2016 (PI: M. Rauch).}
We show some examples of the spectra around Ly$\alpha$ in Figure \ref{fig_spec}, \redc{including HSC J021843-050915}.
Tables \ref{tab_spec} summarizes 50 spectroscopically confirmed LAEs at $z=5.7$ and $6.6$ without severe blending in the IRAC images (see Section \ref{ss_blend}).
Based on spectroscopy in \citet{2018PASJ...70S..15S}, the contamination rate of our $z=5.7$ and $6.6$ samples is $0-30\%$, and appears to depend on the magnitude \citep{2018PASJ...70S..16K}.
We will discuss the effect of the contamination in Section \ref{ss_stack}.

We derive the rest-frame equivalent widths (EWs) of Ly$\m{\alpha}$ ($EW^0_\m{Ly\alpha}$) of our LAEs, in the same manner as \citet{2018PASJ...70S..14S}.
We use colors of $NB718-z$, $NB816-z$, $NB921-y$, and $y-NB973$ for $z=4.9$, $5.7$, $6.6$, and $7.0$ LAEs, respectively.
We assume the redshift of the central wavelength of each NB filter.
\redc{A full description of the calculation is provided in Section 8 in \citet{2018PASJ...70S..14S}.}
\redc{We find that our calculations for most of the LAEs are consistent with previous studies, except for CR7.
The difference in the EW estimates for CR7 comes from different $y$ band magnitudes, probably due to the differences of the instrument, filter, and photometry.
We adopt the estimate in \citet{2015ApJ...808..139S} to compare with the previous studies.}
Note that some of the rest-frame Ly$\m{\alpha}$ EW are lower than $20\ \m{\AA}$, roughly corresponding to the color selection criteria in \citet{2018PASJ...70S..14S}, because of the difference in the color bands used for the EW calculations.
While the $20\ \m{\AA}$ EW threshold in the selection corresponds to the color criteria in $i-NB816$ and $z-NB921$ at $z=5.7$ and $6.6$, respectively, our Ly$\alpha$ EWs are calculated from $NB816-z$, $NB921-y$.
Thus LAEs with $EW^0_\m{Ly\alpha}<20\ \m{\AA}$ are galaxies faint in $i$ ($z$) and bright in $NB816$ and $z$ ($NB921$ and $y$) at $z=5.7$ ($6.6$).
In order to investigate the effect of the AGNs, we also conduct analyses removing LAEs with $\m{log}(L_\m{Ly\alpha}/\m{erg\ s^{-1}})>43.4$, because \citet{2016ApJ...823...20K} reveal that LAEs brighter than $\m{log}(L_\m{Ly\alpha}/\m{erg\ s^{-1}})=43.4$ are AGNs at $z=2.2$.
We find that results do not change, indicating that the effect of the AGNs is not significant.

\subsection{{\sc [Cii]158$\m{\mu m}$} sample}\label{ss_ciisample}
In addition to our HSC LAE samples, we compile previous ALMA and PdBI observations targeting {\sc [Cii]}158$\m{\mu m}$ in galaxies at $z>5$.
We use results of \redc{34} galaxies from \citet{2013ApJ...771L..20K}, \citet{2013ApJ...778..102O}, \citet{2014ApJ...792...34O}, \citet{2015A&A...574A..19S}, \citet{2015MNRAS.452...54M}, \citet{2015Natur.519..327W}, \citet{2015Natur.522..455C}, \citet{2015ApJ...807..180W}, \citet{2016MNRAS.462L...6K}, \citet{2016Sci...352.1559I}, \citet{2016ApJ...829L..11P}, \citet{2017ApJ...836L...2B}, \citet{2017arXiv170604614S}, \citet{2017ApJ...851..145M}, \redc{\citet{2017arXiv171203985C}, and \citet{2018ApJ...854L...7C}}.
\citet{2013ApJ...771L..20K} used PdBI, and the others studies used ALMA.
We take {\sc [Cii]} luminosities, total star formation rates (SFRs), and Ly$\alpha$ EWs from these studies.
The properties of these galaxies are summarized in Table {\ref{tab_cii}}.
\redc{For the {\sc [Cii]} luminosity of Himiko, we adopt the re-analysis result of \citet{2018ApJ...854L...7C}.
Himiko and CR7 overlap with the LAE sample in Section \ref{ss_laesample}.}
We do not include objects with AGN signatures, e.g., HZ5 in \citet{2015Natur.522..455C}, in our sample.

\section{Method}\label{ss_method}
In this section, we estimate rest-frame optical line fluxes of the LAEs by comparing observed SEDs and model SEDs. 

\subsection{Removing Severely Blended Sources}\label{ss_blend}
\begin{figure}
\begin{center}
  \includegraphics[clip,bb=10 25 360 250,width=1.0\hsize]{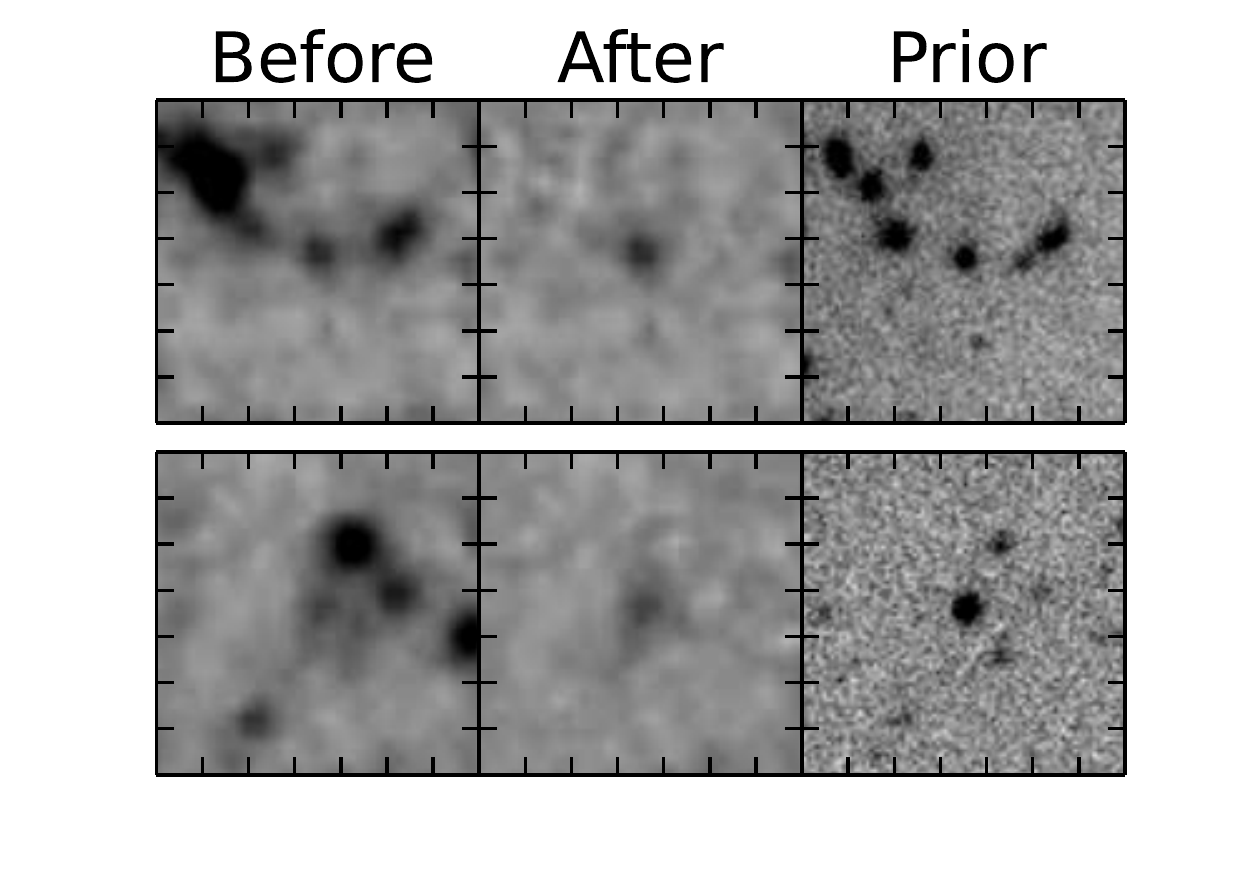}
 \end{center}
   \caption{Images showing examples of the source removal with {\sc t-phot} \citep{2016A&A...595A..97M}.
   The left panels show original images of LAEs at $z=5.7$ in the IRAC $[3.6]$ band.
   The center panels are images after the {\sc t-phot} second pass run (see Section \ref{ss_blend}).
   The sources near the LAE are cleanly removed.
   The prior images are the HSC $grizyNB$ stacked images, presented in the right panels.
   The image size is $14\arcsec\times14\arcsec$.
      \label{fig_snap_tphot}}
\end{figure}

Since point-spread functions (PSFs) of IRAC images are relatively large ($\sim1.\carcsec7$), source confusion and blending are significant for some LAEs.
In order to remove effects of the neighbor sources on the photometry, we firstly generate residual IRAC images where only the LAEs under analysis are left.
We perform a {\tt T-PHOT} second pass run with an option of {\tt exclfile} \citep{2016A&A...595A..97M} to leave the LAEs in the IRAC images.
{\tt T-PHOT} exploits information from high-resolution prior images, such as position and morphology, to extract photometry from lower resolution data where blending is a concern.
As high-resolution prior images in the {\tt T-PHOT} run, we use HSC $grizyNB$ stacked images whose PSFs are $\sim0.\carcsec7$.
The high-resolution image is convolved with a transfer kernel to generate model images for the low-resolution data (here the IRAC images), allowing the flux in each source to vary. 
This model image was in turn fitted to the real low-resolution image.
In this way, all sources are modeled and those close to the LAEs are effectively removed such that these cleaned images can be used to generate reliable stacked images of the LAEs (Figure \ref{fig_snap_tphot}).
We then visually inspect all of our LAEs and exclude 97 objects due to the presence of bad residual features close to the targets that can possibly affect the photometry.
Finally we use the 107, 213, 273, and 20 LAEs at $z=4.9$, $5.7$, $6.6$, and $7.0$ for our analysis, respectively.
\redc{Note that using the HSC images as the prior does not have a significant impact on our photometry, as far as we are interested in the total flux of the galaxy, rather than individual components.
For example, our IRAC color measurement of CR7 is $-1.3\pm0.3$, consistent with that of \citet{2017MNRAS.469..448B}, who use the high-resolution Hubble image ($\m{PSF}\sim0.\carcsec2$) as a prior.}

\subsection{Stacking Analysis}\label{ss_stack}
\begin{figure*}
\begin{center}
  \includegraphics[clip,bb=10 40 850 350,width=1.0\hsize]{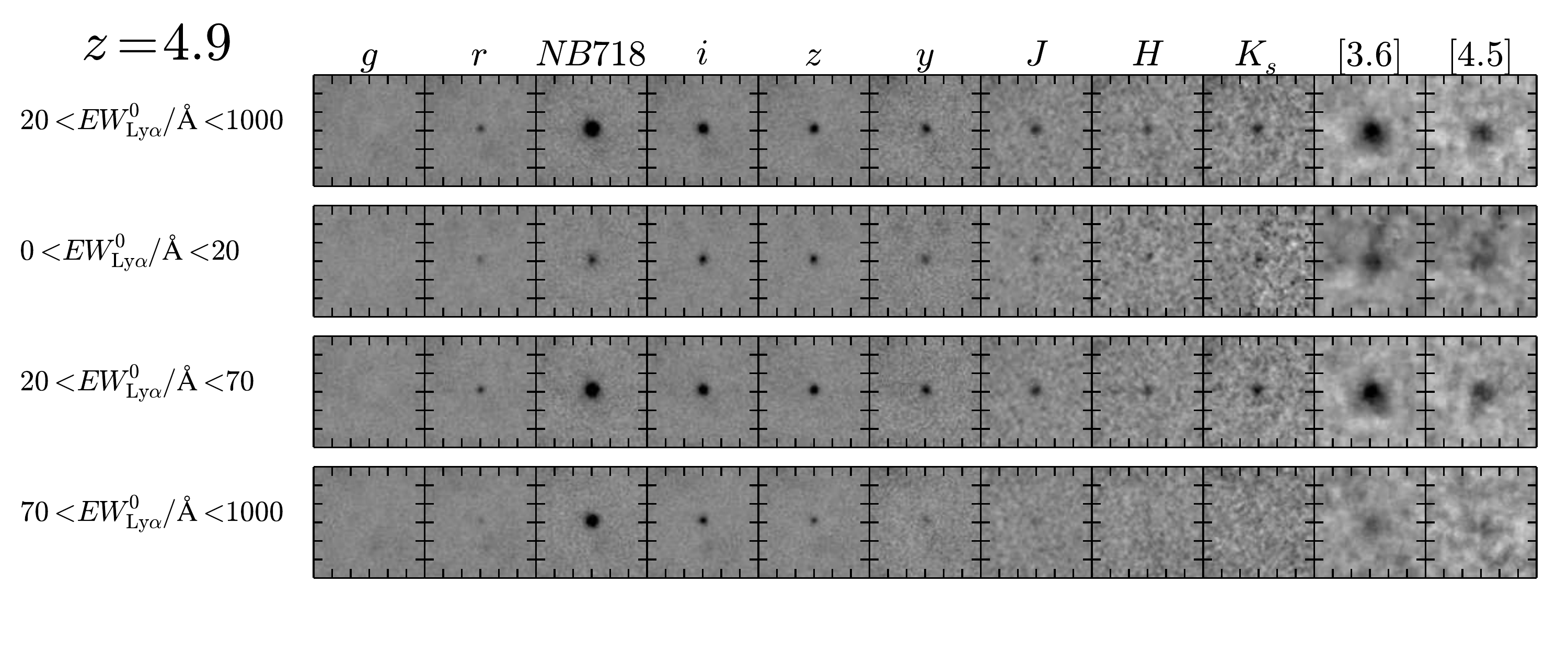}
  \includegraphics[clip,bb=10 40 850 490,width=1.0\hsize]{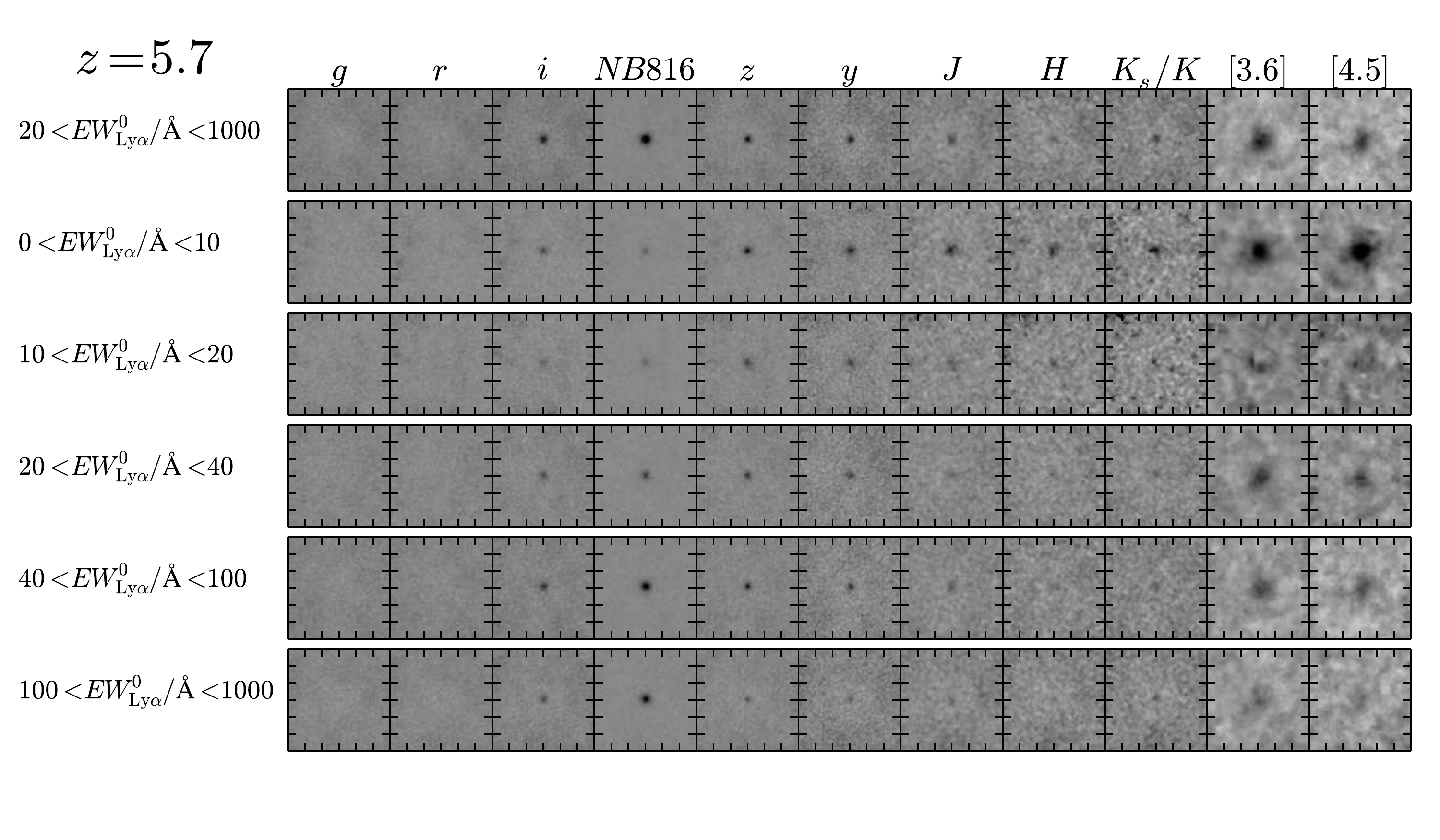}
 \end{center}
   \caption{
   Stacked images of the $z=4.9$ and $5.7$ LAE subsamples in each band. The image size is $12\arcsec\times12\arcsec$.
      \label{fig_snap_z4957}}
\end{figure*}

\begin{figure*}
\begin{center}
  \includegraphics[clip,bb=10 40 850 420,width=1.0\hsize]{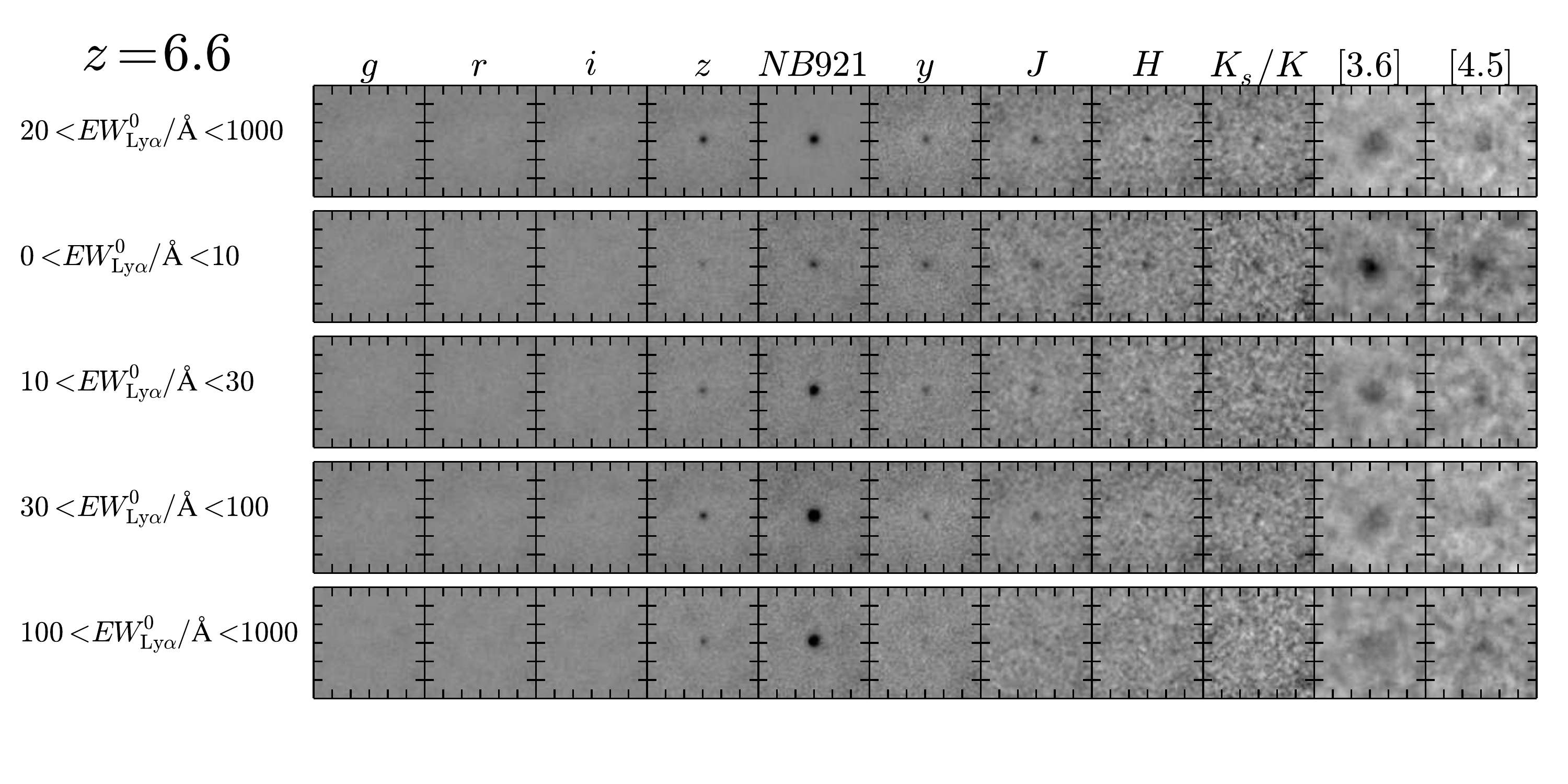}
  \includegraphics[clip,bb=10 40 850 140,width=1.0\hsize]{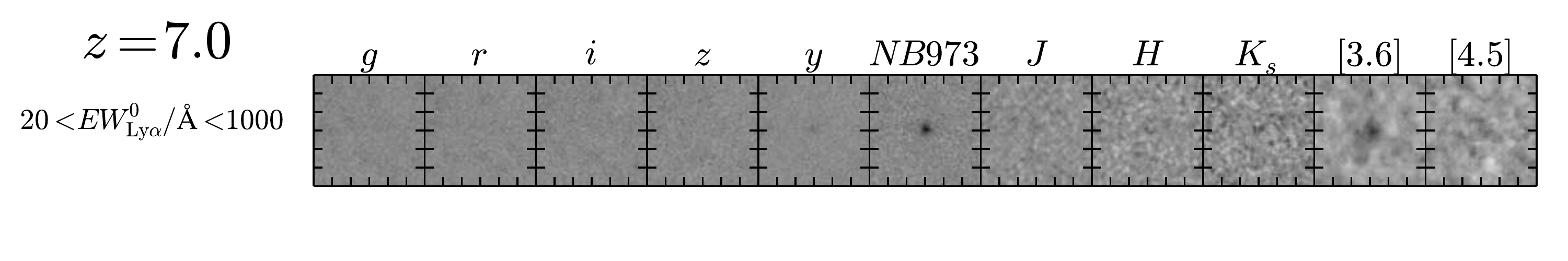}
 \end{center}
   \caption{Same as the Figure \ref{fig_snap_z4957} but for the $z=6.6$ and $7.0$ LAE subsamples.
      \label{fig_snap_z6670}}
\end{figure*}

\begin{figure*}
\begin{center}
  \includegraphics[clip,bb=10 20 580 430,width=0.95\hsize]{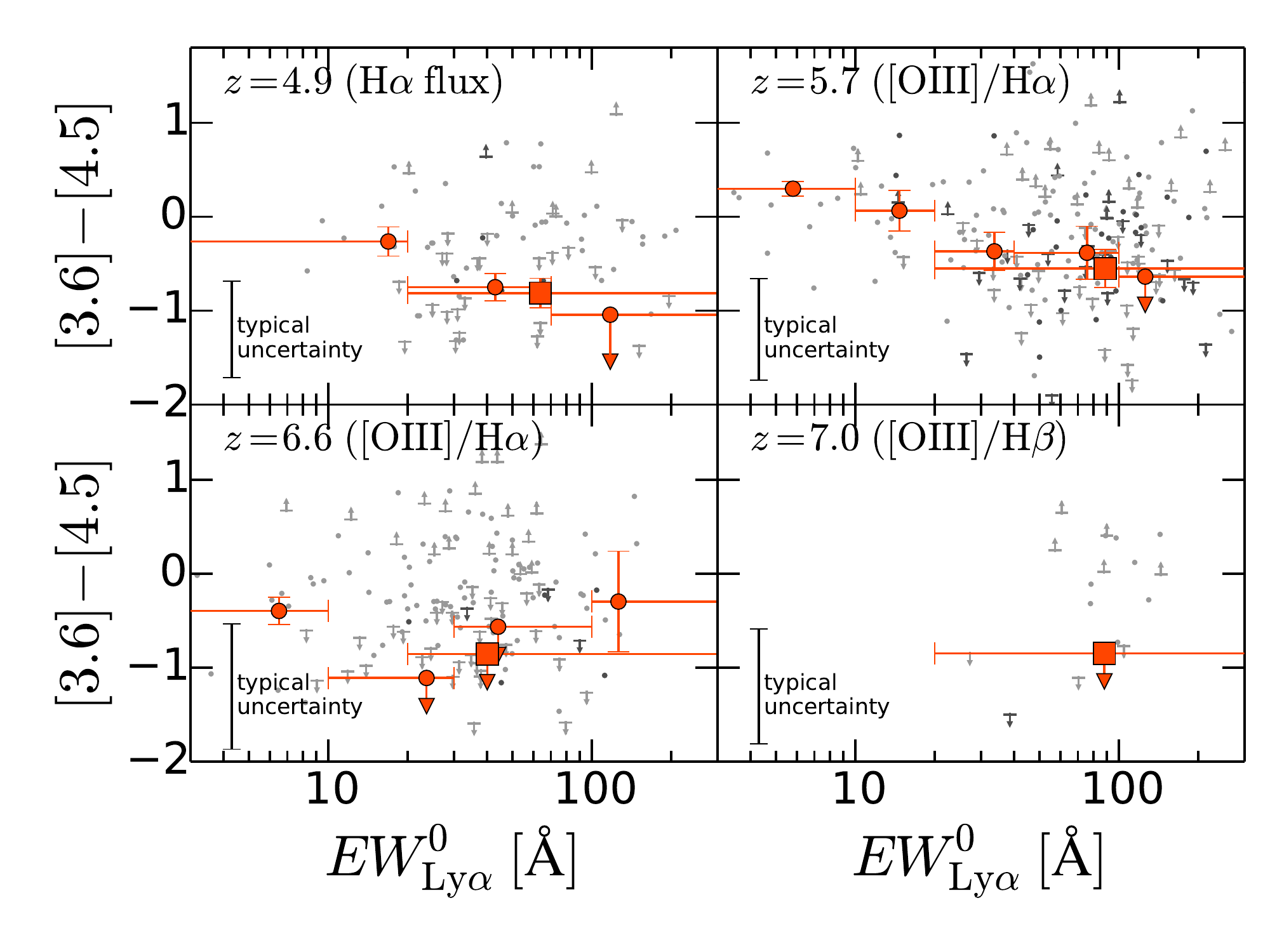}
 \end{center}
   \caption{IRAC $[3.6]-[4.5]$ colors as a function of rest-frame Ly$\m{\alpha}$ EW at $z=4.9$ (upper left), $5.7$ (upper right), $6.6$ (lower left), and $7.0$ (lower right).
   The red filled circles and squares are the results from the stacked images of the subsamples, and the gray dots show the colors of the individual objects detected in the $[3.6]$ and/or $[4.5]$ bands. 
The red squares are the results of the $EW^0_\m{Ly\alpha}>20\ \m{\AA}$ LAE subsample.
   The dark and light gray dots are objects spectroscopically confirmed and not, respectively. 
   The upward and downward arrows represent the $2\sigma$ lower and upper limits, respectively.
      \label{fig_color}}
\end{figure*}

To investigate the connection between the ISM properties and Ly$\alpha$ emission, we divide our LAE samples into subsamples by the Ly$\m{\alpha}$ EW bins at $z=4.9$, $5.7$, and $6.6$.
In addition, we make a subsample of $EW_\m{Ly\alpha}^0>20\ \m{\AA}$ representing a typical LAE sample at each redshift.
Tables \ref{tab_subsample_z49} and \ref{tab_subsample} summarize the EW ranges and number of LAEs in the subsamples at $z=4.9$ and $5.7$, $6.6$, and $7.0$, respectively.
We cut out $12\arcsec\times12\arcsec$ images of the LAEs in HSC $grizyNB718NB816NB921NB973$ ($grizyNB816NB921$), VIRCAM $JHK_s$ (WFCAM $JHK$), and IRAC $[3.6][4.5]$ bands in the UD-COSMOS (UD-SXDS) field.
Then we generate median-stacked images of the subsamples in each bands with {\tt IRAF} task {\tt imcombine}.
Figures \ref{fig_snap_z4957} and \ref{fig_snap_z6670} show the stacked images of the subsamples.
Aperture magnitudes are measured in $3\arcsec$ and $2\arcsec$-diameter circular apertures in the IRAC and the other images, respectively.
To account for flux falling outside these apertures, we apply aperture corrections summarized in Table \ref{table_data}, which are derived from samples of isolated point sources.
We measure limiting magnitudes of the stacked images by making 1000 median-stacked sky noise images, each of which is made of the same number of randomly selected sky noise images as the LAEs in the subsamples.
In addition to this stacking analysis, we measure fluxes of individual LAEs which are detected in the IRAC $[3.6]$ and/or $[4.5]$ bands, but our main results are based on the stacked images.
In Figure \ref{fig_color}, we plot the IRAC colors ($[3.6]-[4.5]$) of the stacked subsamples and individual LAEs.
At $z=4.9$ and $5.7$, the IRAC color decreases with increasing Ly$\m{\alpha}$ EW.
At $z=6.6$, the color decreases with increasing Ly$\m{\alpha}$ EW from $\sim7\ \m{\AA}$ to $\sim30\ \m{\AA}$, then increases from $\sim30\ \m{\AA}$ to $\sim130\ \m{\AA}$.

\redc{We discuss a sample selection effect on the IRAC color results.
Since our sample is selected based on the NB excess, we cannot select low Ly$\alpha$ EW galaxies with UV continua much fainter than the detection limit.
Thus the median UV magnitude is brighter in the lower $EW^0_\m{Ly\alpha}$ subsample (see Tables \ref{tab_subsample_z49} and \ref{tab_subsample}).
We use LAEs of limited UV magnitudes of $-21<M_\m{UV}<-20\ \m{mag}$, and divide them into subsamples based on the Ly$\alpha$ EW.
We stack images of the subsamples, and measure the IRAC colors in the same manner as described above.
We find the similar decreasing trend of the IRAC color with increasing Ly$\m{\alpha}$ EW at $z=5.7$.
At $z=4.9$ and $6.6$, we cannot find the trends due to the small number of the galaxies in the subsamples.
In order to investigate the effects further, a larger LAE sample and/or deep mid-infrared data (e.g., obtained by JWST) are needed.}

We also discuss effects of contamination on the stacked IRAC images.
As explained in Section \ref{ss_laesample}, the contamination fraction in our LAE sample is $0-30\%$, and appears to depend on the magnitude.
Low redshift emitter contaminants do not make the IRAC excess, because no strong emission lines enter in the IRAC bands.
Thus if the LAE sample contains significant contamination, the IRAC color excess becomes weaker.
Here we roughly estimate the effect of the contamination, assuming a flat continuum of the contaminant in the IRAC bands and maximum $30\%$ contamination rate.
If the color excess of LAEs is $[3.6]-[4.5]=-0.5$ (i.e., the flux ratio of $f_{[3.6]}/f_{[4.5]}=1.6$), the $30\%$ contamination makes the mean color excess weaker by $0.1\ \m{mag}$.
Similarly if the color excess of LAEs is $[3.6]-[4.5]=-1.0$ (i.e., the flux ratio of $f_{[3.6]}/f_{[4.5]}=2.5$), the contamination makes the mean color excess weaker by $0.2\ \m{mag}$.
Although we use the median-stack images, which are different from the mean stack and not simple, the maximum effect could be $0.1-0.2\ \m{mag}$.
This effect is comparable to the uncertainties of our $[3.6]-[4.5]$ color measurements.
Thus the effect of the contamination could not be significant.

\redc{In some $z=6.6$ subsamples, LAEs are marginally detected in the stacked images bluer than the Lyman break.
The fluxes in the bluer bands are $\gtrsim7$ times fainter than that in the $y$ band.
These marginal detections could be due to the contamination of the low redshift emitters (e.g., {\sc[Oiii]} emitters), because the $\sim7$ times fainter fluxes in the $gri$ bands can be explained by the $\sim15\%$ contamination with flat continua.
However, we cannot exclude possibilities of the unrelated contamination in the line of sight and/or Lyman continuum leakage.
Larger spectroscopic \redcr{samples are} needed to distinguish these possibilities.}

\begin{deluxetable*}{cccccccccc}
\setlength{\tabcolsep}{0.35cm}
\tablecaption{Summary of the Subsamples at $z=4.9$}
\tablehead{\colhead{Redshift} & \colhead{$EW_\m{Ly\alpha}^\m{min}$} & \colhead{$EW_\m{Ly\alpha}^\m{max}$} & \colhead{$N$} & \colhead{$EW_\m{Ly\alpha}^\m{median}$} & \colhead{$M_\m{UV}^\m{median}$} & \colhead{[3.6]-[4.5]} & $EW^0_\m{H\alpha}$ & $\m{log}\xi_\m{ion}$ & $f_\m{Ly\alpha}$\\
\colhead{(1)}& \colhead{(2)}& \colhead{(3)}& \colhead{(4)} &  \colhead{(5)}& \colhead{(6)}& \colhead{(7)}& \colhead{(8)} & \colhead{(9)}& \colhead{(10)}}
\startdata
$z=4.9$ & 20.0 & 1000.0 & 99 & 63.8 & -20.6 &$-0.81\pm0.16$ &$1390^{+179}_{-447}$ &$25.48^{+0.06}_{-0.06}$ ($25.53^{+0.06}_{-0.06})$ &$0.27^{+0.30}_{-0.24}$ \\ 
$$ & 0.0 & 20.0 & 8 & 16.9 & -21.4 &$-0.26\pm0.16$ &$555^{+332}_{-311}$ &$25.27^{+0.19}_{-0.17}$ ($25.32^{+0.19}_{-0.17})$ &$0.10^{+0.16}_{-0.07}$ \\
$$ & 20.0 & 70.0 & 58 & 43.0 & -20.9 &$-0.75\pm0.14$ &$1490^{+177}_{-175}$ &$25.51^{+0.05}_{-0.05}$ ($25.56^{+0.05}_{-0.05})$ &$0.16^{+0.18}_{-0.15}$ \\
$$ & 70.0 & 1000.0 & 41 & 117.5 & -20.0 &$<-1.04$ &$>1860$ &$>25.50$ ($>25.55$) &$>0.55$
\enddata
\tablecomments{(1) Redshift of the LAE subsample. (2) Lower limit of the rest-frame Ly$\m{\alpha}$ EW of the subsample. (3) Upper limit of the rest-frame Ly$\m{\alpha}$ EW of the subsample. (4) Number of sources in the subsample. (5) Median value of the rest-frame Ly$\m{\alpha}$ EWs in the subsample. (6) Median value of the UV magnitudes in the subsample. (7) IRAC $[3.6]-[4.5]$ color. (8) H$\m{\alpha}$ EWs in the subsample. (9) Ionizing photon production efficiency in units of $\m{Hz\ erg^{-1}}$ with $f_\m{esc}^\m{ion}=0$. The value in the parentheses is the ionizing photon production efficiency with $f_\m{esc}^\m{ion}=0.1$, inferred from this study. (10) Ly$\m{\alpha}$ escape fraction.}
\label{tab_subsample_z49}
\end{deluxetable*}

\begin{deluxetable*}{cccccccccc}
\setlength{\tabcolsep}{0.35cm}
\tablecaption{Summary of the Subsamples at $z=5.7$, $6.6$, and $7.0$}
\tablehead{\colhead{Redshift} & \colhead{$EW_\m{Ly\alpha}^\m{min}$} & \colhead{$EW_\m{Ly\alpha}^\m{max}$} & \colhead{$N$} & \colhead{$EW_\m{Ly\alpha}^\m{median}$} & \colhead{$M_\m{UV}^\m{median}$} & \colhead{[3.6]-[4.5]} & \colhead{[{\sc Oiii}]$\lambda5007/\m{H\alpha}$}  & \colhead{[{\sc Oiii}]$\lambda5007/\m{H\beta}$}  & \colhead{$EW^0_{\rm [OIII]}$} \\
\colhead{(1)}& \colhead{(2)}& \colhead{(3)}& \colhead{(4)} &  \colhead{(5)}& \colhead{(6)}& \colhead{(7)}& \colhead{(8)} & \colhead{(9)} & \colhead{(10)}}
\startdata
$z=5.7$ & 20.0 & 1000.0 & 202 & 88.8 & -19.9 &$-0.55\pm0.20$ &$3.04^{+1.77}_{-1.46}$ &$8.69^{+5.06}_{-4.18}$ & $>460$ \\
$$ & 0.0 & 10.0 & 6 & 5.8 & -22.0 &$0.30\pm0.08$ &$0.45^{+0.12}_{-0.13}$ &$1.28^{+0.33}_{-0.37}$ &\nodata\\
$$ & 10.0 & 20.0 & 5 & 14.7 & -22.0 &$0.06\pm0.22$ &$0.70^{+0.41}_{-0.47}$ &$2.01^{+1.19}_{-1.34}$  &\nodata\\
$$ & 20.0 & 40.0 & 21 & 33.7 & -20.9 &$-0.37\pm0.20$ &$1.84^{+0.63}_{-0.61}$ &$5.26^{+1.80}_{-1.74}$  & $>330$\\
$$ & 40.0 & 100.0 & 107 & 75.7 & -20.2 &$-0.38\pm0.28$ &$2.47^{+3.74}_{-1.99}$ &$7.06^{+10.70}_{-5.69}$ &$>340$\\
$$ & 100.0 & 1000.0 & 74 & 125.9 & -19.4 &$<-0.64$ &$>1.27$ &$>3.63$ &$>490$ \\
$z=6.6$ & 20.0 & 1000.0 & 230 & 40.1 & -20.3 &$<-0.85$ &$>1.18$ &$>3.37$ &$>540$ \\
$$ & 0.0 & 10.0 & 17 & 6.5 & -21.6 &$-0.40\pm0.14$ &$1.00^{+1.26}_{-0.22}$ &$2.86^{+3.60}_{-0.62}$&$>310$ \\
$$ & 10.0 & 30.0 & 92 & 23.6 &  $>-20.3$ &$<-1.11$ &$>2.01$ &$>5.75$ &$>640$ \\
$$ & 30.0 & 100.0 & 148 & 44.1 &  $>-20.3$ &$<-0.56$ &$>0.55$ &$>1.58$&$>390$ \\
$$ & 100.0 & 1000.0 & 16 & 126.2 &  $>-20.3$ &$-0.30\pm0.54$ &$0.79^{+0.62}_{-0.57}$ &$2.27^{+1.78}_{-1.64}$ &\nodata\\
$z=7.0$ & 20.0 & 1000.0 & 20 & 88.1 & -20.4 &$<-0.85$ &$<0.86$ &$<2.46$&\nodata
\enddata
\tablecomments{(1) Redshift of the LAE subsample. (2) Lower limit of the rest-frame Ly$\m{\alpha}$ EW of the subsample. (3) Upper limit of the rest-frame Ly$\m{\alpha}$ EW of the subsample. (4) Number of sources in the subsample. (5) Median value of the rest-frame Ly$\m{\alpha}$ EWs in the subsample. (6) Median value of the UV magnitudes in the subsample. The lower limit indicates that more than half of the LAEs in that subsample are not detected in the rest-frame UV band. (7) IRAC $[3.6]-[4.5]$ color. (8) [{\sc Oiii}]$\lambda5007/\m{H\alpha}$ line flux ratio of the subsample. For the $z=7.0$ subsample, the [{\sc Oiii}]$/\m{H\alpha}$ ratio is calculated from the [{\sc Oiii}]$/\m{H\beta}$ ratio assuming $\m{H\alpha}/\m{H\beta}=2.86$. (9) [{\sc Oiii}]$\lambda5007/\m{H\beta}$ line flux ratio of the subsample. For the $z=5.7$ and $6.6$ subsamples, the [{\sc Oiii}]$/\m{H\beta}$ ratios are calculated from the [{\sc Oiii}]$/\m{H\alpha}$ ratio assuming $\m{H\alpha}/\m{H\beta}=2.86$.
(10) Lower limit of the rest-frame {\sc [Oiii]}5007 EW assuming no emission line in the $[4.5]$ band.}
\label{tab_subsample}
\end{deluxetable*}

\subsection{Model SED}\label{ss_sed}
We generate the model SEDs at $z=4.9$, $5.7$, $6.6$, and $7.0$ using BEAGLE \citep{2016MNRAS.462.1415C}.
In BEAGLE, we use the combined stellar population + photoionization model presented in \citet{2016MNRAS.462.1757G}. 
Stellar emission is based on an updated version of the population synthesis code of \citet{2003MNRAS.344.1000B}, while gas emission is computed with the standard photoionization code CLOUDY\citep{2013RMxAA..49..137F} following the prescription of \citet{2001MNRAS.323..887C}.
The IGM absorption is considered following a model of \cite{2014MNRAS.442.1805I}.  
In BEAGLE we vary the total mass of stars formed, ISM metallicity ($Z_\m{neb}$), ionization parameter ($\m{log}U_\m{ion}$), star formation history, stellar age, and $V$-band attenuation optical depth ($\tau_V$), while we fix the dust-to-metal ratio ($\xi_d$) to $0.3$ \citep[e.g.,][]{2017MNRAS.471.1743D}, and adopt the \citet{2000ApJ...533..682C} dust extinction curve.
The choice of the extinction law does not affect our conclusions, because our SED fittings infer dust-poor populations such as $\tau_V=0.0-0.1$.
Here, we adopt a constant star formation history, and vary the four adjustable parameters of the model in vast ranges, $-2.0<\m{log}(Z_\m{neb}/Z_\odot)<0.2$ (with a step of $0.1\ \m{dex}$), $-3.0<\m{log}U_\m{ion}<-1.0$ (with a step of $0.1\ \m{dex}$), $6.0<\m{log}(\m{Age/yr})<9.1$ (with a step of $0.1\ \m{dex}$), and $\tau_\m{V}=[0,0.05,0.1,0.2,0.4,0.8,1.6,2]$.
The lower limit of the ionization parameter is consistent with recent observations for high redshift galaxies \citep[e.g.,][]{2017PASJ...69...44K}.
The upper limit of the ionization parameter is set to the very high value, because recent observations suggest increase of the ionization parameter toward high redshift \citep{2013ApJ...769....3N}.
The upper limit of the stellar age corresponds to the cosmic age at $z=4.9$ ($9.08\ \m{Gyr}$).
These parameter ranges cover previous results for high redshift LAEs \citep[e.g.,][]{2010ApJ...724.1524O,2010MNRAS.402.1580O}.
We assume that the stellar metallicity is the same as the ISM metallicity, with interpolation of original templates. 
We fix the stellar mass as $M_*=10^{8}\ M_\odot$, which will be scaled later.
In generating model SEDs, we remove emission lines at $4000\ \m{\AA}<\lambda_\m{rest}<7000\ \m{\AA}$, because we estimate line fluxes of the LAEs by measuring the difference between the observed photometry (emission line contaminated) and the model continuum (no emission lines).
We also calculate the Ly$\alpha$ EW of each model SED assuming the Case B recombination without considering the resonance scattering \citep{1989agna.book.....O}, which will be compared with the observed $EW^0_\m{Ly\alpha}$.

\begin{figure*}
\begin{center}
  \includegraphics[clip,bb=10 10 580 430,width=0.93\hsize]{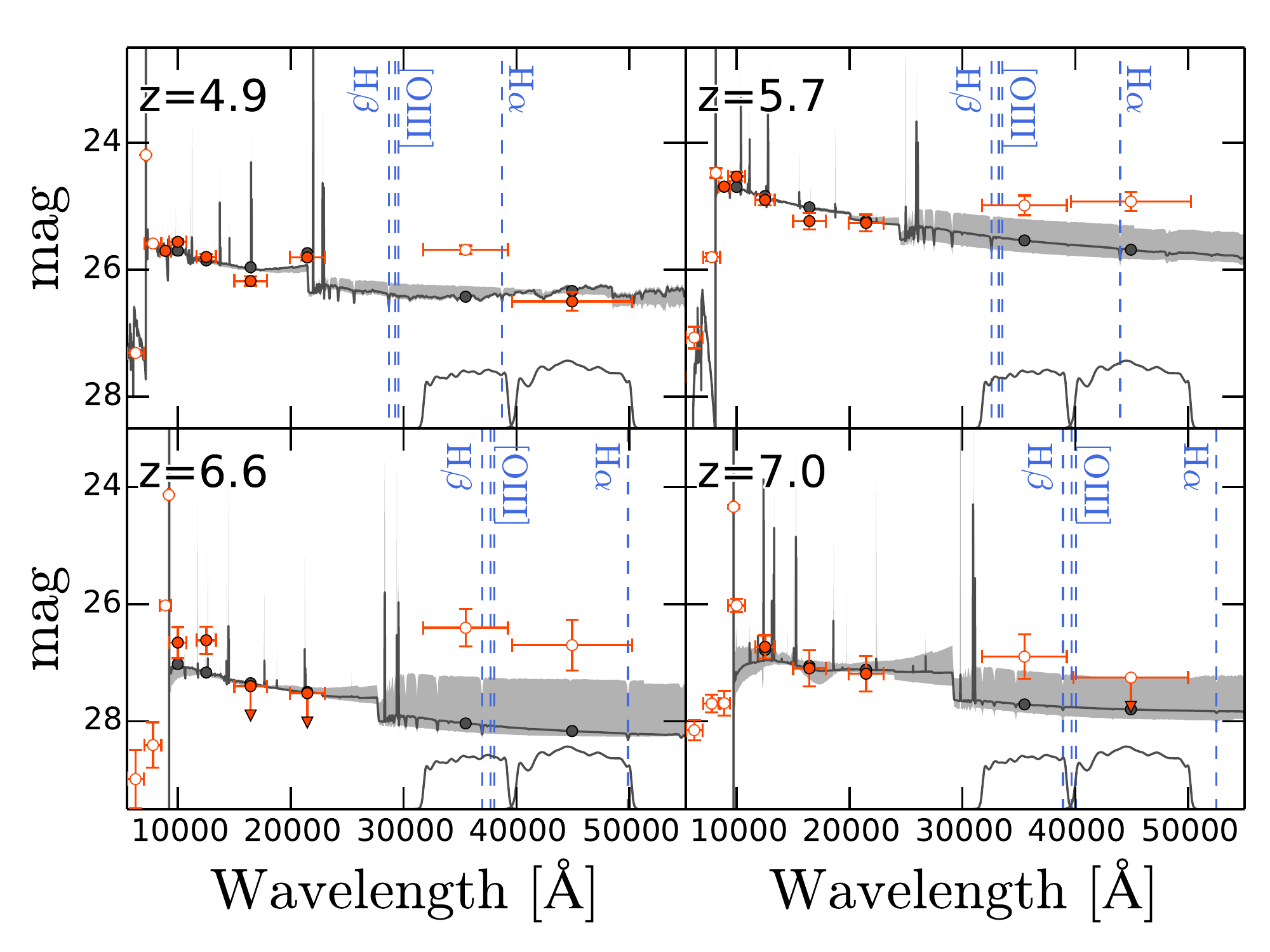}
 \end{center}
   \caption{Examples of the best-fit model SEDs for the subsamples of $z=4.9$, $20<EW^0_\m{Ly\alpha}<1000\ \m{\AA}$ (upper left), $z=5.7$, $10<EW^0_\m{Ly\alpha}<20\ \m{\AA}$ (upper right), $z=6.6$, $100<EW^0_\m{Ly\alpha}<1000\ \m{\AA}$ (lower left), and $z=7.0$, $20<EW^0_\m{Ly\alpha}<1000\ \m{\AA}$ (lower right).
   The red circles represent the magnitudes in the stacked images of each subsample.
   The filled red circles are magnitudes used in the SED fittings.
   We do not use the magnitudes indicated with the red open circles which are affected by the IGM absorption or strong emission lines.
   The dark gray lines with the gray circles show the best-fit model SEDs, without emission lines at $4000<\lambda_\m{rest}<7000\ \m{\AA}$.
   The light gray regions show the $1\sigma$ uncertainties of the best-fit model SEDs.
   We also plot the filter response curves of the IRAC $[3.6]$ and $[4.5]$ bands with the gray curves in each panel.
      \label{fig_sed}}
\end{figure*}

\subsection{Line Flux Estimate}\label{ss_lineflux}
We estimate rest-frame optical emission line fluxes by comparing the stacked SEDs (Section \ref{ss_stack}) with the model SEDs (Section \ref{ss_sed}).
We use 7 ($zyJHK_s[4.5]EW_\m{Ly\alpha}$), 6 ($zyJHK_s(K)EW_\m{Ly\alpha}$), 5 ($yJHK_s(K)EW_\m{Ly\alpha}$), and 4 ($JHK_sEW_\m{Ly\alpha}$) observational data points to constrain the model SEDs at $z=4.9$, $5.7$, $6.6$, and $7.0$, respectively.
Firstly from the all models, we remove models whose Ly$\m{\alpha}$ EWs are lower than the minimum $EW^0_\m{Ly\alpha}$ of each subsample.
We keep models with $EW^0_\m{Ly\alpha}$ higher than the maximum EW of each subsample, because the EWs in the models could be overestimated, as we do not account for the enhanced absorption by dust of resonantly scattered Ly$\m{\alpha}$ photons in the neutral ISM.
Then, the model SEDs are normalized to the fluxes of the stacked images in bands redder than the Ly$\m{\alpha}$ emission and free from the strong rest-frame optical emission lines (i.e., $zyJHK_s[4.5]$, $zyJHK_s(K)$, $yJHK_s(K)$, and $JHK_s$ for $z=4.9$, $5.7$, $6.6$, and $7.0$ LAEs, respectively) by the least square fits.
We then calculate the $\chi^2$ value of each model with these band fluxes, and adopt the least $\chi^2$ model as the best-fit model.

Figure \ref{fig_sed} shows examples of the best-fit SEDs with the observed magnitudes.
The uncertainty of the model is computed with the models in the $1\sigma$ confidence interval.
We calculate the flux differences between the stacked SEDs and the model SEDs in the $[3.6]$ band at $z=4.9$, and $[3.6]$ and $[4.5]$ bands at $z=5.7$, $6.6$, and $7.0$.
The flux differences are corrected for dust extinction with the $\tau_\m{V}$ values in the models, assuming the \citet{2000ApJ...533..682C} extinction curve.

We estimate the H$\m{\alpha}$, H$\m{\beta}$, and [{\sc Oiii}]$\lambda$5007 line fluxes from these flux differences.
Here we consider H$\m{\alpha}$, H$\m{\beta}$, [{\sc Oiii}]$\lambda\lambda$4959,5007, [{\sc Nii}]$\lambda6584$, and [{\sc Sii}]$\lambda\lambda6717,6731$ emission lines, because the other emission lines redshifted into the $[3.6][4.5]$ bands are weak in the metallicity range of $0.02<Z/Z_\odot<2.5$ \citep{2003A&A...401.1063A}.
We use averaged filter throughputs of the $[3.6]$ and $[4.5]$ bands at the wavelength of each redshifted emission line calculated with the redshift distributions of the LAE samples.
We assume the Case B recombination with the electron density of $n_\m{e}=100\ \m{cm^{-3}}$ and the electron temperature of $T_\m{e}=10000\ \m{K}$ \citep[$\m{H\alpha /H\beta}=2.86$;][]{1989agna.book.....O}, and typical line ratios of [{\sc Oiii}]$\lambda$$4959/$[{\sc Oiii}]$\lambda$$5007=0.3$ \citep{2017PASJ...69...44K}, [{\sc Nii}]$/\m{H\alpha}=0.068$, and [{\sc Sii}]$/\m{H\alpha}=0.095$ for sub-solar ($0.2\ Z_\odot$) metallicity (\citealt{2003A&A...401.1063A}, see also \citealt{2017arXiv171000834F}).
Note that these assumptions do not affect our final results, because the statistical uncertainties of the line fluxes or ratios in our study are larger than $10\%$.
For example, recent observations suggest relatively high electron densities of $n_\m{e}\sim100-1000\ \m{cm^{-3}}$ \citep{2015MNRAS.451.1284S,2016ApJ...822...42O,2016ApJ...816...23S,2017ApJ...835...88K}, but our conclusions do not change if we adopt $n_\m{e}=1000\ \m{cm^{-3}}$.
The uncertainties of the emission line fluxes include both the photometric errors and the SED model uncertainties.

\begin{figure}
 \begin{center}
  \includegraphics[clip,bb=0 20 290 280,width=1\hsize]{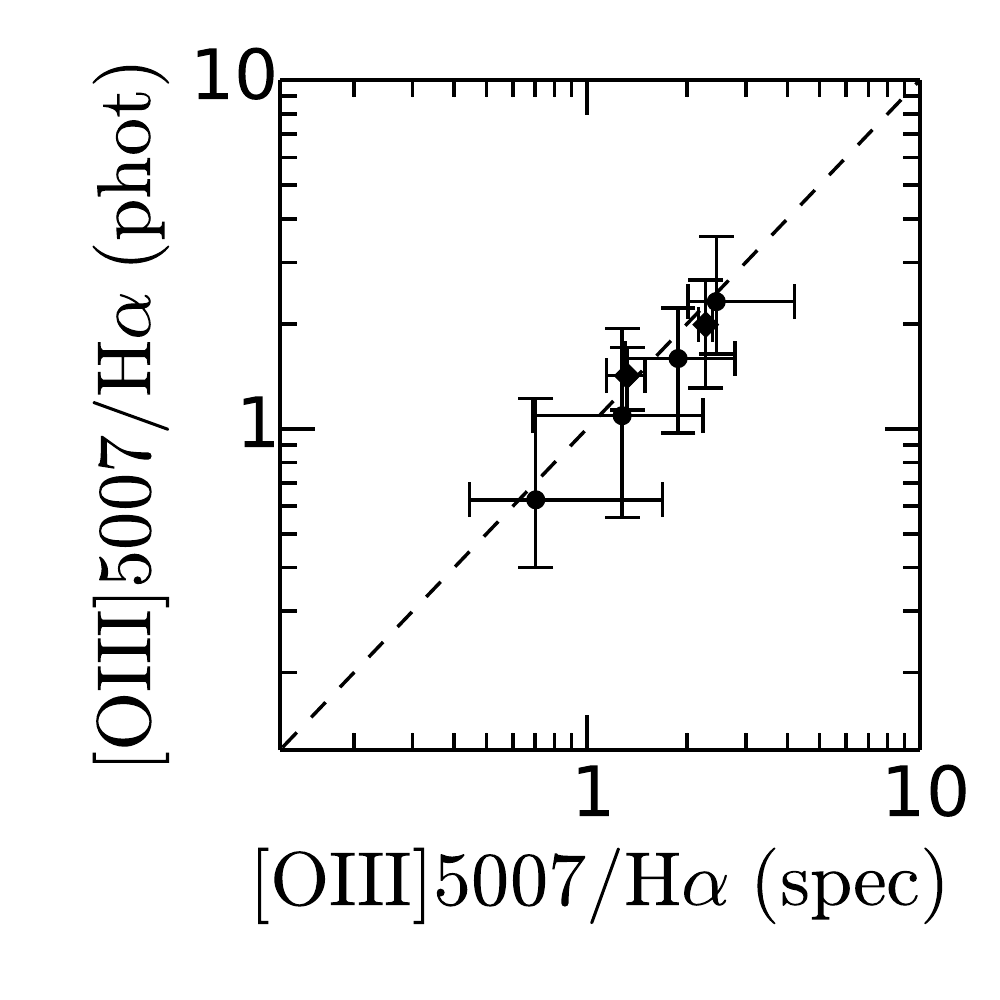}
 \end{center}
   \caption{Comparison of the flux ratios estimated from the spectroscopic and photometric data at $z=1.2-2.2$.
   \redc{The circles (diamonds) represent [{\sc Oiii}]$\lambda$5007/H$\alpha$ flux ratios of galaxies at $z=1.20-1.56$ ($z=2.2$) estimated from the photometric data as a function of those from spectroscopy.
   See Section \ref{ss_lineflux} for more details.}
         \label{fig_testz22}}
\end{figure}

We check the reliability of this flux estimation method.
\redc{We use galaxies at $z=1.2-1.6$ whose [{\sc Oiii}]$\lambda$5007 and H$\alpha$ emission lines are redshifted into the $J_{125}$ and $H_{160}$ bands, respectively.
From the 3D-HST catalogs \citep{2012ApJS..200...13B,2014ApJS..214...24S,2016ApJS..225...27M}, we select 211 galaxies at $z_\m{spec}=1.20-1.56$ in the GOODS-South field with [{\sc Oiii}] and H$\alpha$ emission lines detected at $>3\sigma$ levels.
In addition to the spectroscopic data in \citet{2016ApJS..225...27M}, we estimate the [{\sc Oiii}] and H$\alpha$ fluxes from the broad-band magnitudes following the method described above.
We divide the galaxies \redcr{into} subsamples, and plot the median and $1\sigma$ scatter of the [{\sc Oiii}]/H$\alpha$ ratio of each subsample in Figure \ref{fig_testz22}.
Furthermore, we plot the [{\sc Oiii}]/H$\alpha$ ratios of two LAEs at $z=2.2$, COSMOS-30679 \citep{2013ApJ...769....3N} and COSMOS-12805 \citep{2017PASJ...69...44K}, whose [{\sc Oiii}] and H$\alpha$ lines enter in the $H$ and $K_s$ bands, respectively.
We \redcr{also} measure magnitudes of COSMOS-30679 and COSMOS-12805 in our $grizyJHK_s[3.6][4.5]$ images.}
Although the uncertainties of the ratios estimated from photometry are large, they agree with those from spectroscopy within a factor of $\sim1.5$.
Thus this flux estimation method is valid.

\section{Results}\label{ss_result}
\subsection{Properties of $z=4.9$ LAEs}
\subsubsection{Inferred H$\m{\alpha}$ EW}\label{ss_EWHa}
\begin{figure*}
\begin{center}
  \includegraphics[clip,bb=0 10 580 290,width=0.95\hsize]{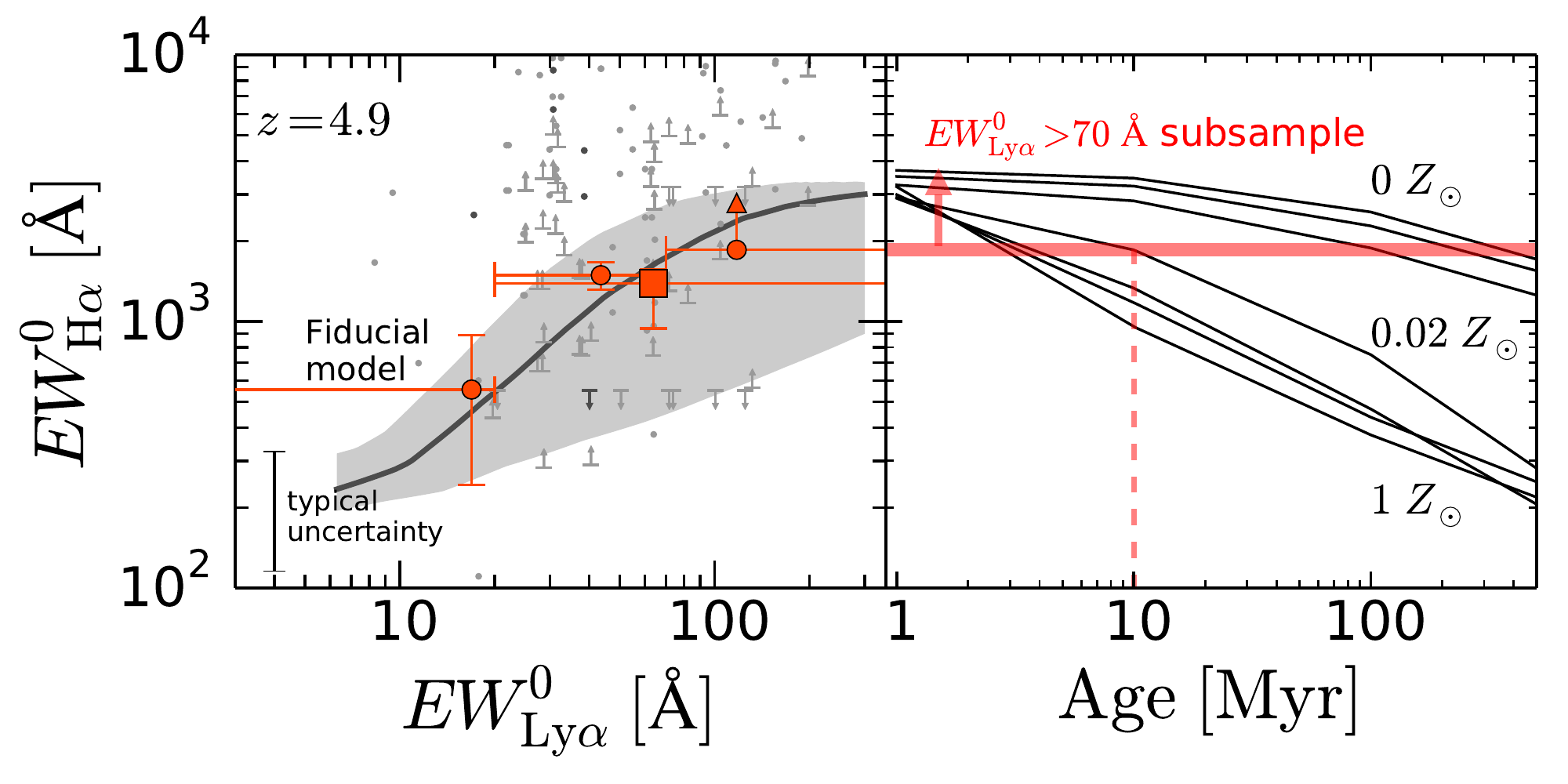}
 \end{center}
   \caption{
   {\bf Left panel:} H$\m{\alpha}$ EWs as a function of Ly$\m{\alpha}$ EWs at $z=4.9$.
   The red square and circles are the results from the stacked images of the subsamples, and the gray dots show the EWs of the individual objects detected in the $[3.6]$ and/or $[4.5]$ bands.
   The red square is the result of the $EW^0_\m{Ly\alpha}>20\ \m{\AA}$ LAE subsample.
   The dark and light gray dots are objects spectroscopically confirmed and not, respectively. 
   The upward and downward arrows represent the $2\sigma$ lower and upper limits, respectively.
   The dark gray curve and the shaded region show the prediction from the fiducial model (see Section \ref{ss_fiducial}).
   {\bf Right panel:} Inferred stellar age and metallicity from the constrained $EW_\m{H\alpha}^0$.
   The red solid line shows the lower limit of $EW_\m{H\alpha}^0\gtrsim2000\ \m{\AA}$ in the $70\ \m{\AA}<EW_\m{Ly\alpha}<1000\ \m{\AA}$ subsample at $z=4.9$.
   The black curves represent $EW_\m{H\alpha}^0$ calculated in \citet{2011MNRAS.415.2920I} with metallicities of $Z=0$, $5\times10^{-6}$, $5\times10^{-4}$, $0.02$, $0.2$, $0.4$, and $1\ Z_\odot$.
   The H$\m{\alpha}$ EW indicates very young stellar age of $<10\ \m{Myr}$ or very low-metallicity of $Z<0.02\ Z_\odot$.
      \label{fig_Ha}}
\end{figure*}

The left panel in Figure \ref{fig_Ha} shows rest-frame H$\m{\alpha}$ EWs ($EW^0_\m{H\alpha}$) as a function of Ly$\m{\alpha}$ EWs at $z=4.9$.
The H$\m{\alpha}$ EW increases from $\sim600\ \m{\AA}$ to $>1900\ \m{\AA}$ with increasing Ly$\m{\alpha}$ EW.
$EW^0_\m{H\alpha}$ of the low-$EW^0_\m{Ly\alpha}$ subsample ($EW^0_\m{Ly\alpha}<20\ \m{\AA}$) is $\sim600\ \m{\AA}$, relatively higher than results of $M_*\sim10^{10}\ M_\odot$ galaxies at $z\sim5$ \citep[$300-400\ \m{\AA}$;][]{2016ApJ...821..122F}, because our galaxies may be less massive ($\m{log}(M_*/M_\odot)\sim8-9$) than the galaxies in \citet{2016ApJ...821..122F}.
On the other hand, the high Ly$\m{\alpha}$ EW ($EW^0_\m{Ly\alpha}>70\ \m{\AA}$) subsample has $EW^0_\m{H\alpha}\gtrsim1900\ \m{\AA}$, which is $\gtrsim5$ times higher than that of the $M_*\sim10^{10}\ M_\odot$ galaxies. 
Based on photoionization model calculations in \citet{2011MNRAS.415.2920I}, this high $EW^0_\m{H\alpha}$ value indicates very young stellar age of $<10\ \m{Myr}$ or very low metallicity of $<0.02\ Z_\odot$ (the right panel in Figure \ref{fig_Ha}).
The individual galaxies are largely scattered beyond the typical uncertainty, probably due to varieties of the stellar age and metallicity.

We compare the H$\alpha$ EWs of the $z=4.9$ LAEs with those of galaxies at other redshifts.
\citet{2014MNRAS.437.3516S} report median H$\alpha$ EWs of $30-200\ \m{\AA}$ for galaxies with $\m{log}(M_*/M_\odot)=9.0-11.5$ at $z=0.40-2.23$.
Our H$\alpha$ EWs are more than two times higher than the galaxies in \citet{2014MNRAS.437.3516S}.
The high EW ($\sim1400\ \m{\AA}$) of our LAEs is comparable to an extrapolation of the scaling relation in \citet{2014MNRAS.437.3516S}, $EW^0_\m{H\alpha}/\m{\AA}\sim7000(M_*/M_\odot)^{-0.25}(1+z)^{1.72}$, for $z=4.9$ and $\m{log}(M_*/M_\odot)=8.15$ (see Section \ref{ss_mainsequence}).
This good agreement indicates that this scaling relation may hold at $z\sim5$ and the lower stellar mass.
\citet{2012ApJ...757L..22F} estimate H$\alpha$ EWs of galaxies at $z\sim1$ to be $EW_\m{H\alpha}=10-100\ \m{\AA}$, which are lower than ours.
The high H$\alpha$ EWs of our $z=4.9$ LAEs are comparable to those of galaxies at $z\sim6.7$ \citep{2014ApJ...784...58S}.

\subsubsection{Ly$\m{\alpha}$ Escape Fraction}\label{ss_fLya}
We estimate the Ly$\m{\alpha}$ escape fraction, $f_\m{Ly\alpha}$, which is the ratio of the observed $\m{Ly\alpha}$ luminosity to the intrinsic one, by comparing Ly$\alpha$ with H$\alpha$.
Because Ly$\alpha$ photons are resonantly scattered by neutral hydrogen (HI) gas in the ISM, the Ly$\alpha$ escape fraction depends on kinematics and distribution of the ISM, as well as the metallicity of the ISM.
The Ly$\m{\alpha}$ escape fraction can be estimated by the following equation:
\begin{equation}
f_\m{Ly\alpha}=\frac{L^\m{obs}_\m{Ly\alpha}}{L^\m{int}_\m{Ly\alpha}}=\frac{L^\m{obs}_\m{Ly\alpha}}{8.7L^\m{int}_\m{H\alpha}},
\end{equation}
where subscripts ``int" and ``obs" refer to the intrinsic and observed luminosities, respectively.
Here we assume the Case B recombination \citep{1971MNRAS.153..471B}.
The intrinsic H$\alpha$ luminosities are derived from the dust-corrected H$\alpha$ fluxes, estimated in Section \ref{ss_lineflux}.

\begin{figure}
 \begin{center}
  \includegraphics[clip,bb=10 10 350 280,width=1\hsize]{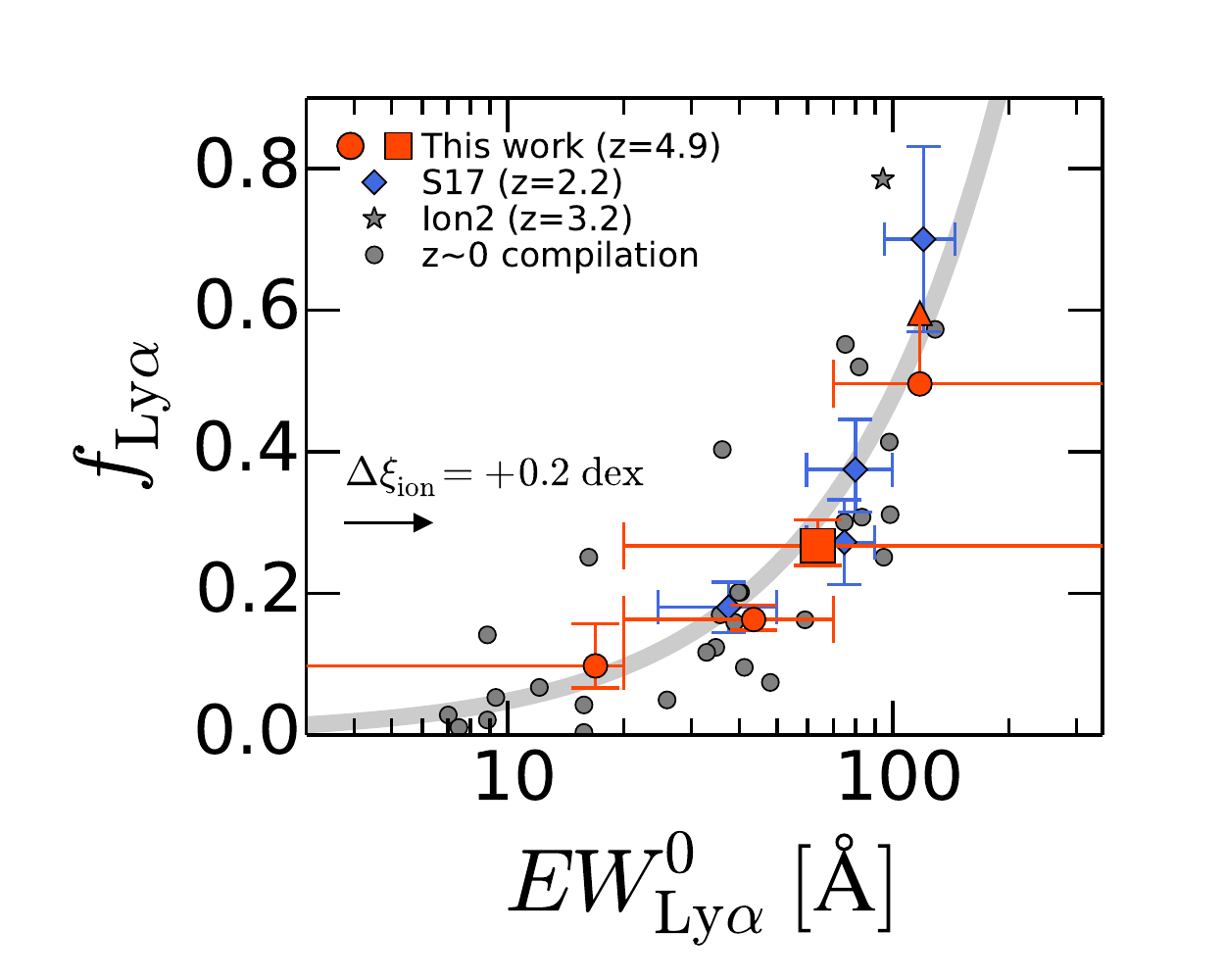}
 \end{center}
   \caption{
Ly$\m{\alpha}$ escape fractions of the LAEs at $z=4.9$ as a function of Ly$\alpha$ EW.
   The red squares and circle show the results of the subsamples divided by $EW_\m{Ly\alpha}^0$, and the upward arrow represents the $2\sigma$ lower limit.
   We plot the Ly$\m{\alpha}$ escape fractions of $z=2.2$ LAEs in \citet{2017MNRAS.466.1242S} with the blue diamonds.
   The gray star and circles are the Ly$\m{\alpha}$ escape fractions of ``Ion2" at $z=3.2$    \citep{2016A&A...585A..51D,2016ApJ...825...41V} and local galaxies \citep{2009MNRAS.399.1191C,2015ApJ...809...19H,2016ApJ...820..130Y,2005ApJ...619L..35H,2009ApJ...706..203O,2014ApJ...782....6H,2014ApJ...797...11O}, respectively.
  \redc{The gray curve represents Equation (\ref{eq_fescLya}).}
   The black arrow indicates the shift in $EW_\m{Lya}^0$ ($\propto L_\m{Ly\alpha}/L_\m{UV}$), which is expected for a higher $\xi_\m{ion}$ ($\propto L_\m{H\alpha}/L_\m{UV}$) with constant $f_\m{Ly\alpha}$ ($\propto L_\m{Ly\alpha}/L_\m{H\alpha}$).
         \label{fig_fLya}}
\end{figure}

We plot the estimated Ly$\m{\alpha}$ escape fractions as a function of Ly$\m{\alpha}$ EW in Figure \ref{fig_fLya}.
The Ly$\m{\alpha}$ escape fraction increases from $\sim10\%$ to $>50\%$ with increasing $EW_\m{Ly\alpha}^0$ from $20\ \m{\AA}$ to $100\ \m{\AA}$, whose trend is identified for the first time at $z=4.9$.
In addition, the escape fractions at $z=4.9$ agree very well with those at $z=2.2$ at given $EW_\m{Ly\alpha}^0$ \citep{2017MNRAS.466.1242S}.
\citet{2017MNRAS.466.1242S} suggest a possible non-evolution of the $f_\m{Ly\alpha}-EW_\m{Ly\alpha}^0$ relation from $z\sim0$ to $z=2.2$.
We confirm this redshift-independent $f_\m{Ly\alpha}-EW_\m{Ly\alpha}^0$ relation up to $z=4.9$.
\redc{In Figure \ref{fig_fLya}, we also plot the following relation \redcr{in \citet{2018arXiv180308923S}}, which fits our's and previous results:}
\begin{equation}
\redcr{f_\m{Ly\alpha}=0.0048(EW_\m{Ly\alpha}/\m{\AA}).\label{eq_fescLya}}
\end{equation}
We will discuss implications of these results in Section \ref{ss_reio}.

\begin{figure*}
\begin{center}
  \begin{minipage}{0.55\hsize}
 \begin{center}
  \includegraphics[clip,bb=0 15 420 290,width=1\hsize]{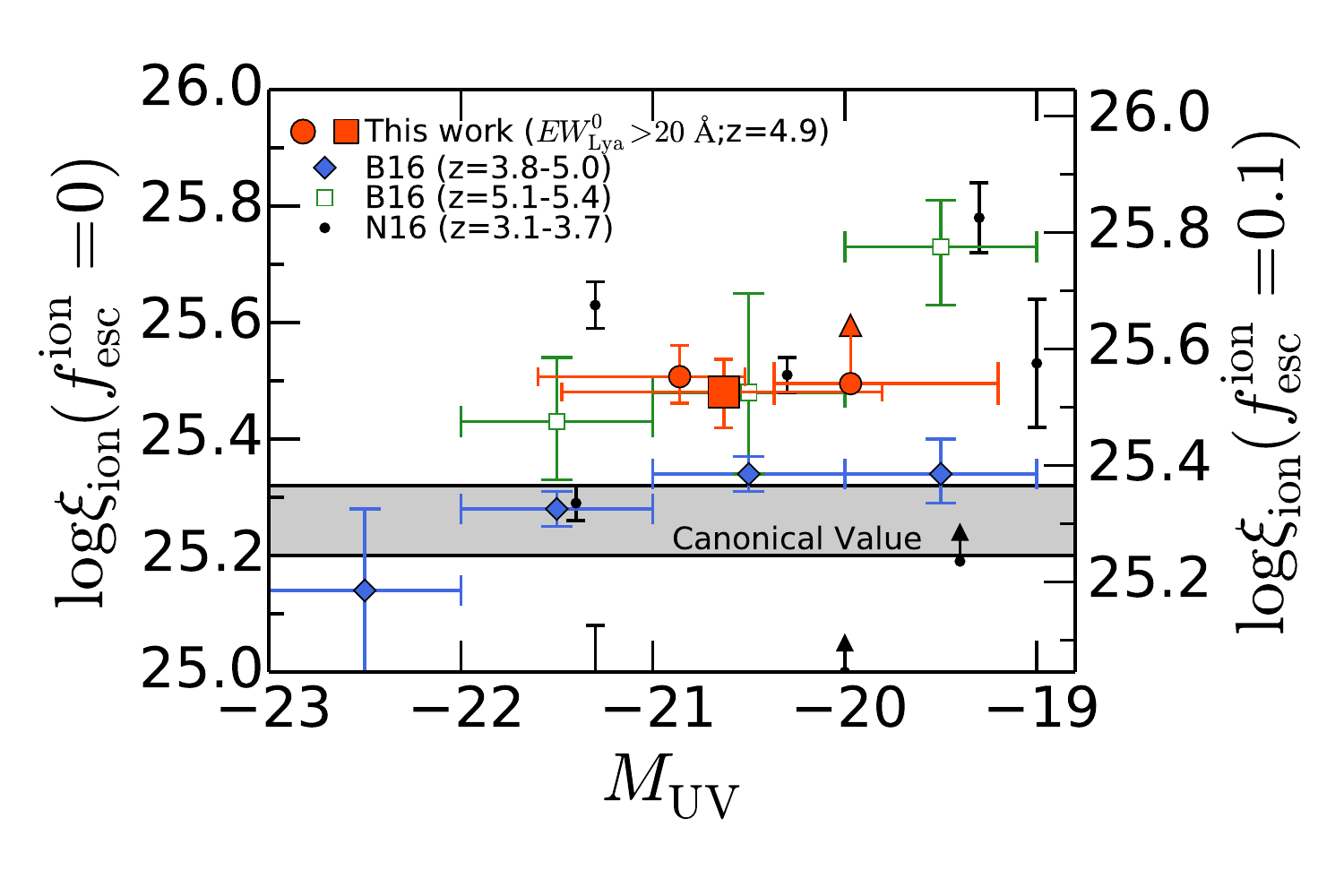}
 \end{center}
 \end{minipage}
 \begin{minipage}{0.44\hsize}
 \begin{center}
  \includegraphics[clip,bb=0 5 345 290,width=1\hsize]{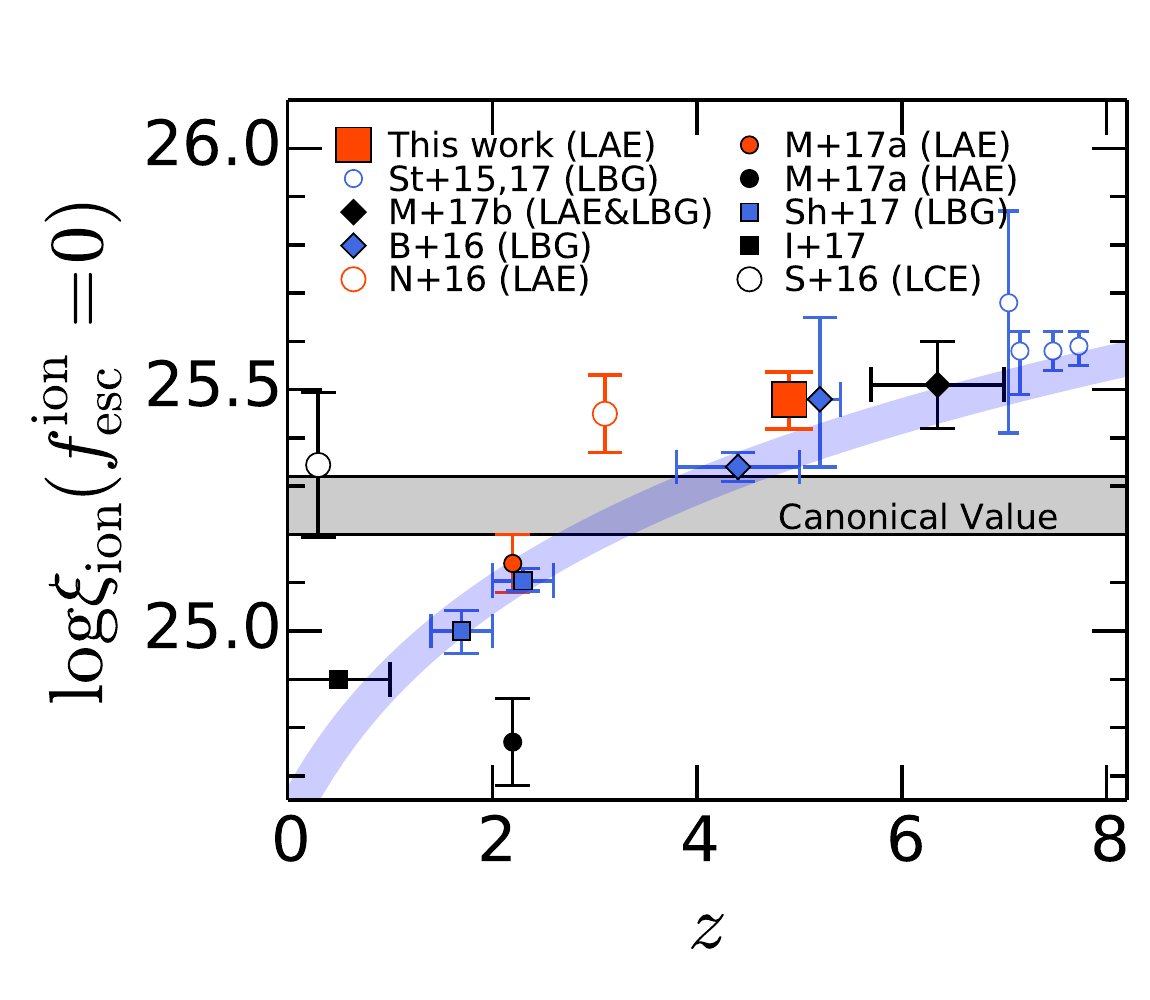}
 \end{center}
 \end{minipage}
 \\
 \end{center}
   \caption{
   {\bf Left panel:} Inferred ionizing photon production efficiencies of the LAEs at $z=4.9$ as a function of UV magnitude.
   The left and right axes represent the efficiencies with the ionizing photon escape fractions of $0$ and $10\%$, respectively.
   The red circles and square show the results of the subsamples divided by $EW_\m{Ly\alpha}^0$, and the upward arrow represents the $2\sigma$ lower limit.
   The $\xi_\m{ion}$ values of LBGs at $z=3.8-5.0$ in \citet{2016ApJ...831..176B} are represented as the blue diamonds.
   For references, we plot the $\xi_\m{ion}$ values of LBGs at $z=5.1-5.4$ \citep{2016ApJ...831..176B} and LAEs at $z=3.1-3.7$ \citep{2016ApJ...831L...9N} with the green open squares and black circles, respectively.
   The gray shaded region indicates typically assumed $\xi_\m{ion}$ \citep[see Table 2 in][]{2016ApJ...831..176B}.
 \redc{{\bf Right panel:} Redshift evolution of $\xi_\m{ion}$.
 The red square denotes $\xi_\m{ion}$ of our $EW^0_\m{Ly\alpha}>20\ \m{\AA}$ LAE subsample.
 We also plot results of \citet[][blue open circle]{2015MNRAS.454.1393S,2017MNRAS.464..469S}, \citet[][black diamond]{2017MNRAS.472..772M}, \citet[][blue diamons]{2016ApJ...831..176B}, \citet[][red open circle]{2016ApJ...831L...9N},  \citet[red and black squares for LAEs and H$\alpha$ emitters (HAEs), respectively]{2017MNRAS.465.3637M}, \citet[][blue square]{2017arXiv171100013S}, \citet[][black square]{2017MNRAS.467.4118I}, and \citet[][black open circle for Lyman continuum emitters (LCEs)]{2016A&A...591L...8S}.
 The blue curve represents the redshift evolution of $\xi_\m{ion}\propto(1+z)$ \citep{2017MNRAS.465.3637M}.}
         \label{fig_xi}}
\end{figure*}

\subsubsection{Ionizing Photon Production Efficiency}\label{ss_xi}
We estimate the ionizing photon production efficiencies of the $z=4.9$ LAEs from their H$\m{\alpha}$ fluxes and UV luminosities.
The definition of the ionizing photon production efficiency is as follows:
\begin{equation}
\xi_\m{ion}=\frac{N(\m{H^0})}{L_\m{UV}},
\end{equation}
where $N(\m{H^0})$ is the production rate of the ionizing photon which can be estimated from the H$\m{\alpha}$ luminosity using a conversion factor by \citet{1995ApJS...96....9L},
\begin{eqnarray}
L^\m{int}_\m{H\alpha}[\m{erg\ s^{-1}}]&=&1.36\times10^{-12}N_\m{obs}(\m{H^0})[\m{s}^{-1}]\notag\\
&=&1.36\times10^{-12}(1-f_\m{esc}^\m{ion})N(\m{H^0})[\m{s}^{-1}],\label{eq_nh0}
\end{eqnarray}
where $f_\m{esc}^\m{ion}$ is the ionizing photon escape fraction, and $N_\m{obs}(\m{H^0})$ is the ionizing photon production rate with $f_\m{esc}^\m{ion}=0$.

The left panel in Figure \ref{fig_xi} shows estimated $\xi_\m{ion}$ values as a function of UV magnitude.
We calculate the values of the $\xi_\m{ion}$ in two cases; $f_\m{esc}^\m{ion}=0$, following previous studies such as \citet{2016ApJ...831..176B}, and $f_\m{esc}^\m{ion}=0.1$, inferred from our analysis in Section \ref{ss_fesc}.
The ionizing photon production efficiency is estimated to be $\m{log}\xi_\m{ion}/\m{[Hz\ erg^{-1}]}=25.48^{+0.06}_{-0.06}$ for the $EW^0_\m{Ly\alpha}>20\ \m{\AA}$ subsample with $f_\m{esc}^\m{ion}=0$.
This value is systematically higher than those of LBGs at the similar redshift and UV magnitude \citep[$\m{log}\xi_\m{ion}/\m{[Hz\ erg^{-1}]}\simeq25.3$;][]{2016ApJ...831..176B} and the canonical values ($\m{log}\xi_\m{ion}/\m{[Hz\ erg^{-1}]}\simeq25.2-25.3$) by $60-100\%$.
These higher $\xi_\m{ion}$ in our LAEs may be due to the younger age (see Section \ref{ss_EWHa}) or higher ionization parameter \citep{2013ApJ...769....3N,2014MNRAS.442..900N}.

\redc{We also compare $\xi_\m{ion}$ of our LAEs with studies at different redshifts in the right panel in Figure \ref{fig_xi}.
Our estimates for the $z=4.9$ LAEs are comparable to those of LBGs at higher redshift, $z=5.1-5.4$ \citep{2016ApJ...831..176B}, and of bright galaxies at $z\sim5.7-7.0$ \citep{2017MNRAS.472..772M}.
Our estimates are higher than those of galaxies at $z\sim2$ \citep{2017MNRAS.465.3637M,2017arXiv171100013S}.}

\subsubsection{Ionizing Photon Escape Fraction}\label{ss_fesc}
\begin{figure*}
\begin{center}
  \begin{minipage}{0.45\hsize}
 \begin{center}
  \includegraphics[clip,bb=0 10 345 290,width=1\hsize]{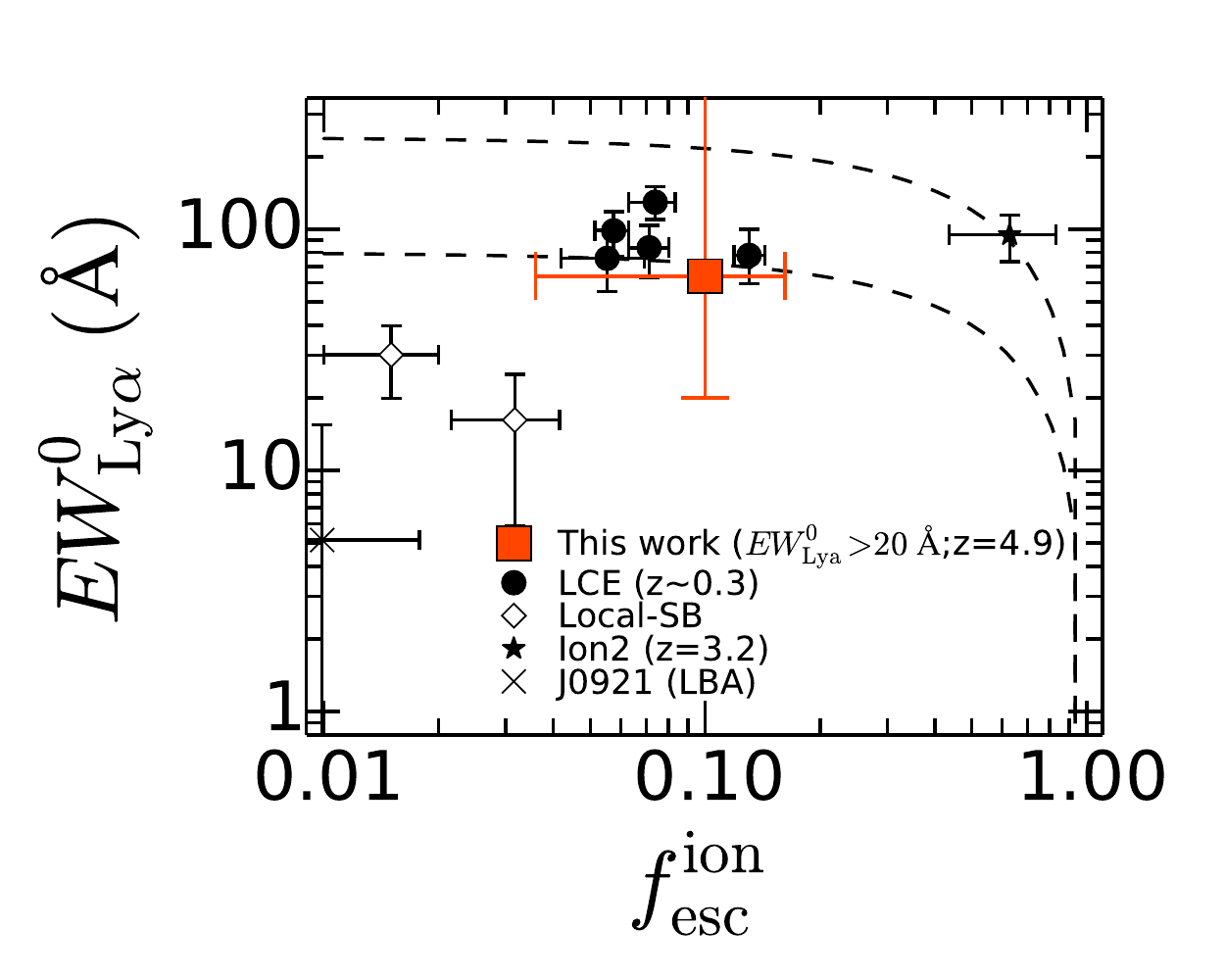}
 \end{center}
 \end{minipage}
 \begin{minipage}{0.45\hsize}
 \begin{center}
  \includegraphics[clip,bb=0 10 345 290,width=1\hsize]{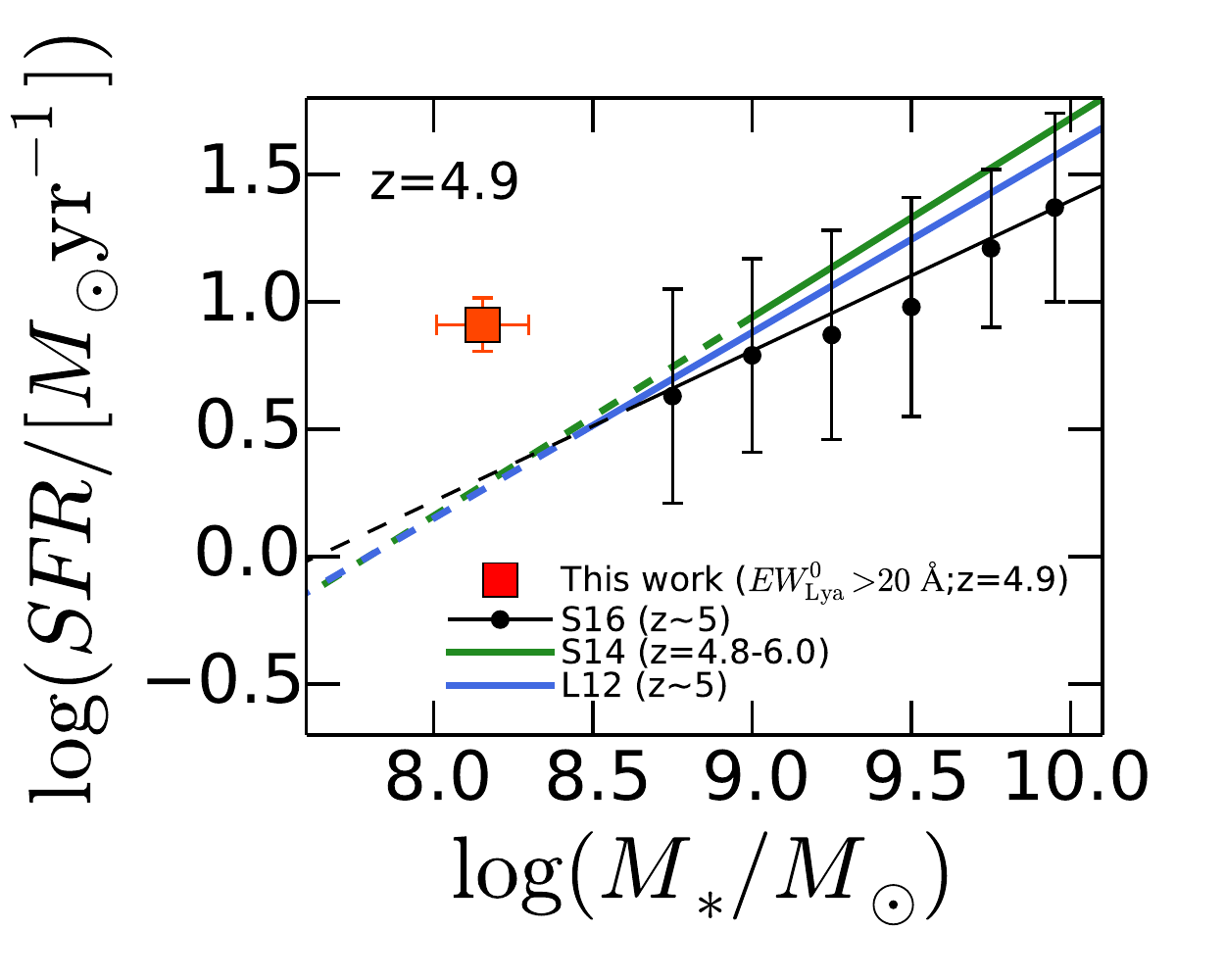}
 \end{center}
 \end{minipage}
 \\
 \end{center}
   \caption{
   {\bf Left panel:} Inferred ionizing photon escape fraction ($f_\m{esc}^\m{ion}$) of the LAEs at $z=4.9$.
 The red square denotes the escape fraction of the $EW^0_\m{Ly\alpha}>20\ \m{\AA}$ LAE subsample.
 The black circles, open diamonds, the black star, and the cross are the results of LCEs \citep{2016Natur.529..178I,2016MNRAS.461.3683I,2017A&A...597A..13V}, local star-burst galaxies \citep{2016ApJ...823...64L}, ``Ion2" at $z=3.2$ \citep{2016A&A...585A..51D,2016ApJ...825...41V}, and a Lyman break analog \citep[LBA;][]{2014Sci...346..216B}, respectively.
 The upper (lower) dashed curve is a theoretical prediction for the same attenuation in the Ly$\alpha$ and in the Lyman continuum emission, $EW_\m{Ly\alpha}=EW^\m{SFH}_\m{Ly\alpha}\times(1-f^\m{ion}_\m{esc})$, for an instantaneous burst (constant) star formation history of $EW^\m{SFH}_\m{Ly\alpha}=240\ \m{\AA}$ ($80 \m{\AA}$), following \citet{2017A&A...597A..13V}.
  {\bf Right panel:} 
 Stellar mass and SFR of the $z=4.9$ LAEs.
   The red square is the result of the subsample with $EW^0_\m{Ly\alpha}>20\ \m{\AA}$.
   The black circles with the black line show the result of \citet{2015ApJ...799..183S}.
   The green and blue lines are the results of \citet{2014ApJ...791L..25S} and \citet{2012ApJ...752...66L}, respectively.
   The dashed lines represent extrapolations from the ranges these studies investigate.
         \label{fig_MS}}
\end{figure*}

We estimate the ionizing photon escape fraction of the $z=4.9$ LAEs from the H$\alpha$ flux and the SED fitting result, following a method in \citet{2010ApJ...724.1524O}.
We can measure the ionizing photon production rate with the zero escape fraction, $N_\m{obs}(\m{H^0})$, from Equation (\ref{eq_nh0}).
On the other hand, we can estimate $N(\m{H^0})$ from the SED fitting.
Thus the ionizing photon escape fraction is
\begin{equation}
f_\m{esc}^\m{ion}=1-\frac{N_\m{obs}(\m{H^0})}{N(\m{H^0})}.
\end{equation}
We estimate $f_\m{esc}^\m{ion}$ only for the subsample of $EW^0_\m{Ly\alpha}>20\ \m{\AA}$, whose SED is well determined.
We plot $f_\m{esc}^\m{ion}$ of our $z=4.9$ LAEs in the left panel of Figure \ref{fig_MS}.
The estimated escape fraction is $f_\m{esc}^\m{ion}\sim0.10$, which is comparable to local Lyman continuum emitters \citep{2016Natur.529..178I,2016MNRAS.461.3683I,2017A&A...597A..13V,2017MNRAS.469.3252P}.
Note that this estimate largely depends on the stellar population model.
\citet{2016MNRAS.456..485S} report that binary star populations produce a higher number of ionizing photons, exceeding the single-star population flux by $50-60\%$.
Thus we take the $60\%$ systematic uncertainty into account, resulting in the escape fraction of $f_\m{esc}^\m{ion}=0.10\pm0.06$.
The validity of this method will be tested in future work.

\begin{figure*}
 \begin{center}
  \includegraphics[clip,bb=0 0 850 290,width=1\hsize]{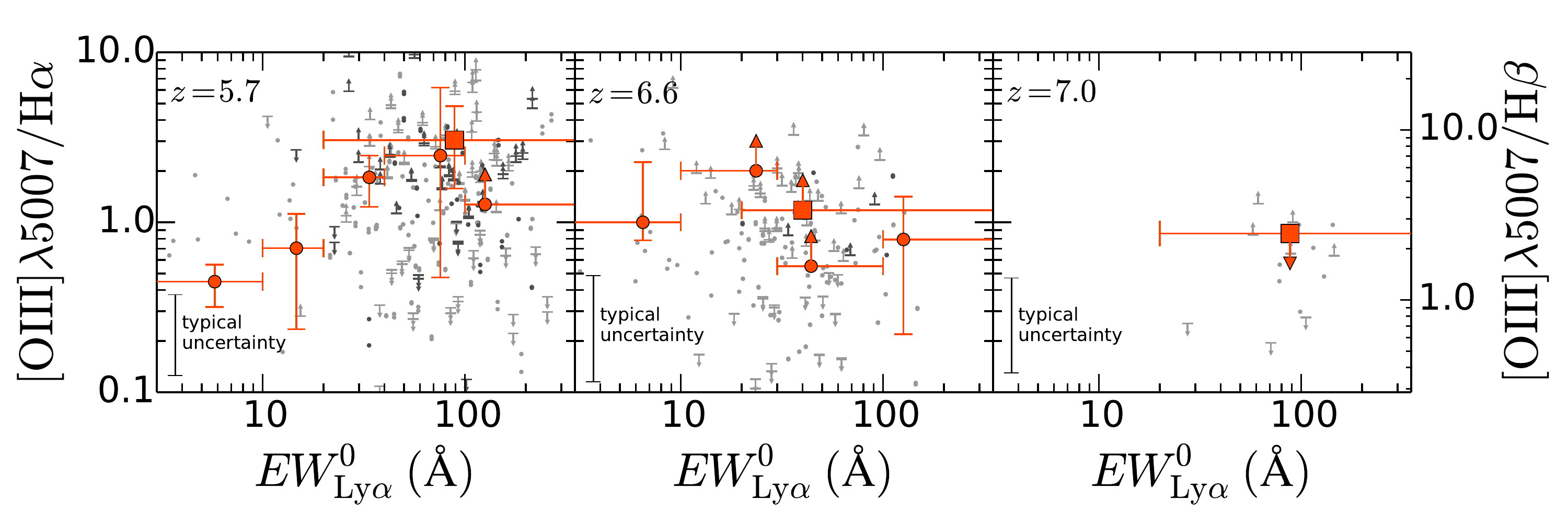}
 \end{center}
   \caption{Inferred {\sc [Oiii]$\lambda$5007}$/\m{H\alpha}$ and {\sc [Oiii]$\lambda$5007}$/\m{H\beta}$ flux ratios as a function of rest-frame Ly$\m{\alpha}$ EW at $z=5.7$ (left), $6.6$ (center), and $7.0$ (right).
   The red squares and circles are the results from the stacked images of the subsamples, and the dark and light gray dots show the ratios of the individual objects detected in the $[3.6]$ and/or $[4.5]$ bands which are spectroscopically confirmed and not, respectively. 
   The upward and downward arrows represent $2\sigma$ lower and upper limits, respectively.
      \label{fig_oiiiha}}
\end{figure*}

\subsubsection{Star Formation Main Sequence}\label{ss_mainsequence}
We can derive the SFR and the stellar mass, $M_*$, from the SED fitting.
The SFR, stellar mass, and the specific SFR are estimated to be 
\begin{eqnarray}
\m{log}SFR&=&0.91\pm0.14,\\
\m{log}M_*&=&8.15\pm0.10,\\
\m{log}(SFR/M_*)&=&-7.24\pm0.18,
\end{eqnarray}
respectively, for the $EW^0_\m{Ly\alpha}>20\ \m{\AA}$ subsample, where $SFR$, $M_*$, and $SFR/M_*$ are in units of $M_\odot\ \m{yr^{-1}}$, $M_\odot$, and $\m{yr^{-1}}$, respectively.
We plot the result in in the right panel of Figure \ref{fig_MS}.
At the fixed stellar mass, the SFR of the LAEs is higher than the extrapolation of the relations measured with LBGs \citep{2015ApJ...799..183S,2014ApJ...791L..25S,2012ApJ...752...66L}.
Thus the LAEs may have the higher SFR than other galaxies with similar stellar masses, as also suggested by \citet{2010ApJ...724.1524O} at $z\sim6-7$ and \citet{2016ApJ...817...79H} at $z\sim2$.
However, some studies infer that LAEs are located on the main sequence \citep[e.g.,][]{2017MNRAS.468.1123S,2017arXiv170709373K} at $z\sim2-3$, so further investigation is needed.

\subsection{Properties of $z=5.7$, $6.6$, and $7.0$ LAEs }\label{ss_z576670}
\subsubsection{{\sc[Oiii]}$\lambda$$\mathrm{5007/H\alpha}$ and {\sc[Oiii]}$\lambda$$\mathrm{5007/H\beta}$ Ratios} \label{ss_oiiiha}
We plot the [{\sc Oiii}]$\lambda5007/\m{H\alpha}$ ([{\sc Oiii}]$\lambda5007/\m{H\beta}$) flux ratios of the subsamples and individual LAEs at $z=5.7$ and $6.6$ ($z=7.0$) in Figure \ref{fig_oiiiha}.
The [{\sc Oiii}]$/\m{H\alpha}$ ratios of the subsamples are typically $\sim1$, but vary as a function of $EW^0_\m{Ly\alpha}$.
At $z=5.7$, the ratio increases from $\sim0.5$ to $2.5$ with increasing Ly$\m{\alpha}$ EW from $EW^0_\m{Ly\alpha}=6\ \m{\AA}$ to $80\ \m{\AA}$.
On the other hand at $z=6.6$, the ratio increases with increasing $EW^0_\m{Ly\alpha}$ from $7\ \m{\AA}$ to $20\ \m{\AA}$, then decreases to $\sim130\ \m{\AA}$, showing the turn-over trend at the $2.3\sigma$ confidence level.
The [{\sc Oiii}]$/\m{H\alpha}$ ratio depends on the ionization parameter and metallicity.
The low [{\sc Oiii}]$/\m{H\alpha}$ ratio in the high-EW subsample at $z=6.6$, whose ionization parameter is expected to be high, indicates the low-metallicity in the high-EW subsample.
At $z=7.0$, the {\sc[Oiii]}$\mathrm{/H\beta}$ ratio is lower than $2.8$.
The ratios of individual galaxies are largely scattered, which may be due to varieties of the ionization parameter and metallicity.

\begin{figure}
\begin{center}
  \includegraphics[clip,bb=0 0 330 290,width=1\hsize]{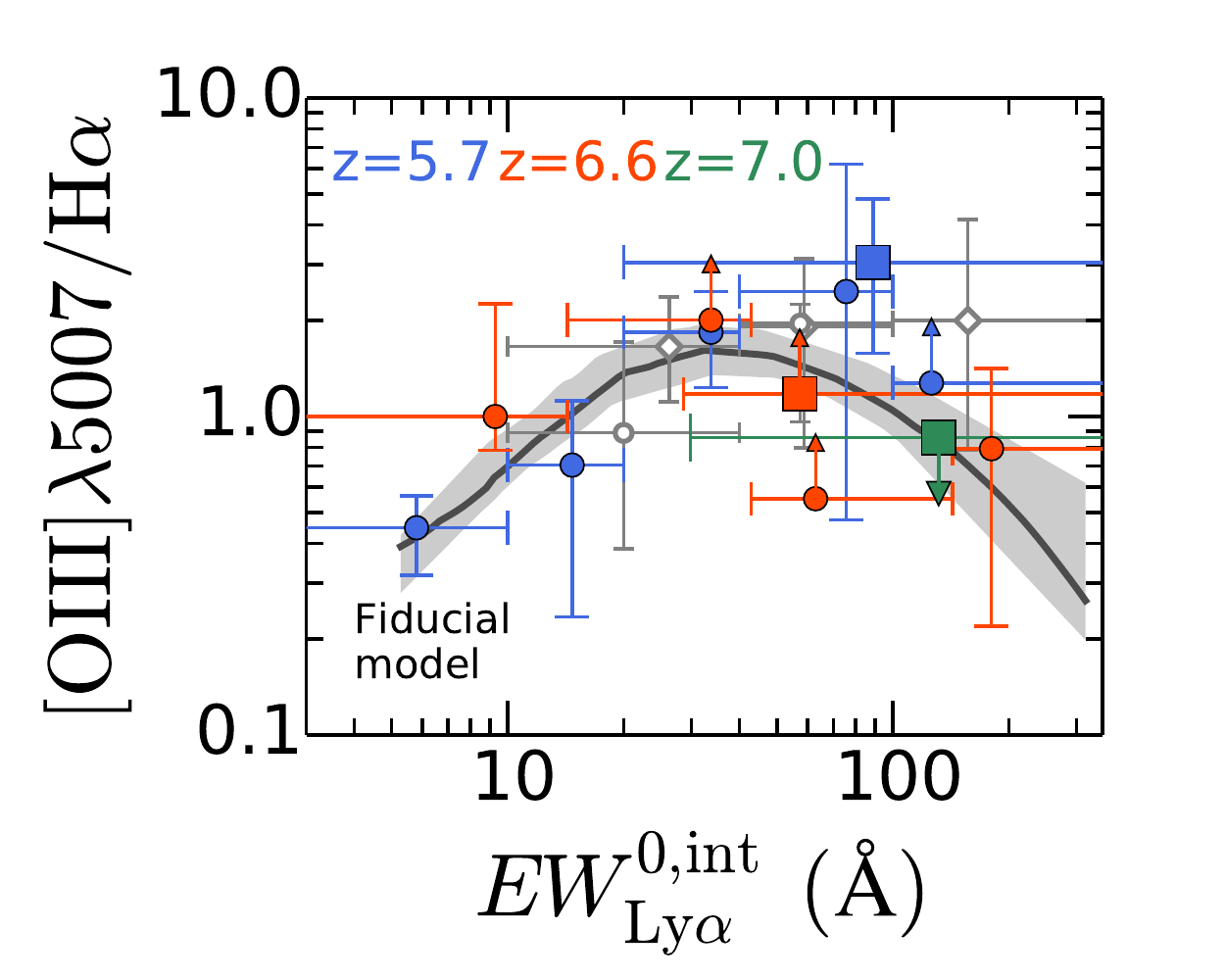}
 \end{center}
   \caption{Same as Figure \ref{fig_oiiiha} but plotted in one figure.
   The blue, red, and green circles and squares are the {\sc [Oiii]$\lambda$5007}$/\m{H\alpha}$ flux ratios at $z=5.7$, $6.6$, and $7.0$, respectively.
   The squares represent the results of the $EW_\m{Ly\alpha}>20\ \m{\AA}$ subsamples.
   The open gray diamonds and circles are the ratios of $z=2.5$ and $0.3$ galaxies \citep{2016ApJ...832..171T,2011ApJ...738..136C}, respectively.
   We plot the median and the $1\sigma$ scatters of the ratios in $EW^0_\m{Ly\alpha}$ subsamples.
   We also plot the fitting result of the $(Z,\m{log}U,\m{Age})-EW^0_\m{Ly\alpha}$ relations with the dark gray curve with the shaded region representing the $1\sigma$ uncertainty.
   See Section \ref{ss_metal} for more details about the fitting.
      \label{fig_oiiiha_all}}
\end{figure}

We compare the [{\sc Oiii}]$/\m{H\alpha}$ flux ratios at $z=5.7$, $6.6$, and $7.0$.
The [{\sc Oiii}]/H$\alpha$ ratio of the $z=7.0$ subsample is estimated from the [{\sc Oiii}]/H$\beta$ ratio assuming the Case B recombination ($\m{H\alpha}/\m{H}\beta=2.86$).
Here we use the intrinsic Ly$\alpha$ EW, $EW^\m{0,int}_\m{Ly\alpha}$, which is corrected for the IGM absorption.
\citet{2018PASJ...70S..16K}, \citet{2017ApJ...844...85O}, and \citet{2014ApJ...797...16K} measure the IGM transmission, $T_\m{Ly\alpha}^\m{IGM}$, at $z=6.6$, $7.0$, and $7.3$, relative to the one at $z=5.7$, as $T_\m{Ly\alpha,z=6.6}^\m{IGM}/T_\m{Ly\alpha,z=5.7}^\m{IGM}=0.70$, $T_\m{Ly\alpha,z=7.0}^\m{IGM}/T_\m{Ly\alpha,z=5.7}^\m{IGM}=0.62$, and $T_\m{Ly\alpha,z=7.3}^\m{IGM}/T_\m{Ly\alpha,z=5.7}^\m{IGM}=0.29$, respectively.
Thus we estimate the intrinsic rest-frame Ly$\alpha$ EW, $EW_\m{Ly\alpha}^\m{0,int}$, from the observed rest-frame Ly$\alpha$ EW, $EW_\m{Ly\alpha}^\m{0}$ at a given redshift, by interpolating these measurements as follows:
\begin{eqnarray}
z<5.7:&&EW_\m{Ly\alpha}^\m{0,int}=EW_\m{Ly\alpha}^\m{0},\label{eq_IGM_57}\\
5.7<z<6.6:&&EW_\m{Ly\alpha}^\m{0,int}=EW_\m{Ly\alpha}^\m{0}/(2.90-0.33z),\label{eq_IGM_5766}\\
6.6<z<7.0:&&EW_\m{Ly\alpha}^\m{0,int}=EW_\m{Ly\alpha}^\m{0}/(1.20-0.08z),\label{eq_IGM_6670}\\
7.0<z<7.3:&&EW_\m{Ly\alpha}^\m{0,int}=EW_\m{Ly\alpha}^\m{0}/(9.54-1.27z),\label{eq_IGM_7073}\\
7.3<z:&&EW_\m{Ly\alpha}^\m{0,int}=EW_\m{Ly\alpha}^\m{0}/0.29.\label{eq_IGM_73}
\end{eqnarray}
The typical uncertainty of these corrections is $20\%$ \citep{2018PASJ...70S..16K}.

The [{\sc Oiii}]$/\m{H\alpha}$ ratios of the $z=5.7$, $6.6$, and $7.0$ LAEs are presented in Figure \ref{fig_oiiiha_all}.
We do not find significant redshift evolution of the ratio from $z=5.7$ to $7.0$ after the IGM correction.
We also plot the ratios of galaxies at $z=2.5$ and $0.3$ from \cite{2016ApJ...832..171T} and \citet{2011ApJ...738..136C}, respectively, by making subsamples based on the Ly$\alpha$ EW.
The ratios of the $z=0-2.5$ galaxies are also comparable to our results at $z=5.7-7.0$ \citep[see also][]{2017arXiv170906092W}.

\subsubsection{Metallicity-$EW_\m{Ly\alpha}$ Anti-correlation}\label{ss_metal}
We investigate physical quantities explaining our observed [{\sc Oiii}]$/\m{H\alpha}$ ratios as a function of Ly$\alpha$ EW.
We simply parameterize the metallicity, $Z_\m{neb}$, the ionization parameter, $U_\m{ion}$, and the stellar age with the Ly$\alpha$ EW in units of $\m{\AA}$ as follows:
\begin{eqnarray}
&&\m{log}Z_\m{neb}=a_Z(\m{log}EW^\m{0,int}_\m{Ly\alpha})^2+b_Z,\label{eq_paramZ}\\
&&\m{log}U_\m{ion}=a_U\m{log}EW^\m{0,int}_\m{Ly\alpha}+b_U,\label{eq_paramU}\\
&&\m{log}\m{Age}=a_A\m{log}EW^\m{0,int}_\m{Ly\alpha}+b_A\label{eq_paramA},
\end{eqnarray}
where $Z_\m{neb}$ and $\m{Age}$ are in units of $Z_\odot$ and $\m{yr}$, respectively.
We find that the quadratic function of Equation (\ref{eq_paramZ}) can describe the observed [{\sc Oiii}]$/\m{H\alpha}$ results better than a linear function.
With these equations, we calculate the metallicity, the ionization parameter, and the stellar age for given Ly$\alpha$ EW.
BEAGLE can predict [{\sc Oiii}]$/\m{H\alpha}$ ratio for the parameter sets of $(Z_\m{neb}, U_\m{ion}, \m{Age})$.
We fit our observational results of the [{\sc Oiii}]$/\m{H\alpha}$ ratios with this model, and constrain the parameters in Equation (\ref{eq_paramZ})-(\ref{eq_paramA}) using the Markov Chain Monte Carlo (MCMC) parameter estimation technique.
Here we assume the electron density of $n_\m{e}=100\ \m{cm^{-1}}$, and this assumption does not have significant impacts on our analysis.
Because the critical density of {[\sc Oiii]}$\lambda$5007 is very high, $6.4\times10^5\ \m{cm^{-3}}$, the [{\sc Oiii}]$/\m{H\alpha}$ ratio does not significantly change in the observed range of the electron density \citep[e.g., $n_\m{e}\sim100-1000\ \m{cm^{-3}}$;][]{2016ApJ...822...42O,2015MNRAS.451.1284S,2016ApJ...816...23S,2017ApJ...835...88K}.

\begin{figure}
\begin{center}
  \includegraphics[clip,bb=0 0 300 585,width=1\hsize]{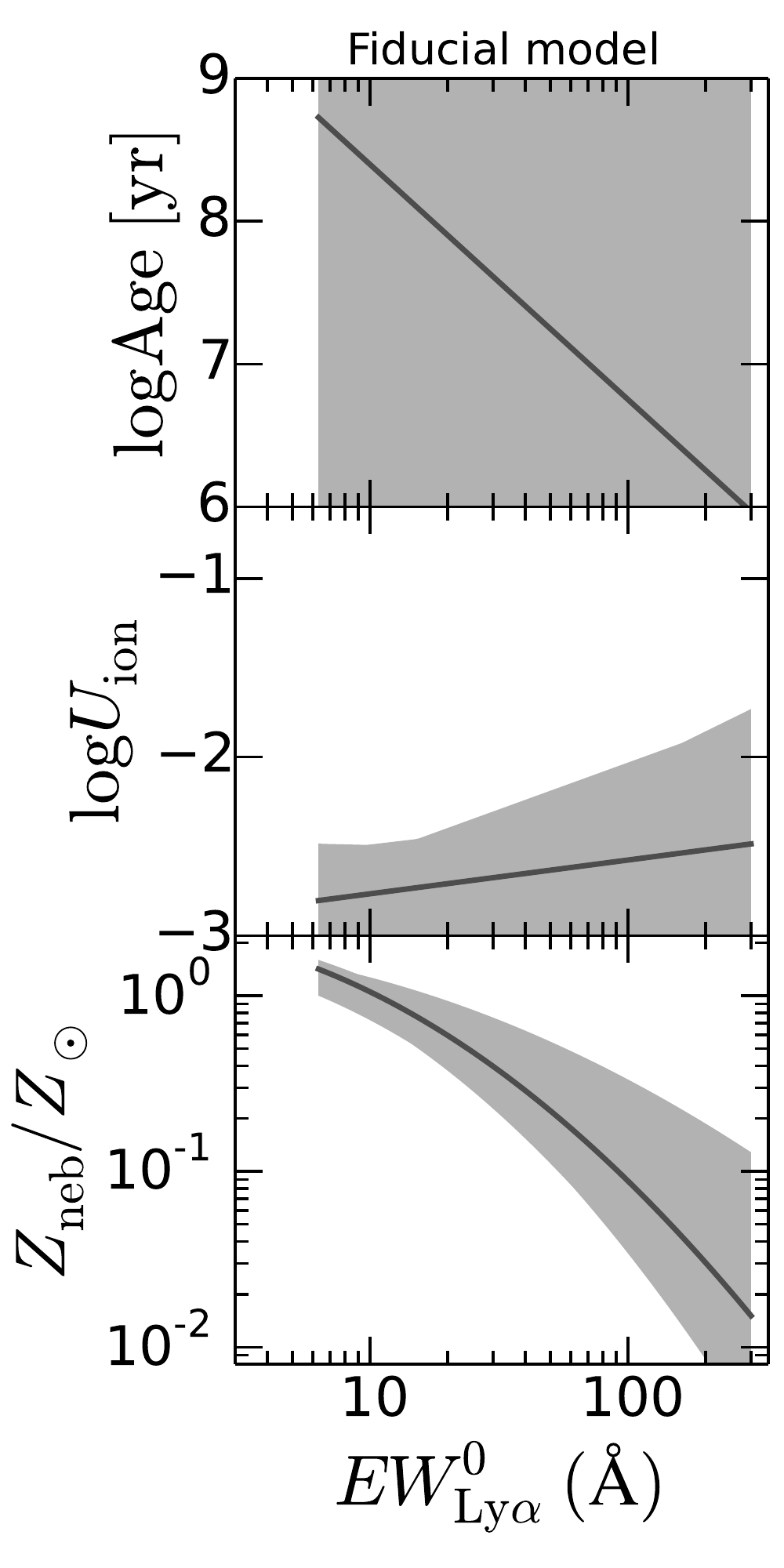}
 \end{center}
   \caption{The allowed parameter ranges of the stellar age (the upper panel), the ionization parameter (the center panel), and the metallicity (the lower panel) constrained with the {\sc [Oiii]$\lambda$5007}$/\m{H\alpha}$ ratios (Figure \ref{fig_oiiiha_all}).
   The dark gray curves with the shaded regions show the best-fit relations and their $1\sigma$ uncertainties, respectively.
   The metallicity-Ly$\alpha$ EW relation is constrained well, while the ionization parameter and stellar age are not.
      \label{fig_params}}
\end{figure}

The results are presented in Figure \ref{fig_params}.
The best-fit relations with $1\sigma$ errors are
\begin{eqnarray}
&&\m{log}Z_\m{neb}=-0.36^{+0.17}_{-0.11}(\m{log}EW^\m{0,int}_\m{Ly\alpha})^2+0.38^{+0.10}_{-0.19},\label{eq_paramZres}\\
&&\m{log}U_\m{ion}=0.19^{+0.52}_{-0.41}\m{log}EW^\m{0,int}_\m{Ly\alpha}-2.96^{+0.61}_{-0.54},\\
&&\m{log}\m{Age}=-1.65^{+3.59}_{-0.09}\m{log}EW^\m{0,int}_\m{Ly\alpha}+10.04^{+0.14}_{-5.52}\label{eq_paramAres}.
\end{eqnarray}
Although the ionization parameter and the stellar age are not well determined, we have constrained the metallicity-Ly$\alpha$ EW relation well.
The result suggests an anti-correlation between the metallicity and the Ly$\alpha$ EW, implying the very metal-poor ISM ($\sim0.03\ Z_\odot$) in the galaxies with $EW^\m{0,int}_\m{Ly\alpha}\sim200\ \m{\AA}$ \citep[see also;][]{2007A&A...468..877N,2017MNRAS.465.1543H}.
This anti-correlation is supported by results of \citet{2016ApJ...822...29F} based on the rest-frame UV absorption lines, \redc{and consistent with a lower limit for the metallicity of $z=2.2$ LAEs in \citet{2012ApJ...745...12N}.}

\begin{figure}
\begin{center}
  \includegraphics[clip,bb=0 0 420 360,width=1\hsize]{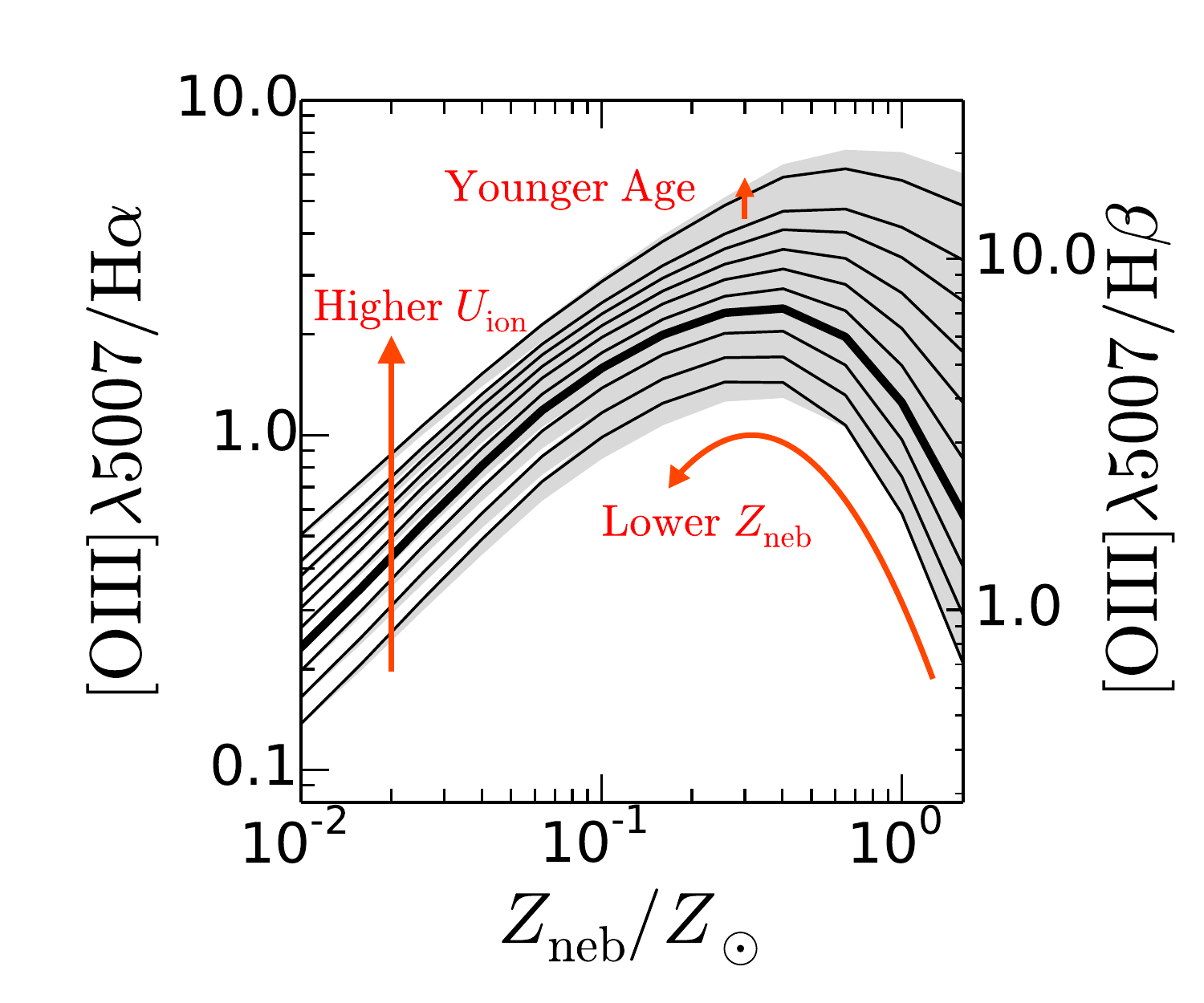}
 \end{center}
   \caption{Predicted {\sc [Oiii]$\lambda$5007}/$\m{H\alpha}$ flux ratio as a function of metallicity.
   The thick black curve is the predicted ratio with the photoionization model of $\m{log}U_\m{ion}=-2.4$ and $\m{log}(\m{Age/yr})=7$.
   The dark gray curves shows the ratios with the models of $-3.0<\m{log}U_\m{ion}<-1.0$ and $\m{log}(\m{Age/yr})=7$ with a $0.2\ \m{dex}$ step in the ionization parameter.
   The light gray shaded region represents ratios with $6<\m{log}(\m{Age/yr})<9$.
   The {\sc [Oiii]}/$\m{H\alpha}$ ratio strongly depends on the metallicity and the ionization parameter, but not so strongly on the stellar age.
      \label{fig_model}}
\end{figure}

\begin{figure*}
\begin{center}
  \begin{minipage}{0.48\hsize}
 \begin{center}
  \includegraphics[clip,bb=15 10 340 290,width=1\hsize]{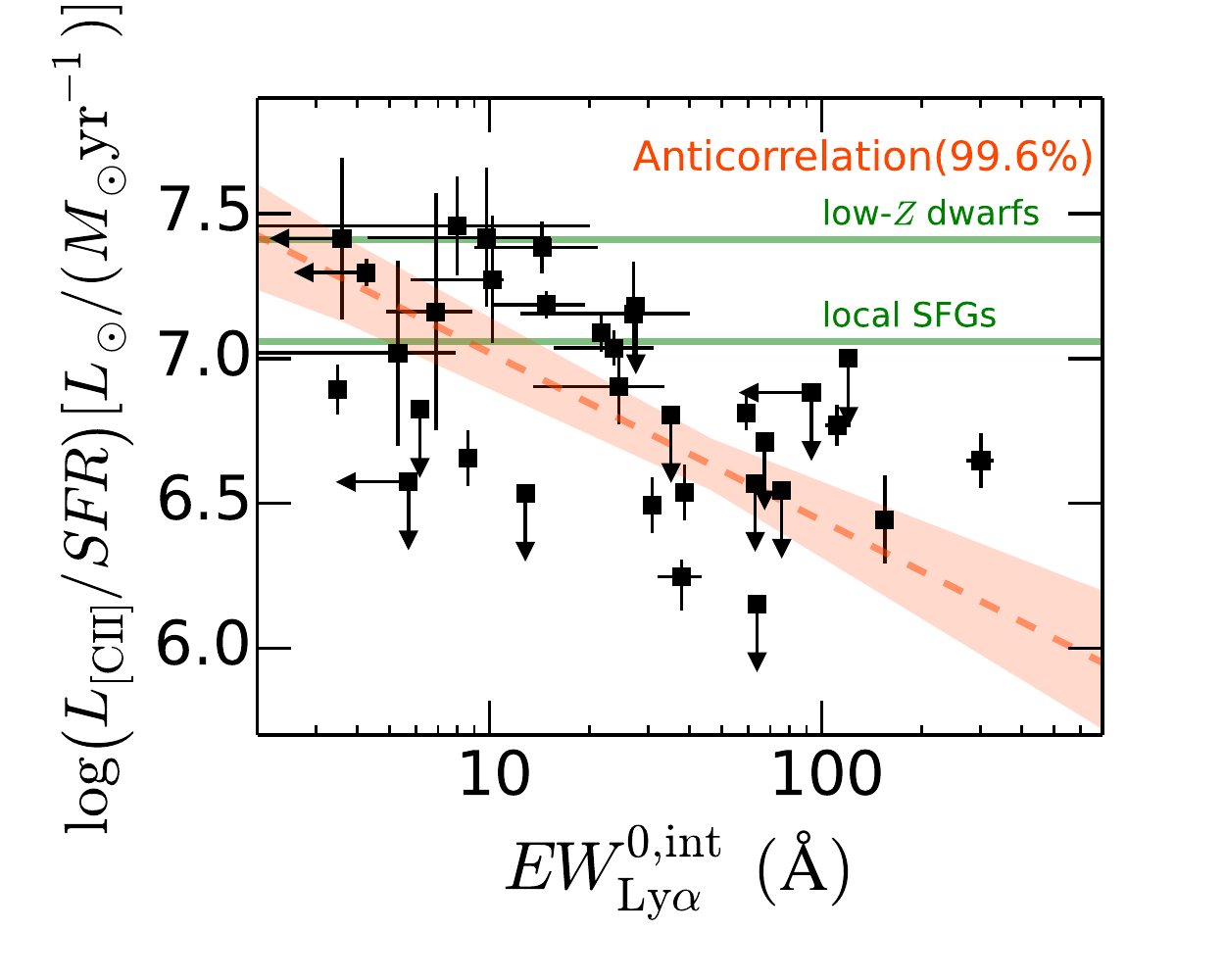}
 \end{center}
 \end{minipage}
 \begin{minipage}{0.48\hsize}
 \begin{center}
  \includegraphics[clip,bb=15 10 340 290,width=1\hsize]{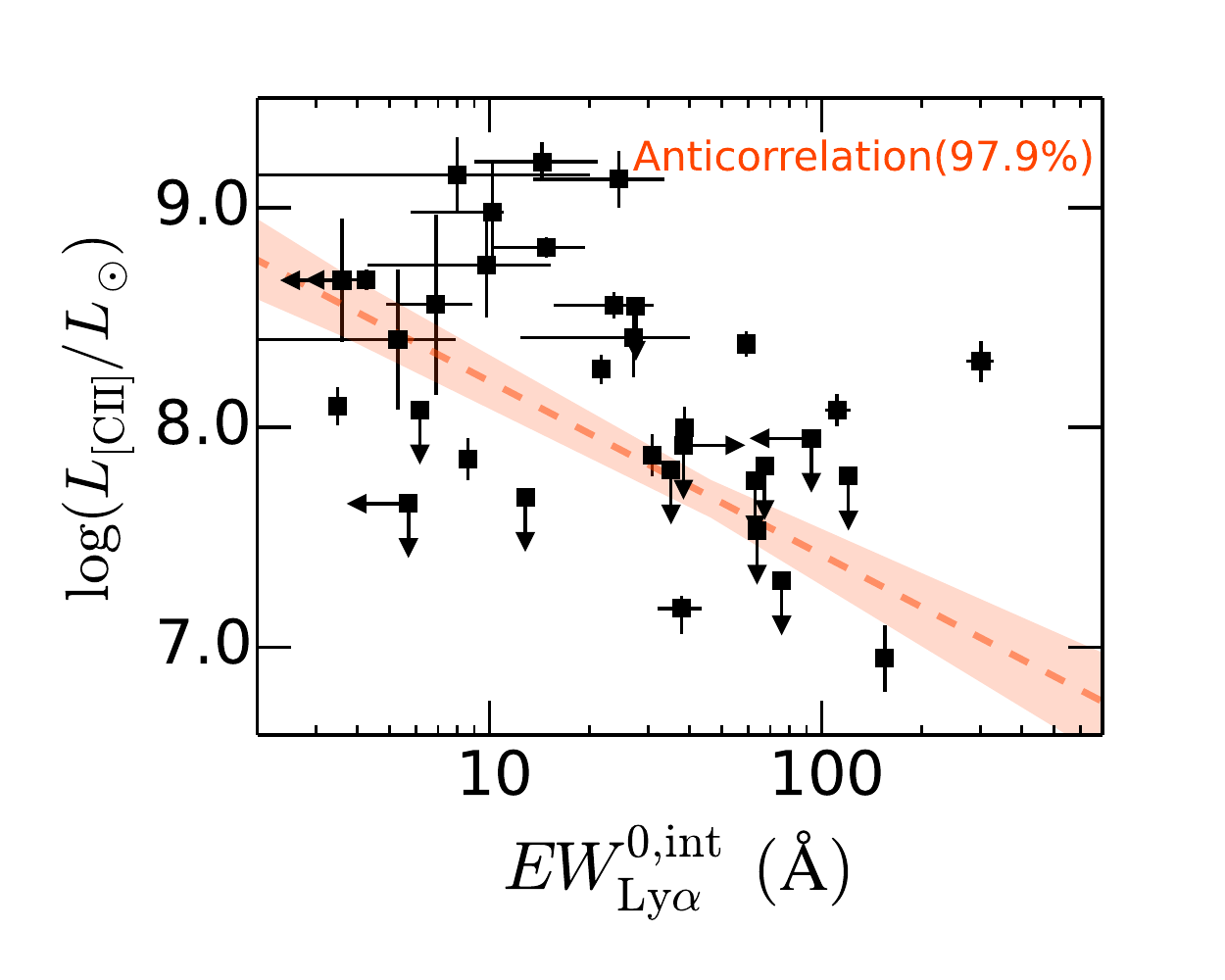}
 \end{center}
 \end{minipage}
 \end{center}
   \caption{
   {\bf Left panel:} Ratio of the {\sc [Cii]} luminosity to the SFR as a function of rest-frame Ly$\m{\alpha}$ EW.
   We plot the results of the previous ALMA and PdBI observations of $z>5$ galaxies (see Table \ref{tab_cii}).
   The SFR is the total star formation rate as $SFR=SFR_\m{UV}+SFR_\m{IR}$. 
   We find the anti-correlation in the $L_\m{[CII]}/SFR-EW^\m{0,int}_\m{Ly\alpha}$ plane at the $99.6\%$ confidence level.
   The green horizontal lines show the $L_\m{[CII]}/SFR$ ratios for low-metallicity dwarf galaxies and local star-forming galaxies in \citet{2014A&A...568A..62D} for $SFR=10\ \m{M_\odot yr^{-1}}$.
   The red-dashed line and the shaded region denote the best-fit $L_\m{[CII]}/SFR-EW^{0,\m{int}}_\m{Ly\alpha}$ relation.
   {\bf Right panel:} Same as the left panel but for the {\sc [Cii]} luminosity. The confidence level of the anti-correlation is $97.9\%$.
      \label{fig_cii}}
\end{figure*}

These results can be understood as follows.
We find the turn-over trend of the [{\sc Oiii}]$/\m{H\alpha}$ ratio with increasing $EW^\m{0,int}_\m{Ly\alpha}$.
This turn-over trend can be reproduced only by the metallicity change, if we assume that the quantities of $(Z_\m{neb}, U_\m{ion}, \m{Age})$ are simple monotonic functions of $EW^\m{0,int}_\m{Ly\alpha}$.
Figure \ref{fig_model} shows the BEAGLE calculations of [{\sc Oiii}]$/\m{H\alpha}$ as a function of metallicity, in parameter ranges of $-2.0<\m{log}(Z_\m{neb}/Z_\odot)<0.2$, $-3.0<\m{log}U_\m{ion}<-1.0$, and $6.0<\m{log}(\m{Age/yr})<9.1$.
At fixed ionization parameter and stellar age, the ratio increases with decreasing metallicity from $\sim1$ to $0.4\ Z_\odot$, and then decreases with metallicity from $\sim0.4$ to $0.01\ Z_\odot$, making the turn-over trend similarly seen in the [{\sc Oiii}]/H$\alpha$-$EW^{0,\m{int}}_\m{Ly\alpha}$ plane \citep[see also; ][]{2006A&A...459...85N,2008A&A...488..463M}.
On the other hand, the ratio monotonically increases with increasing ionization parameter.
The ratio does not significantly depend on the stellar age ($<0.1\ \m{dex}$), since the number of the ionizing photon saturates at $\m{Age}\gtrsim10\ \m{Myr}$.
Thus we can constrain the $Z_\m{neb}-EW^\m{0,int}_\m{Ly\alpha}$ relation from the observed [{\sc Oiii}]$/\m{H\alpha}$ ratios.

\subsection{{\sc [Cii]158}$\m{\mu m}$-Ly$\alpha$ Relation}\label{ss_cii}
In Figure \ref{fig_cii}, we plot the observed ratios of the {\sc [Cii]} luminosity to SFR, $L_\m{[CII]}/SFR$, and {\sc [Cii]} luminosities as functions of Ly$\alpha$ EW corrected for the IGM absorption, $EW^\m{0,int}_\m{Ly\alpha}$.
\redc{We conduct the Kendall correlation test using the {\tt cenken} function in the NADA library from the R-project statistics package, and find anti-correlations in both $L_\m{[CII]}/SFR-EW^\m{0,int}_\m{Ly\alpha}$ and $L_\m{[CII]}-EW^\m{0,int}_\m{Ly\alpha}$ planes at $2.9\sigma$ ($99.6\%$) and $2.3\sigma$ ($97.9\%$) significance levels, respectively.}
The $L_\m{[CII]}/SFR$ ratio of local star-forming galaxies is $\mathrm{log}(L_\m{[CII]}/SFR)/[\m{L_\odot/(M_\odot\ yr^{-1})}]\simeq7$ \citep{2014A&A...568A..62D}.
We find that the typical $L_\m{[CII]}/SFR$ ratio of the galaxies at $z>5$ with $EW^\m{0,int}_\m{Ly\alpha}\sim100\ \m{\AA}$ is lower than those of the local star-forming galaxies by a factor of $\sim3$, indicating the {\sc [Cii]} deficit.
Thus we statistically confirm the {\sc [Cii]} deficit in high $EW^\m{0,int}_\m{Ly\alpha}$ galaxies for the first time.
We discuss physical origins of the $L_\m{[CII]}/SFR-EW^\m{0,int}_\m{Ly\alpha}$ anti-correlation in Section \ref{ss_fiducial}.
In Figure \ref{fig_cii}, we also plot the following power-law functions:
\begin{eqnarray}
&&\m{log}(L_\m{[CII]}/SFR)=-0.58\ \m{log}EW_\m{Ly\alpha}^{0,\m{int}}+7.6,\\
&&\m{log}L_\m{[CII]}=-0.79\ \m{log}EW_\m{Ly\alpha}^{0,\m{int}}+9.0,
\end{eqnarray}
where $L_\m{[CII]}$, $SFR$, and $EW_\m{Ly\alpha}^{0,\m{int}}$ are in units of $L_\odot$, $M_\odot\ \m{yr^{-1}}$, and $\m{\AA}$, respectively.
\redc{These anti-correlations cannot be explained only by the SFR difference of the galaxies, because there is no significant trend between $EW_\m{Ly\alpha}^{0,\m{int}}$ and SFR in our sample.
We divide our sample into subsamples of $EW^{0,\m{int}}_\m{Ly\alpha}=0-10$, $10-100$, and $100-1000\ \m{\AA}$, and find that median SFRs of subsamples are comparable (within a factor of $\sim2$).}
\redc{\citet{2017arXiv171203985C} also report the anti-correlation in the $L_\m{[CII]}/SFR-EW^\m{0,int}_\m{Ly\alpha}$ plane, although the slope is shallower than ours.
As discussed in \citet{2017arXiv171203985C}, their shallower slope is probably due to the fact that they use individual subcomponents extracted from galaxies.
Using subcomponents allows us to investigate physical properties of each clump, but could make the correlation weaker if accuracy of the measurements are not enough.}

\section{Discussion}\label{ss_discussion}

\subsection{Fiducial Model Reproducing the Ly$\alpha$, {\sc [Oiii]}, H$\alpha$, and {\sc [Cii]}}\label{ss_fiducial}
We have constrained the relations of $Z_\m{neb}-EW^\m{0,int}_\m{Ly\alpha}$, $U_\m{ion}-EW^\m{0,int}_\m{Ly\alpha}$, and $\m{Age}-EW^\m{0,int}_\m{Ly\alpha}$ from the {[\sc Oiii]}/H$\alpha$ ratios in Section \ref{ss_metal}.
Hereafter, we call this model ``the fiducial model".
In this section, we investigate whether the fiducial model can also reproduce our other observational results.

In Section \ref{ss_EWHa}, we find that the H$\alpha$ EW positively correlates with the Ly$\alpha$ EW at $z=4.9$.
The H$\alpha$ EW depends on the metallicity and the stellar age, as shown in the right panel in Figure \ref{fig_Ha}.
Since we do not have a good constraint on the stellar age, as shown in Figure \ref{fig_params}, we assume $\m{log(Age/yr)}=6.5$ and $6<\m{log(Age/yr)}<7$ as the best value and the uncertainty, respectively, which are typical for LAEs \citep{2010ApJ...724.1524O,2010MNRAS.402.1580O} and are consistent with Equation (\ref{eq_paramA}).
We derive the parameter sets of $(Z_\m{neb},U_\m{ion},\m{Age})$ given $EW^\m{0,int}_\m{Ly\alpha}$ from this fiducial model, and calculate H$\alpha$ EWs using BEAGLE.
In the left panel in Figure \ref{fig_Ha}, we plot the prediction of the fiducial model with the dark gray curve.
We find that the fiducial model agrees well with the observed $EW^0_\m{H\alpha}-EW^0_\m{Ly\alpha}$ relation.

\begin{figure}
 \begin{center}
  \includegraphics[clip,bb=15 10 340 290,width=1\hsize]{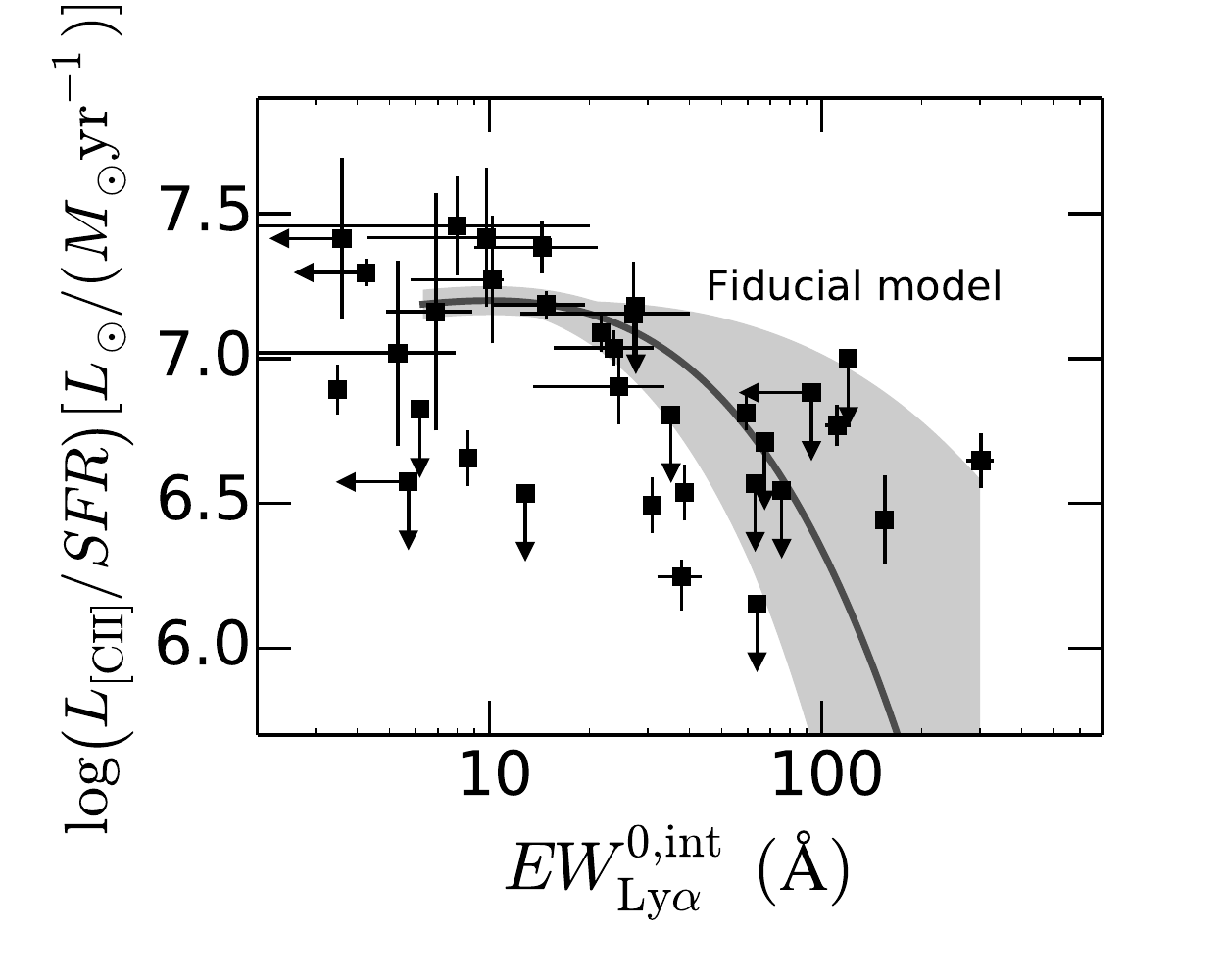}
 \end{center}
   \caption{Same as the left panel in Figure \ref{fig_cii} but with the prediction from the fiducial model.
   The dark gray curve and the shaded region represent the prediction from the fiducial model and its $1\sigma$ uncertainty, respectively, with the $L_\m{[CII]}/SFR-Z_\m{neb}$ relation from \citet[][Equation (\ref{eq_V15})]{2015ApJ...813...36V}.
   See Section \ref{ss_fiducial} for more details.
         \label{fig_ciimodel}}
\end{figure}

The ratio of $L_\m{[CII]}/SFR$ negatively correlates with the Ly$\alpha$ EW, as shown in Section \ref{ss_cii}.
\citet{2015ApJ...813...36V} present the following formula describing their simulation results:
\begin{eqnarray}\label{eq_V15}
&&\m{log}(L_\m{[CII]}/SFR)=7.0+0.2\m{log}SFR+0.021\m{log}Z_\m{neb}\notag\\
&&\ \ \ \ \ \ \ +0.012\m{log}SFR\ \m{log}Z_\m{neb}-0.74(\m{log}Z_\m{neb})^2,
\end{eqnarray}
where $L_\m{[CII]}$, $SFR$, and $Z_\m{neb}$ are in units of $L_\odot$, $M_\odot\ \m{yr^{-1}}$, and $Z_\odot$, respectively \citep[see also][]{2018A&A...609A.130L}.
By substituting Equation (\ref{eq_paramZres}) to this equation, we can obtain an $L_\m{[CII]}/SFR-EW^\m{0,int}_\m{Ly\alpha}$ relation.
Here we assume the typical SFR of our sample, $SFR=10\ \m{M_\odot\ \m{yr}^{-1}}$, but the choice of the SFR does not have a significant impact on the discussion.
We plot this $L_\m{[CII]}/SFR-EW^\m{0,int}_\m{Ly\alpha}$ relation (i.e., the fiducial model) with the dark gray curve in Figure \ref{fig_ciimodel}.
The fiducial model can nicely explains the $L_\m{[CII]}/SFR-EW^\m{0,int}_\m{Ly\alpha}$ anti-correlation, indicating that the {\sc [Cii]} deficit in high Ly$\alpha$ EW galaxies may be due to the low metallicity.

There are two other possibilities for the $L_\m{[CII]}/SFR-EW^\m{0,int}_\m{Ly\alpha}$ anti-correlation.
One is the density bounded nebula in high-redshift galaxies.
Recently there is growing evidence that high-redshift galaxies have high ionization parameters with intense radiation \citep{2014MNRAS.442..900N}.
Such intense radiation ionizes {\sc Cii} and {\sc Hi} in the {\sc Hi} and photo-dissociation region, making the density bounded nebula, decreasing the {\sc [Cii]} emissivity, and increasing the transmission of Ly$\alpha$ \citep[see also discussions in][]{2015ApJ...813...36V}.
Thus if the ionizing parameter positively correlates with the Ly$\alpha$ EW, we could explain the $L_\m{[CII]}/SFR-EW^\m{0,int}_\m{Ly\alpha}$ anti-correlation by this density bounded nebula scenario.
The other is a very high density of photo dissociation region (PDR).
The critical density of the {\sc[Cii]} line is $\sim3000\ \m{cm^{-3}}$.
A very high density PDR ($>3000\ \m{cm^{-3}}$) enhances more rapid collisional de-excitations for the forbidden {\sc[Cii]} line, and decreases the {\sc[Cii]} emissivity.
\redcrr{Indeed, \citet{2013ApJ...774...68D,2014ApJ...788L..17D} report an anti-correlation between the {\sc[Cii]} to FIR luminosity ratio ($L_\m{[CII]}/L_\m{FIR}$) and the FIR luminosity surface density ($\Sigma_\m{FIR}$) for local starburst galaxies, which may be due to high ionization parameters or collisional de-excitations in high $\Sigma_\m{FIR}$ galaxies \citep{2016ApJ...826..112S}.}
Although we find that the $Z_\m{neb}-EW_\m{Ly\alpha}^{0,\m{int}}$ relation can explain the $L_\m{[CII]}/SFR-EW^\m{0,int}_\m{Ly\alpha}$ anti-correlation, these two scenarios are still possible.

\begin{figure*}
\begin{center}
  \includegraphics[clip,bb=40 30 600 330,width=0.95\hsize]{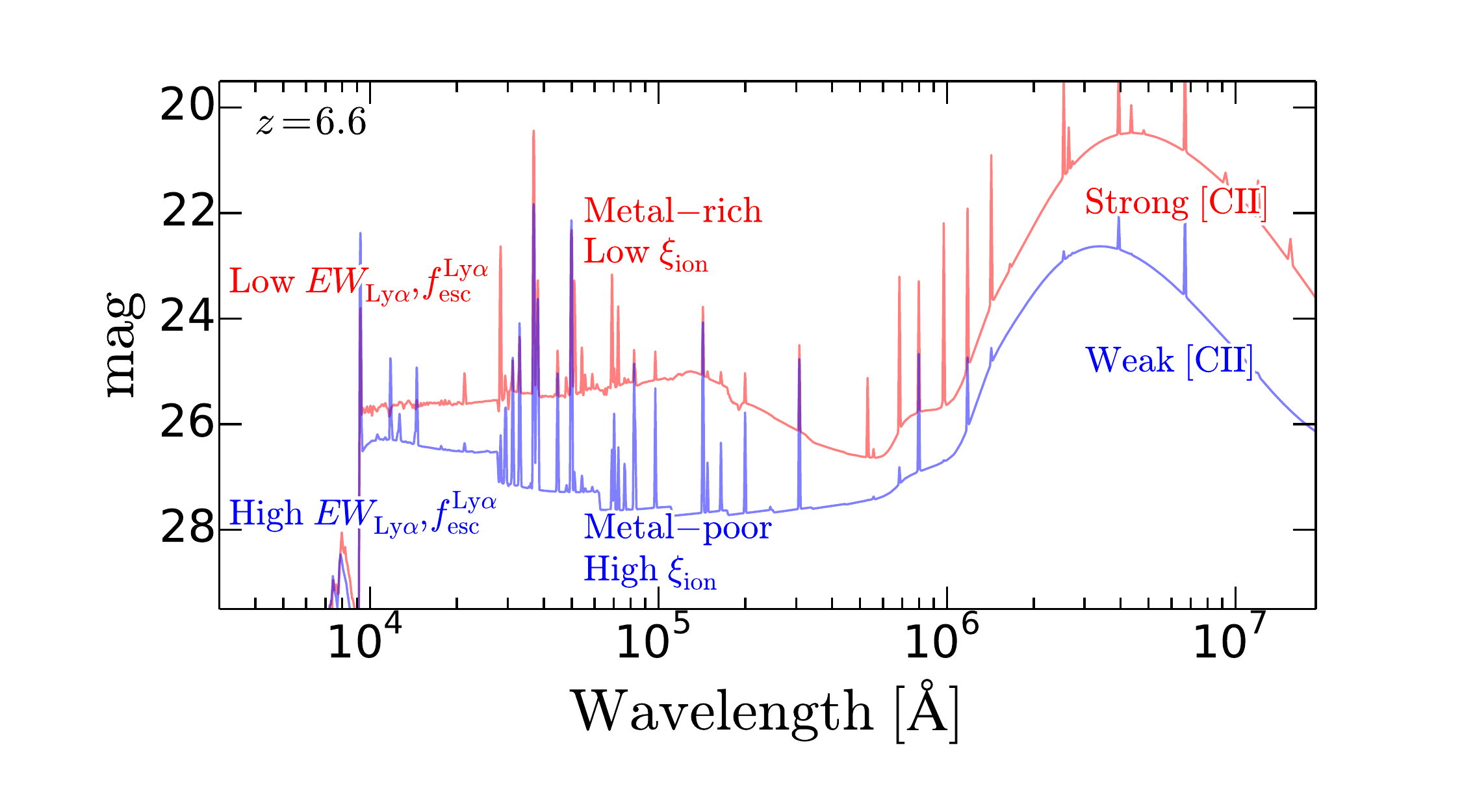}
 \end{center}
   \caption{Schematic figure summarizing our findings.
   We find that galaxies with high (low) Ly$\alpha$ EWs have the high (low) Ly$\alpha$ escape fractions, and are metal-poor (metal-rich) with the high (low) ionizing photon production efficiencies and the weak (strong) {\sc [Cii]} emission.
      \redc{The blue and red curves show model SEDs of star forming galaxies with ($\m{log}(Z_\m{neb}/Z_\odot), \m{log}U_\m{ion}, \m{log(Age/yr)})=(-2,-2,6)$ and $(0,-3,8)$ generated by BEAGLE, respectively.}
   \label{fig_sedEM_result}}
\end{figure*}

Nevertheless, we find that the predictions from the fiducial model with the $Z_\m{neb}-EW_\m{Ly\alpha}^{0,\m{int}}$ anti-correlation agree well with our observational results of the H$\alpha$ EW and $L_\m{[CII]}/SFR$.
These good agreements suggest a picture that galaxies with high (low) Ly$\alpha$ EWs have the high (low) Ly$\alpha$ escape fractions, and are metal-poor (metal-rich) with the high (low) ionizing photon production efficiencies and the weak (strong) {\sc [Cii]} emission (Figure \ref{fig_sedEM_result}).

\begin{figure}
 \begin{center}
  \includegraphics[clip,bb=0 5 360 300,width=1\hsize]{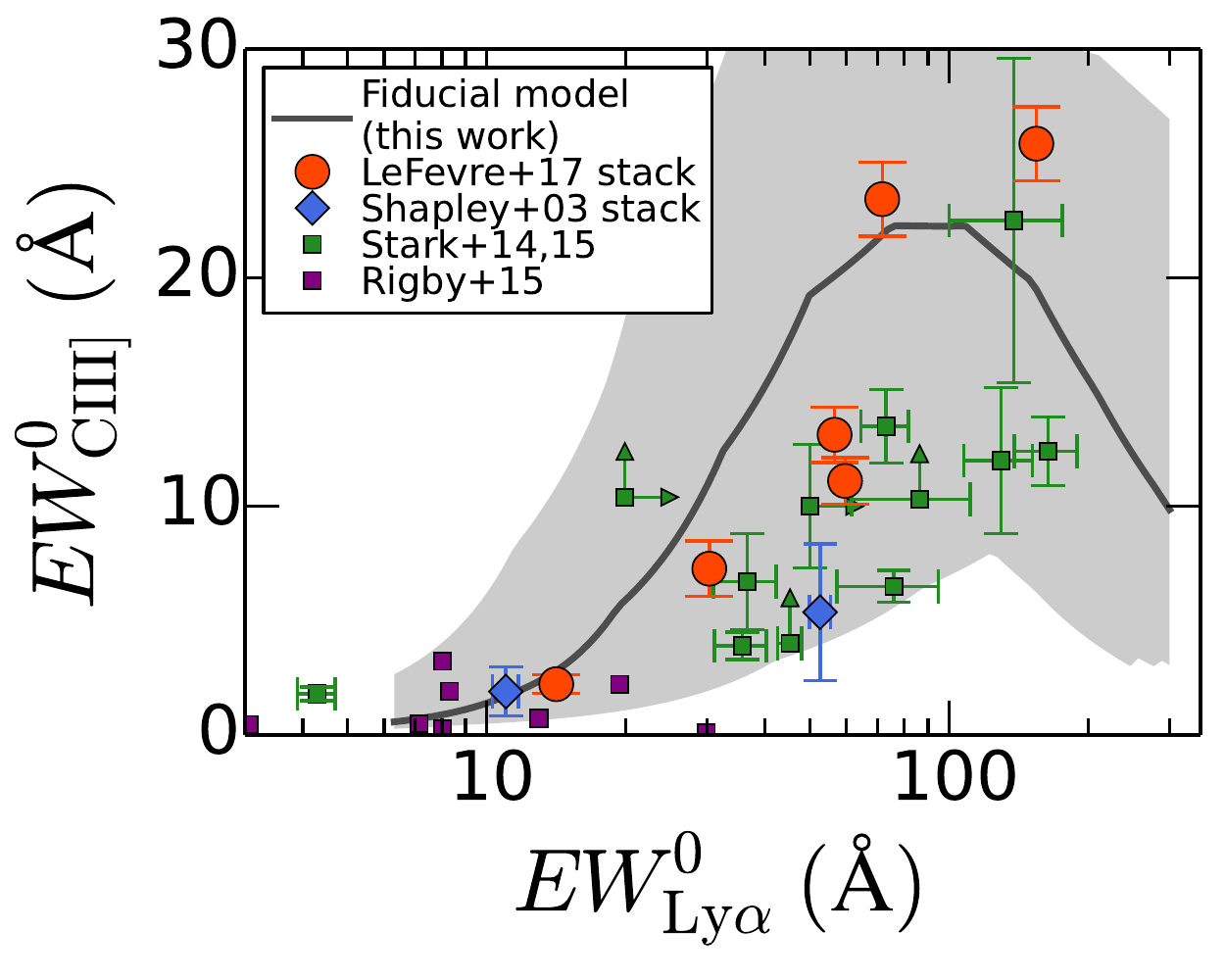}
 \end{center}
   \caption{Predicted {\sc Ciii]}$\lambda\lambda$1907,1909 EW as a function of Ly$\alpha$ EW.
      The dark gray curve and the shaded region represent the prediction from the fiducial model and its $1\sigma$ uncertainty, respectively.
      The other symbols show results from observations.
      The red circle and blue diamond are the stacked results from \citet{2017arXiv171010715L} and \citet{2003ApJ...588...65S}, respectively.
      Data points of individual galaxies are represented with the green and purple squares \citep{2014MNRAS.445.3200S,2015ApJ...814L...6R}.
         \label{fig_ciii}}
\end{figure}

\subsection{Predicted {\sc Ciii]}$\lambda\lambda$1907,1909 EWs of LAEs}
The {\sc Ciii]}$\lambda\lambda$1907,1909 lines are believed to be the second most-frequent emission lines in the UV rest-frame spectra of SFGs after Ly$\alpha$.
Recent observations with the VIMOS Ultra-Deep Survey (VUDS) and the MUSE Hubble Ultra Deep Field Survey allow us to investigate statistical properties of the {\sc Ciii]} emission at $1<z<4$ \citep{2017arXiv170903990N,2017arXiv171010715L,2017arXiv171006432M}.
These studies reveal that the {\sc Ciii]} EW ($EW^0_\m{CIII]}$) positively correlates with Ly$\alpha$ EW \citep{2014MNRAS.445.3200S,2017arXiv171010715L}.
Here we investigate whether the fiducial model can reproduce the observed correlation between $EW^0_\m{CIII]}$ and $EW^0_\m{Ly\alpha}$.
Since the {\sc Ciii]} EW depends on the metallicity, ionization parameter, and stellar age, we can predict the {\sc Ciii]} EW with Equations (\ref{eq_paramZres})-(\ref{eq_paramAres}) using BEAGLE.

Figure \ref{fig_ciii} shows predicted {\sc Ciii]} EWs with observational results.
Although the uncertainty is large due to the poor constrains on the ionization parameter and stellar age, the prediction reproduces the positive correlation at $10\ \m{\AA}<EW_\m{Ly\alpha}^0<100\ \m{\AA}$.
Thus LAEs with $EW_\m{Ly\alpha}^0\sim100\ \m{\AA}$ would be strong {\sc Ciii]} emitters.
Beyond $EW_\m{Ly\alpha}^0\sim100\ \m{\AA}$, the {\sc Ciii]} EW decreases due to low carbon abundance, suggested by \citet{2017arXiv170903990N}.

\subsection{Implication for Cosmic Reionization}\label{ss_reio}
In Section \ref{ss_fLya}, we find that the $\m{Ly\alpha}$ escape fraction given the $\m{Ly\alpha}$ EW, $f_\m{Ly\alpha}(EW^0_\m{Ly\alpha})$, does not change significantly with redshift from $z=0$ to $4.9$.
Thus the ratio of the EW to the escape fraction, $EW^0_\m{Ly\alpha}/f_\m{Ly\alpha}(EW^0_\m{Ly\alpha})$, also does not change with redshift.
Since the $\m{Ly\alpha}$ EW and the $\m{Ly\alpha}$ escape fraction are proportional to the ratio of the $\m{Ly\alpha}$ luminosity to the UV luminosity ($EW_\m{Ly\alpha}\propto L_\m{Ly\alpha}/L_\m{UV}$), and of the $\m{Ly\alpha}$ luminosity to the $\m{H\alpha}$ luminosity ($f_\m{Ly\alpha}\propto L_\m{Ly\alpha}/L_\m{H\alpha}$), the ratio of the EW to the escape fraction is proportional to the ionizing photon production efficiency, as follows:
\begin{equation}
\frac{EW^0_\m{Ly\alpha}}{f_\m{Ly\alpha}(EW^0_\m{Ly\alpha})}\propto\frac{L_\m{Ly\alpha}/L_\m{UV}}{L_\m{Ly\alpha}/L_\m{H\alpha}}\propto\frac{L_\m{H\alpha}}{L_\m{UV}}\propto\xi_\m{ion}.
\end{equation}
Thus the redshift-independent $f_\m{Ly\alpha}-EW^0_\m{Ly\alpha}$ relation indicates that the ionizing photon production efficiency depends on the Ly$\m{\alpha}$ EW, but not on the redshift.
If this is redshift-independent even at $z>5$, galaxies in the reionization epoch have $\xi_\m{ion}$ values comparable to those of LAEs at lower redshift with similar Ly$\alpha$ EWs.
Thus this redshift-independent $f_\m{Ly\alpha}-EW^0_\m{Ly\alpha}$ relation justifies studies of low-redshift analogs to understand physical properties of the ionizing sources at the epoch of the cosmic reionization.


We discuss the contributions of star-forming galaxies to the cosmic reionization based on the results of the $z=4.9$ LAEs.
If we assume that the faint star-forming galaxies at the reionization epoch have similar properties to the $EW_\m{Ly\alpha}^0>20\ \m{\AA}$ LAEs at $z=4.9$, the ionizing photon production efficiency is $\m{log}\xi_\m{ion}/\m{[Hz\ erg^{-1}]}\simeq25.53^{+0.06}_{-0.06}$ and the ionizing photon escape fraction is $f_\m{esc}^\m{ion}\sim0.10$.
Based on recent UV LFs measurements, the ionizing photon budget is explained only by star-forming galaxies if $\m{log}f^\m{ion}_\m{esc}\xi_\m{ion}/\m{[Hz\ erg^{-1}]}=24.52^{+0.14}_{-0.07}$ \citep{2018ApJ...854...73I}.
Our $z=4.9$ LAE results suggest $\m{log}f^\m{ion}_\m{esc}\xi_\m{ion}/\m{[Hz\ erg^{-1}]}=24.53$, indicating that the photon budget can be explained only by the star-forming galaxies, with minor contribution from faint AGNs \citep[see also;][]{2016MNRAS.457.4051K,2017ApJ...847L..15O}.

\section{Summary}\label{ss_summary}
We have investigated ISM properties from 1,092 LAEs at $z=4.9$, $5.7$, $6.6$ and $7.0$, using wide and deep mid-infrared images obtained in SPLASH and stellar-synthesis and photoionization models.
The deep {\it Spitzer} data constrain the strengths of the rest-frame optical emission lines which are not accessible from the ground telescopes at $z>4$.
In addition, we study the connection between the Ly$\alpha$ emission and [{\sc Cii}]158$\m{\mu m}$ emission using ALMA and PdBI {\sc [Cii]} observations targeting 34 galaxies at $z=5.148-7.508$ in the literature.
Our major findings are summarized below.

\begin{enumerate}

\item
The H$\m{\alpha}$ EW increases with increasing Ly$\m{\alpha}$ EW at $z=4.9$.
The H$\m{\alpha}$ EW of the $0\ \m{\AA}<EW_\m{Ly\alpha}^0<20\ \m{\AA}$ subsample is $\sim600\ \m{\AA}$, relatively higher than the results of $M_*\sim10^{10}\ M_\odot$ galaxies.
On the other hand, the H$\m{\alpha}$ EW of the $EW_\m{Ly\alpha}^0>70\ \m{\AA}$ subsample is higher than $\sim1900\ \m{\AA}$, indicating the very young stellar age of $<10\ \m{Myr}$ or the very low metallicity of $<0.02\ Z_\odot$.
{\it Figure \ref{fig_Ha}; Section \ref{ss_EWHa}}

\item
We find that the Ly$\alpha$ escape fraction, $f_\m{Ly\alpha}$, positively correlates with the Ly$\alpha$ EW, $EW^0_\m{Ly\alpha}$, at $z=4.9$.
This $f_\m{Ly\alpha}-EW^0_\m{Ly\alpha}$ relation does not show redshift evolution at $z=0-4.9$, indicating that the ionizing photon production efficiency depends on the Ly$\alpha$ EW, but not on the redshift.
This result justifies the studies of low-redshift analogs to understand physical properties of the ionizing sources at the epoch of the cosmic reionization.
{\it Figure \ref{fig_fLya}; Sections \ref{ss_fLya} \& \ref{ss_reio}}

\item
The ionizing photon production efficiency of the $EW^0_\m{Ly\alpha}>20\ \m{\AA}$ LAE subsample is $\m{log}\xi_\m{ion}/[\m{Hz\ erg^{-1}}]=25.5$, significantly higher than those of LBGs at the similar redshift and UV magnitude, as well as than the canonical values by $60-100\ \%$.
The ionizing photon escape fraction is estimated to be $f_\m{esc}^\m{ion}\sim0.1$.
From our measured $\xi_\m{ion}$ and $f_\m{esc}^\m{ion}$, we find that the ionizing photon budget for the reionization can be explained by the star-forming galaxies, if they have similar properties to our LAEs at $z=4.9$.
{\it Figure \ref{fig_xi}; Sections \ref{ss_xi}, \ref{ss_fesc} \& \ref{ss_reio}}

\item
We estimate [{\sc Oiii}]$\lambda$$5007/\m{H\alpha}$ flux ratios of the LAEs as a function of Ly$\m{\alpha}$ EW at $z=5.7$ and $6.6$.
At $z=5.7$, the ratio increases from $0.5$ to $2.5$ with increasing Ly$\m{\alpha}$ EW from $EW^0_\m{Ly\alpha}=6\ \m{\AA}$ to $80\ \m{\AA}$.
On the other hand at $z=6.6$, the ratio increases with increasing $EW^0_\m{Ly\alpha}$ from $7\ \m{\AA}$ to $20\ \m{\AA}$, then decreases to $130\ \m{\AA}$, showing the turn-over trend at the $2.3\sigma$ confidence level.
The [{\sc Oiii}]$\lambda$$5007/\m{H\beta}$ flux ratio of the $z=7.0$ LAEs is lower than $2.8$.
All of the observed [{\sc Oiii}]$/\m{H\alpha}$ and [{\sc Oiii}]$/\m{H\beta}$ ratios can be understood by the anti-correlation between the metallicity and the Ly$\alpha$ EW.
High Ly$\alpha$ EW ($\sim200\ \m{\AA}$) subsamples are expected to be very metal-poor, $Z_\m{neb}\sim0.03\ Z_\odot$.
{\it Figures \ref{fig_oiiiha}, \ref{fig_oiiiha_all}, \& \ref{fig_params}; Section \ref{ss_z576670}}

\item
We find anti-correlations in both $L_\m{[CII]}/SFR-EW^0_\m{Ly\alpha}$ and $L_\m{[CII]}-EW^0_\m{Ly\alpha}$ planes at the $99.6\%$ and $97.9\%$ confidence levels, respectively.
This is the first time to statistically confirm the {[\sc Cii]} deficit in high $EW_\m{Ly\alpha}^0$ galaxies.
{\it Figure \ref{fig_cii}; Section \ref{ss_cii}}

\item
We find that the fiducial model with the $Z_\m{neb}-EW_\m{Ly\alpha}^\m{0,int}$ anti-correlation can explain the results of the $EW^0_\m{H\alpha}-EW^0_\m{Ly\alpha}$ and $L_\m{[CII]}/SFR-EW^\m{0,int}_\m{Ly\alpha}$ relations.
These good agreements suggest a picture that galaxies with high (low) Ly$\alpha$ EWs have the high (low) Ly$\alpha$ escape fractions, and are metal-poor (metal-rich) with the high (low) ionizing photon production efficiencies and the weak (strong) {\sc [Cii]} emission.
{\it Figures \ref{fig_Ha}, \ref{fig_oiiiha_all}, \& \ref{fig_ciimodel}; Section \ref{ss_fiducial}}

\end{enumerate}

\acknowledgments
We thank the anonymous referee for a careful reading and valuable comments that improved the clarity of the paper.
We are grateful to Richard Ellis, Anna Feltre, Max Gronke, Toshihiro Kawaguchi, Kimihiko Nakajima, Kazuhiro Shimasaku, and David Sobral for their useful comments and discussions.
We thank Stefano Carniani for providing their data points.

The Hyper Suprime-Cam (HSC) collaboration includes the astronomical communities of Japan and Taiwan, and Princeton University.  The HSC instrumentation and software were developed by the National Astronomical Observatory of Japan (NAOJ), the Kavli Institute for the Physics and Mathematics of the Universe (Kavli IPMU), the University of Tokyo, the High Energy Accelerator Research Organization (KEK), the Academia Sinica Institute for Astronomy and Astrophysics in Taiwan (ASIAA), and Princeton University.  Funding was contributed by the FIRST program from Japanese Cabinet Office, the Ministry of Education, Culture, Sports, Science and Technology (MEXT), the Japan Society for the Promotion of Science (JSPS),  Japan Science and Technology Agency  (JST),  the Toray Science  Foundation, NAOJ, Kavli IPMU, KEK, ASIAA,  and Princeton University.

The Pan-STARRS1 Surveys (PS1) have been made possible through contributions of the Institute for Astronomy, the University of Hawaii, the Pan-STARRS Project Office, the Max-Planck Society and its participating institutes, the Max Planck Institute for Astronomy, Heidelberg and the Max Planck Institute for Extraterrestrial Physics, Garching, The Johns Hopkins University, Durham University, the University of Edinburgh, Queen's University Belfast, the Harvard-Smithsonian Center for Astrophysics, the Las Cumbres Observatory Global Telescope Network Incorporated, the National Central University of Taiwan, the Space Telescope Science Institute, the National Aeronautics and Space Administration under Grant No. NNX08AR22G issued through the Planetary Science Division of the NASA Science Mission Directorate, the National Science Foundation under Grant No. AST-1238877, the University of Maryland, and Eotvos Lorand University (ELTE).

This paper makes use of software developed for the Large Synoptic Survey Telescope. We thank the LSST Project for making their code available as free software at http://dm.lsst.org.

This work is supported by World Premier International Research Center Initiative (WPI Initiative), MEXT, Japan, and KAKENHI  (15H02064) Grant-in-Aid for Scientific Research (A) through Japan Society for the Promotion of Science (JSPS).
Y.H. acknowledges support from the Advanced Leading Graduate Course for Photon Science (ALPS) grant and the JSPS through the JSPS Research Fellowship for Young Scientists.
JC and SC acknowledge support from the European Research Council (ERC) via an Advanced Grant under grant agreement no. 321323-NEOGAL.
ST acknowledge support from the ERC Consolidator Grant funding scheme (project ConTExt, grant number 648179). The Cosmic Dawn Center is funded by the Danish National Research Foundation.

\bibliographystyle{apj}
\bibliography{apj-jour,reference}

\end{document}